\documentclass{elsart}

\usepackage[square,comma]{natbib}
\usepackage{graphicx}
\usepackage{pxfonts}
\usepackage{lineno}

\usepackage{amssymb}
\usepackage{footnote}
\usepackage{multirow}
\usepackage{varwidth}
\usepackage{array}
\usepackage{epstopdf}

\journal{}

\begin{document}

\thispagestyle{empty}
\begin{Large}
\textbf{DEUTSCHES ELEKTRONEN-SYNCHROTRON}

\textbf{\large{Ein Forschungszentrum der
Helmholtz-Gemeinschaft}\\}
\end{Large}

DESY 14-137

August 2014

\begin{eqnarray}
\nonumber
\end{eqnarray}
\begin{center}
\begin{Large}
\textbf{Perspectives of Imaging of Single Protein Molecules with the Present Design of the European XFEL. \\ $~~$ \\ - Part I - \\  $~~$ \\ X-ray Source, Beamlime Optics and Instrument Simulations. $~~$ \\ $~~$ \\}
\end{Large}

\begin{large}
Svitozar Serkez, Vitali Kocharyan, Evgeni Saldin, Igor Zagorodnov,
\end{large}
\textsl{\\Deutsches Elektronen-Synchrotron DESY, Hamburg}

\begin{large}
$~$ \\ Gianluca Geloni,
\end{large}
\textsl{\\European XFEL GmbH, Hamburg}

\begin{large}
$~$ \\ and Oleksandr Yefanov
\end{large}
\textsl{\\Center for Free-Electron Laser Science, Hamburg}

\begin{eqnarray}
\nonumber
\end{eqnarray}
ISSN 0418-9833
\begin{eqnarray}
\nonumber
\end{eqnarray}
\begin{large}
\textbf{NOTKESTRASSE 85 - 22607 HAMBURG}
\end{large}
\end{center}
\clearpage
\newpage

\begin{frontmatter}



\title{Perspectives of Imaging of Single Protein Molecules with the Present Design of the European XFEL. \\ $~~$ \\ $~~$ \\ - Part I - \\ $~~$ \\ X-ray Source, Beamlime Optics and Instrument Simulations.}


\author[DESY]{Svitozar Serkez,}
\author[DESY]{Vitali Kocharyan,}
\author[DESY]{Evgeni Saldin,}
\author[DESY]{Igor Zagorodnov,}
\author[XFEL]{Gianluca Geloni,}
\author[CFEL]{and Oleksandr Yefanov}

\address[DESY]{Deutsches Elektronen-Synchrotron (DESY), Hamburg, Germany}
\address[XFEL]{European XFEL GmbH, Hamburg, Germany}
\address[CFEL]{Center for Free-Electron Laser Science, Hamburg, Germany}

\newpage

\begin{abstract}
The Single Particles, Clusters and Biomolecules (SPB) instrument at the European XFEL is located behind the SASE1 undulator, and aims to support imaging and structure determination of biological specimen between about $0.1 \mu$m and $1 \mu$m size. The instrument is designed to work at photon energies from $3$ keV up to $16$ keV. This wide operation range is a cause for challenges to the focusing optics.  In particular, a long propagation distance of about $900$ m between x-ray source and sample leads to a large lateral photon beam size at the optics. The beam divergence is the most important parameter for the optical system, and is largest for the lowest photon energies and for the shortest pulse duration (corresponding to the lowest charge). Due to the large divergence of nominal X-ray pulses with duration shorter than $10$ fs, one suffers diffraction from mirror aperture, leading to a 100-fold decrease in fluence at photon energies around 4 keV, which are ideal for imaging of single biomolecules. The nominal SASE1 output power is about 50 GW. This is very far from the level required for single biomolecule imaging, even assuming perfect beamline and focusing efficiency. Here we demonstrate that the parameters of the accelerator complex and of the SASE1 undulator offer an opportunity to optimize the SPB beamline for single biomolecule imaging with minimal additional costs and time. Start to end simulations from the electron injector at the beginning of the accelerator complex up to the generation of diffraction data indicate that one can achieve diffraction without diffraction with about $0.5$ photons per Shannon pixel at near-atomic resolution with $10^{13}$ photons in a $4$ fs pulse at $4$ keV photon energy and in a $100$ nm focus, corresponding to a fluence of $10^{23} \mathrm{ph}/\mathrm{cm}^2$. This result is exemplified using the RNA Pol II molecule as a case study.

\end{abstract}

%
%
%
\end{frontmatter}



\section{\label{sec:intro} Introduction}

The main expectation, but also the main challenge emerging from applications of XFEL sources to life science is the determination of 3D structures of biomolecules and of their complexes from diffraction images of single particles. Crucial parameters are pulse duration, photon energy and fluence. In order to achieve "diffraction before destruction" one requires pulses containing enough photons to produce measurable diffraction patterns and short enough to outrun radiation damage \cite{HAJD}-\cite{SEIB}. The highest diffraction signals are achieved at the longest wavelength that supports a given resolution, which is desired to be near-atomic. These consideration suggest that the ideal  photon energy range for single biomolecule imaging spans between $3$ keV and $5$ keV \cite{BERG}.

The European XFEL facility, which is presently under construction phase in the Hamburg area, has been designed in such a way that electron bunches can be fed simultaneously at two undulator lines, SASE1 and SASE2 \cite{ETDR,TSCH}.  Additionally, the design of the European XFEL includes one line in the soft X-ray range, SASE3, which follows SASE1 and can produce X-rays either using the spent electron beam from SASE1 or using fresh bunches when the lasing in SASE1 is inhibited \cite{BRIN}. Owing to the high repetition rate of the superconducting driving electron accelerator of European XFEL, these three lines can lase effectively at the same time\footnote{With obvious compromises on the X-ray beam parameters due to the fact that the three lines are fed with electron bunches with the same characteristics}.

The SASE1 undulator beamline will be equipped with an instrument dedicated to bioimaging experiments, named Single Particles, Clusters and Biomolecules (SPB) instrument \cite{MANC,MANT}. It will deal with imaging and structure determination of biological specimen between $0.1 \mu$m and $1 \mu$m size. It is designed to work in the photon energy range spanning between $3$ keV and $16$ keV. This wide operation range is a source of challenges for the focusing optics.  In particular, due to the large propagation distance between X-ray source and focus (about $900$ m), one has a large lateral beam size at the Kirkpatrick-Baez (KB) optics position, which is used for focusing the X-ray beam on the sample. The requirements on the mirror clear aperture becomes most stringent at lower photon energies between $3$ keV and $5$ keV \cite{MANT}.  As remarked above, in order to fulfill "diffraction before destruction" requirements, one needs pulses shorter than $10$ fs. In order to produce these electron pulses exploiting the nominal mode of operation, one needs to use very low charge. However, lower charges are related to smaller emittance, finally resulting in a tight radiation source size and a large divergence. To be specific, at $4$ keV such divergence amounts to $5 \mu$rad. At $3$ keV it would increase up to $6 \mu$rad.  Due to the large divergence of the nominal X-ray pulses one suffers major diffraction effects, leading to a $100$-fold decrease in fluence at photon energies around 4 keV, ideal for single biomolecule imaging.  The nominal SASE1 output power is about 50 GW \cite{TSCH,SY1}. This is very far from the level required for single biomolecule imaging, even assuming perfect beamline and focusing efficiency. To conclude, single biomolecule imaging capabilities at the European XFEL could only be possible in the low photon energy range, but cannot be reached exploiting nominal electron bunch distributions and nominal mode operation for the undulator.

In this study we take advantage of the same method already exploited in \cite{10OU} to get around this obstacle.  We propose a configuration of X-ray source which combines self-seeding \footnote{After the successful operation of the self-seeding setup at the LCLS \cite{AMAN},  it was decided that the same self-seeding scheme should be enabled at the European XFEL. Such scheme is expected to be implemented starting from the early stage of operation of the facility.} and undulator tapering techniques in order to increase the SASE1 output peak power from the baseline value of $50$ GW, up to more than $1.5$ TW. We propose to overcome the issue of SPB focusing efficiency based on a special mode of operation of the accelerator complex, which will allow to reach the required fluence with the present design of SPB beamline optics. Specifically, we propose to take advantage of a slotted foil (also known as emittance spoiler) setup \cite{EMM1}-\cite{DING} in the last electron bunch compressor of the accelerator to achieve, simultaneously, X-ray pulse duration control down to ultra-short duration (a few fs), and small angular divergence\footnote{The use of the slotted foil setup allows one to control the photon bunch duration independently of the electron beam duration, and therefore of the charge. Since higher charge is related to larger emittance, and larger emittance is related to smaller divergence of the photon beam, one can effectively optimize the angular divergence of the photon beam, without sacrifying the duration control.} (about $2 \mu$rad).

Computer simulation and modeling are used as a powerful tool both for design and understanding potential issues along the bio-imaging beamline. In particular, we consider the problem of modeling a single-protein imaging experiment from the generation of the electron bunch, up to the simulation of the sample reconstruction. This problem can be naturally separated in two parts. Part I, described in this work, deals with the simulation of the experiment up to the production of diffraction data, while Part II, which will be published in a separate work, will deal with the data processing leading to the sample reconstruction.

In this article, the electron beam evolution is first followed from the photocathode throughout the accelerator, including the slotted foil setup in the second bunch compressor, up to the entrance of the SASE1 undulator. As a next step we perform X-ray source simulations.  Combining self-seeding \cite{SELF}-\cite{ASYM}, \cite{AMAN} and tapering \cite{TAP1}-\cite{LAST} we demonstrate that X-ray beams can be delivered in $4$ fs-long pulses carrying about $6$ mJ energy each, at photon energies around $4$ keV. The radiation is then transported to the interaction region through the SPB beamline X-ray optics. We show that for the $100$ nm-scaled focus one is expected to achieve a fluence up to  $10^{23}$ ph/cm$^2$.  Simulations of noisy diffraction patterns and interpretation of measurable data follow. We confirm that with such high fluence, using a baseline $20$ cm by $20$ cm Adaptive Gain Integrating Pixel Detector (AGIPD)  \cite{AGIP},  one should be able to reconstruct a molecule of about 10 nm size with a resolution better than $0.4$ nm collecting about a few ten thousands patterns. Explicit simulations are presented using the 15 nm-size RNA Pol II molecule as a case study.

\section{\label{sec:expr} Experiment requirements}

Imaging of single molecules at near-atomic resolution is expected to result in a significant advance in structural biology. In fact, one could obtain structural information of large macromolecular assemblies that cannot crystallize, like membrane proteins. In order perform single molecule imaging, a straightforward  "diffraction before destruction" method has been proposed \cite{HAJD}-\cite{BERG}. First, a great number of single molecules with the same structure are injected into vacuum and interact with ultrashort X-ray pulses. After interacting with a single XFEL pulse a molecule is completely destroyed. Therefore one molecule can only yield a single measurement. The diffraction patterns arising from elastically scattered photons are recorded, all with unknown orientation, by a 2D detector placed downstream of the interaction point before destruction takes place, and contain information in the reciprocal space from which the object can be reconstructed. This process is repeated until a sufficient number of images are recorded to grant reconstruction of the original molecule. Next, the relative orientations of the different images is determined, so that a 3D diffraction pattern can be assembled in the reciprocal space \cite{HULD}-\cite{IKED}. Finally, the 3D electron density of the molecule is obtained by the 3D diffraction pattern with the help of a phase retrieval method.

An important parameter of the problem is the number of scattered photons per effective Shannon pixel. For biological material, the photon count per shot per pixel of solid angle $\Omega_p$, averaged over shells of wavenumber $q$, can be estimated by \cite{GUIN}:

\begin{eqnarray}
\langle N_p \rangle = F \cdot P \cdot r_e^2 N_\mathrm{atom} |f|^2 \Omega_p~,
\label{Npave}
\end{eqnarray}
where $F$ is the photon fluence, $N_\mathrm{atom}$ is the number of non-hydrogen atoms in the molecule, $P \sim 1/2$ is the polarization factor, and $f \sim 7$ is the average atomic scattering factor (e.g. between $C$ and  $O$).  It is assumed that the positions of the atoms are completely uncorrelated, which is approximately true at resolutions approaching the atomic scale. Diffraction from an object of maximum width $w$ produces speckles of size $2\pi/w$ in the reciprocal space. One can therefore define the Shannon interval for an intensity pattern $\Delta q_s = \pi/w$ \cite{CHAS}\footnote{See section \ref{sub:noise} for more details}. In this case we have $\Omega_p \sim \lambda^2/(2\pi)^2 (\Delta q_s)^2 = \lambda^2/(2w)^2$ for a particle of width $w$, and Eq. (\ref{Npave}) becomes

\begin{eqnarray}
\langle N_p \rangle  = F \cdot P \cdot r_e^2 N_\mathrm{atom} |f|^2 \frac{\lambda^2}{4w^2}~.
\label{Npave2}
\end{eqnarray}
This number is proportional to $\lambda^2$, where $\lambda$ is the radiation wavelength. As a result, lower photon energies result in a stronger diffraction  signal. However, the possible decrease in photon energy is limited by the resolution that one needs to achieve. A balance may be reached in the photon energy range between $3$ keV and $5$ keV. One can estimate the minimum number of required photons per XFEL pulse for successful imaging of single biomolecules. One should preliminarily observe that the size of the X-ray beam should be adapted to the size of the sample, in order to optimize the signal level. This is done with the help of suitable focusing optics. The FWHM focal spot size should be roughly between $5$ and $10$ times larger than the sample size. For example, a spot size of 100 nm is good for sample sizes of 10 - 20 nm \cite{BERG}. Using the model described above we find that a biomolecule of around $15$ nm diameter, with $N_\mathrm{atom} \sim 30000$, requires a pulse fluence of about $10^{13} \mathrm{ph}/(100 ~\mathrm{nm})^2$, for an average of $\langle N_p \rangle \sim 1.5$ photons per Shannon pixel at a photon energy of $4$ keV. This signal level is  higher than what is required by method of pattern orientation determination discussed in Section \ref{sec:test}.

From this discussion it is clear that the imaging of single molecules is based on the use of extremely intense and extremely short X-ray pulses to limit radiation-induced changes during the exposure. Only in this way high-quality structural data can be collected before the sample structure itself changes significantly. Damage-induced errors in the diffraction pattern due to atomic motion are described by the so-called $R$ factor, which is function of pulse duration, intensity and photon energy. $R$ provides information on the extent to which the elastically scattered radiation is perturbed by X-ray-induced damage,  and provides a direct assessment of the data quality ($R = 0$ is ideal, a larger value of $R$ means larger errors). Following \cite{ETDR} we consider this damage acceptable if $R$ factor is smaller than $15 \%$. In its turn, this requires a pulse duration shorter than $4$ fs at photon energy $4$ keV \cite{ETDR}. It should be noted, however, that if the X-ray intensity increases above a certain threshold, almost every electron will be stripped from every atom in the sample. In this case, as remarked in \cite{CHAS}: `the turn-off time of diffraction will not be necessarily limited by atomic motion, but the variability in the atomic scattering factors at undisplayed atomic position. This has no spatial correlation, and leads to a diffuse background'. It was recently shown that the electronic density will not be considerably altered by the scattering process up to a fluence of about $10^{13} \mathrm{photons}/(100 \mathrm{nm})^2$ at a photon energy $3$ keV and pulse duration $5$ fs \cite{LORE}. In the same reference it is also shown how a further increase in the fluence will rapidly increase the background contribution to the diffraction pattern. The role of impact ionization in molecules has also been analyzed in \cite{BEAT}, where it is provided an estimate of the pulse duration required to suppress impact ionization during the x-ray pulse. These considerations show that, for example, $10^{13}$ photons focused down to a 100 nm spot size at a photon energy of $4$ keV can be used to achieve near-atomic resolution in single biomolecule imaging for an X-ray pulse duration shorter than about $4$ fs. It is expected that at this timescale, which is shorter than the Auger decay times of light atoms \cite{SHOR}, the radiation damage of the sample through photoionization and subsequent impact ionization processes will be suppressed due to hollow-ion formation \cite{SON1}.

Therefore, one concludes that the key metric  for optimizing a photon source for single biomolecule imaging is the peak power. Ideally, the peak power in our case of interest should be more than 1 TW. In order to motivate this number with an example, we note that $10^{13}$ photons at $4$ keV correspond to an energy of about $6$ mJ which yields, in $4$ fs, a peak power of about 1.5 TW. It is worthwhile to mention that 1 TW at 4 keV gives the same signal per Shannon pixel as 27 TW at 12 keV (assuming a fixed pulse duration).

\section{\label{sec:concept} Concept of an X-ray source for the SPB beamline}

\begin{figure}
\begin{center}
\includegraphics[trim = 50 190 10 70, clip, width=1.0\textwidth]{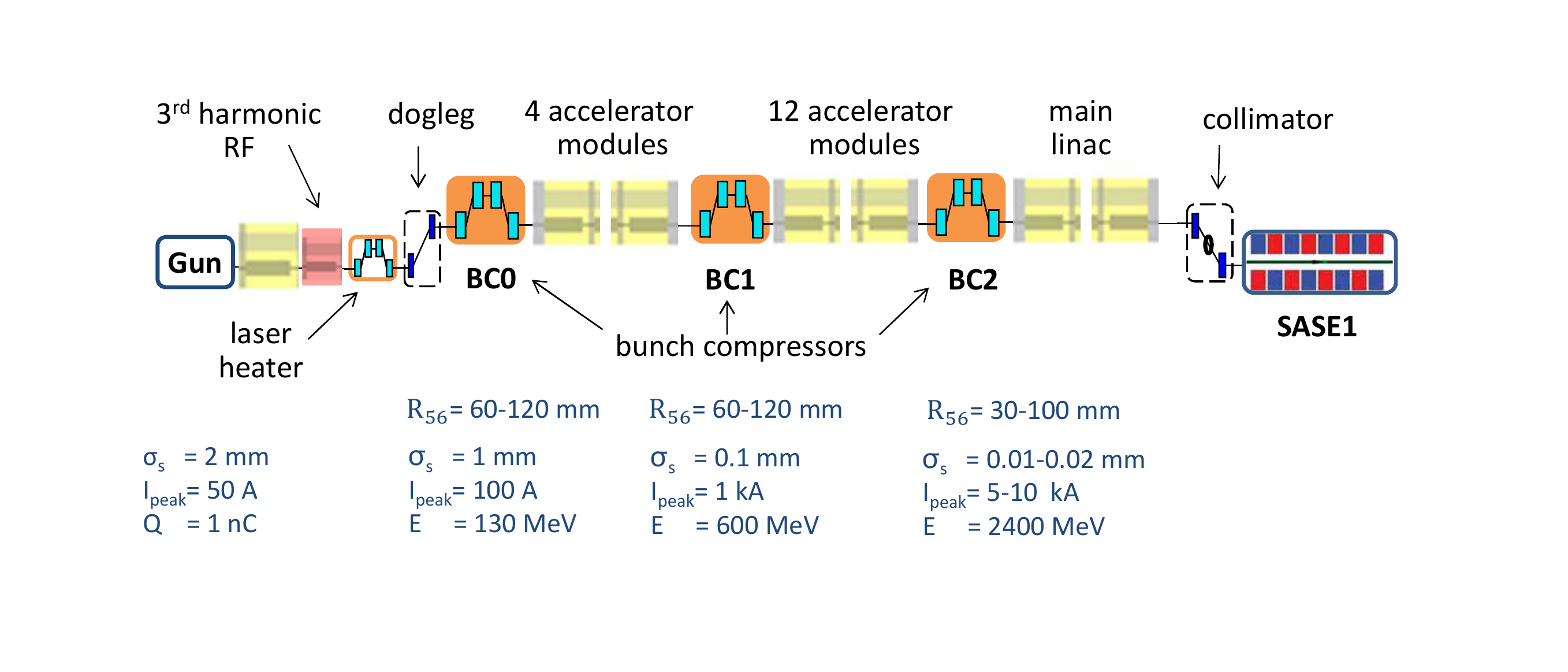}
\end{center}
\caption{Sketch of the European XFEL bunch compression system.} \label{XFELplot}
\end{figure}

The SPB instrument at the European XFEL will be located at the SASE1 undulator line. Fig. \ref{XFELplot} shows this line from the injector up to the SASE1 undulator. Our scheme for an X-ray source suitable for the SPB instrument is heavily based on the use of a slotted spoiler foil in the last bunch compressor chicane. A proposal \cite{EMM1,EMM2} and an experimental verification \cite{DING} have been made in order to generate femtosecond x-ray pulses at the LCLS by using a slotted spoiler foil located in the center of the last bunch compressor. The method takes advantages of the high sensitivity of the FEL gain process to the transverse emittance of the electron bunch. By spoiling the emittance of most of the beam while leaving a short unspoiled temporal slice, one can produce an x-ray FEL pulse much shorter than in the case when the original electron bunch is sent through the undulator.

\begin{figure}
\begin{center}
\includegraphics[trim = 20 600 0 50, clip, width=0.75\textwidth]{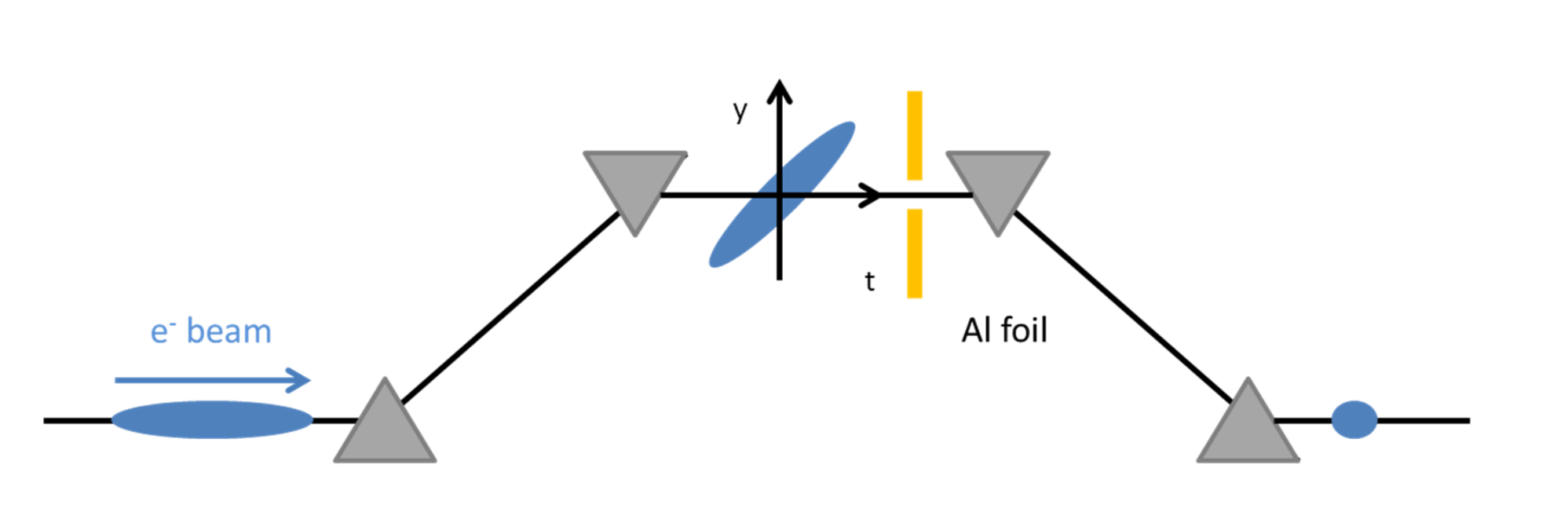}
\end{center}
\caption{Sketch of an electron bunch at the center of  the BC2 magnetic bunch compressor chicane.} \label{slot1}
\end{figure}

\begin{figure}
\begin{center}
\includegraphics[trim = 0 550 0 25, clip, width=0.75\textwidth]{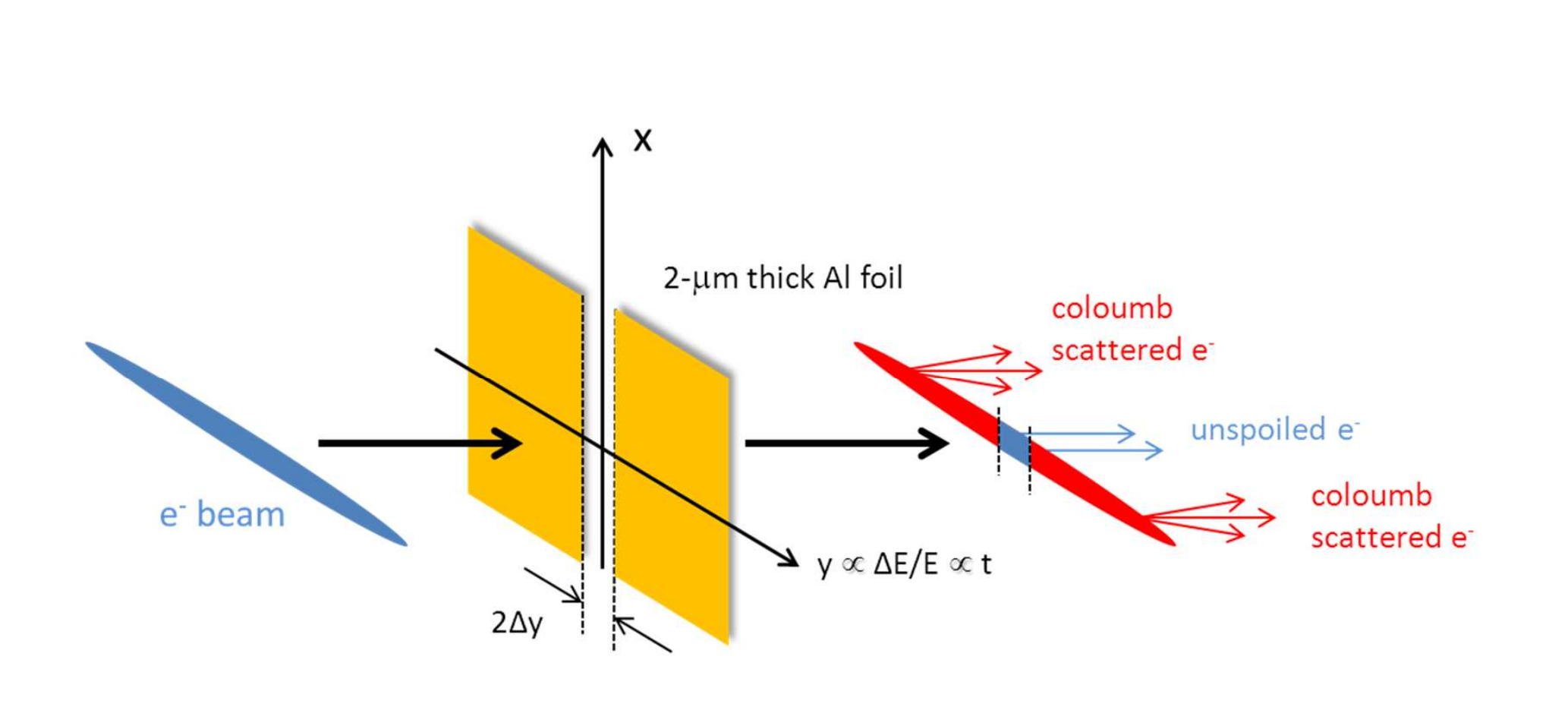}
\end{center}
\caption{ The slotted foil at chicane center generates a narrow, unspoiled beam center (adapted from \cite{EMM2})} \label{slot2}
\end{figure}

Fig. \ref{slot1} shows a sketch of the slotted  foil  at the center of the last bunch compressor BC2 at the European XFEL. The last linac section before the third bunch compressor BC2 is set at an off-crest accelerating rf phase, so that the beam energy at the entrance of BC2 is correlated with time. Due to chromatic dispersion, this chirp transforms into in a $y-t$ bunch tilt in the chicane. At the center of BC2, i.e. at the point of maximum tilt, a thin foil is placed in the path of the beam. The foil has a narrow slot at its center, oriented as shown in Fig. \ref{slot2}. Coulomb scattering of the electrons passing through the foil increases the horizontal and vertical emittances of most of the beam, but leaves a thin unspoiled slice where the beam passes through the slit, Fig. \ref{slot2}.

Spoiling the emittance of most of the beam by a factor $\sim 5$ strongly suppresses the FEL gain, while the short, unspoiled temporal slice produces an x-ray FEL pulse much shorter than the FWHM electron bunch duration. If a very narrow slit is used, uncorrelated energy spread and betatron beam size dominate the output slice length \cite{EMM1,EMM2}, and one obtains a nonlinear growth of the x-ray pulse length versus the slot width. When the slit becomes larger, the growth becomes linear and the x-ray pulse length is mainly determined by the width of the slit \cite{DING}. The minimum duration of the unspoiled slice of the electron bunch measured at the LCLS is about 3 fs \cite{DING}. We use current profile, normalized emittance, energy spread profile, electron beam energy spread and wakefields from \cite{S2ER}. The electron beam charge is 1 nC, and the peak current is 10 kA. We performed numerical simulations with $2\cdot 10^5$ macroparticles to study the performance of the slotted spoiler with the help of the tracking code ELEGANT \cite{ELEG}.

\begin{figure}
\begin{center}
\includegraphics[clip, width=0.75\textwidth]{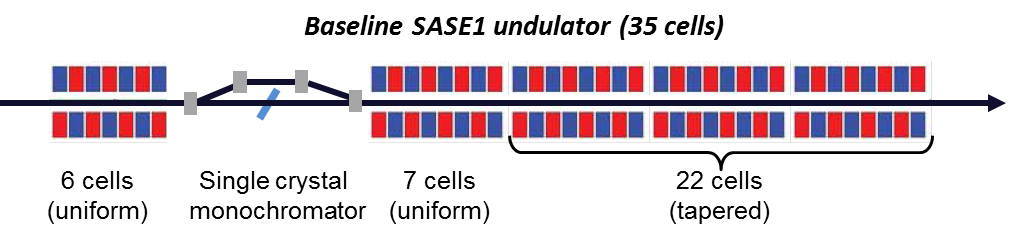}
\end{center}
\caption{Scheme for a 1 TW-power level undulator source for the SPB beamline. Self-seeding and undulator tapering greatly improve the FEL efficiency. X-ray pulse length control is obtained using a slotted foil in the last bunch compressor BC2. The magnetic chicane accomplishes three tasks by itself.  It creates an offset for single crystal monochromator, it removes the electron microbunching produced in the upstream undulator, and it acts as a magnetic delay line.}
\label{layout}
\end{figure}

A design of a self-seeding setup based on the undulator system for the European XFEL is sketched in Fig. \ref{layout}. The method for generating high power x-ray pulses exploits a combination of a self-seeding scheme with an undulator tapering technique. Tapering consists in a slow reduction of the field strength of the undulator in order to preserve the resonance wavelength, while the kinetic energy of the electrons decreases due to the FEL process. The undulator taper could be simply implemented at discrete steps from one undulator segment to the next, by changing the undulator gap. Highly monochromatic pulses generated with the self-seeding technique make the tapering much more efficient than in the SASE case.

Here we study a scheme for generating 1 TW-level X-ray pulses in the SASE1 tapered undulator. We optimize our setup based on start-to end simulations for 14 GeV electron beam with 1 nC charge compressed up to 10 kA peak current. In this way, the output power of the SASE1 undulator could be increased from the value of 100 GW in the SASE regime\footnote{There is an increase in the SASE saturation power with respect to the nominal mode of operation, due to the increase in peak current from the nominal 5 kA to our case, where 10 kA are obtained with particular settings of the bunch formation system.} to about 1.5 TW at the photon energy range around 4 keV.

\begin{figure}
\begin{center}
\includegraphics[clip, width=0.75\textwidth]{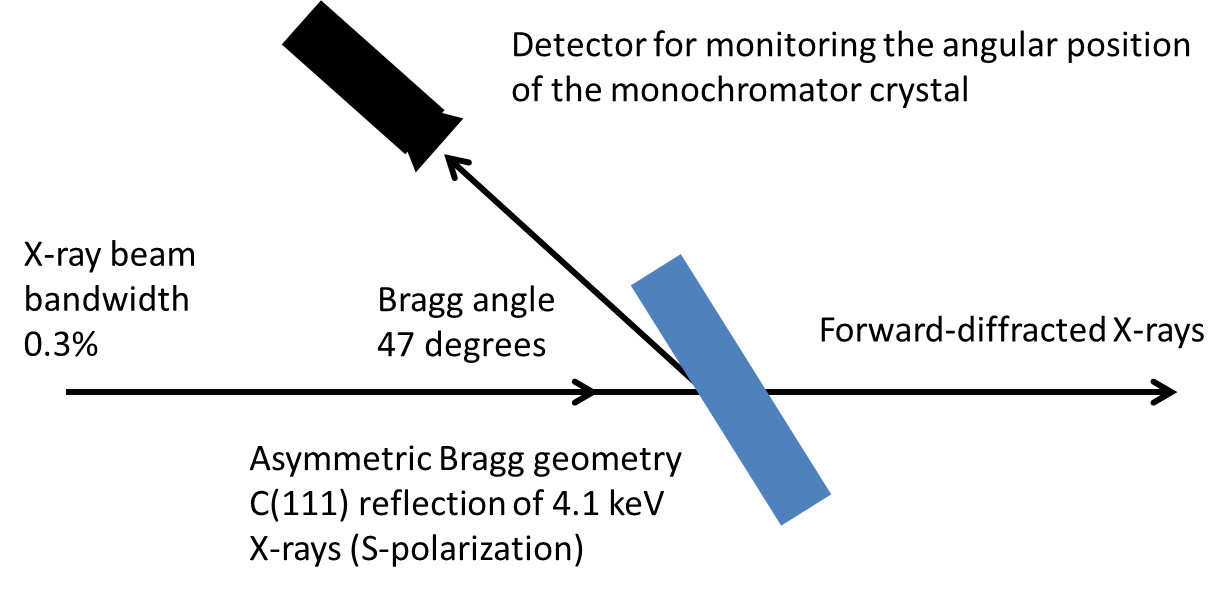}
\end{center}
\caption{Schematic of the single crystal monochromator for operation in the photon energy range around 4 keV.} \label{cryg}
\end{figure}
\begin{figure}
\begin{center}
\includegraphics[clip, width=0.30\textwidth]{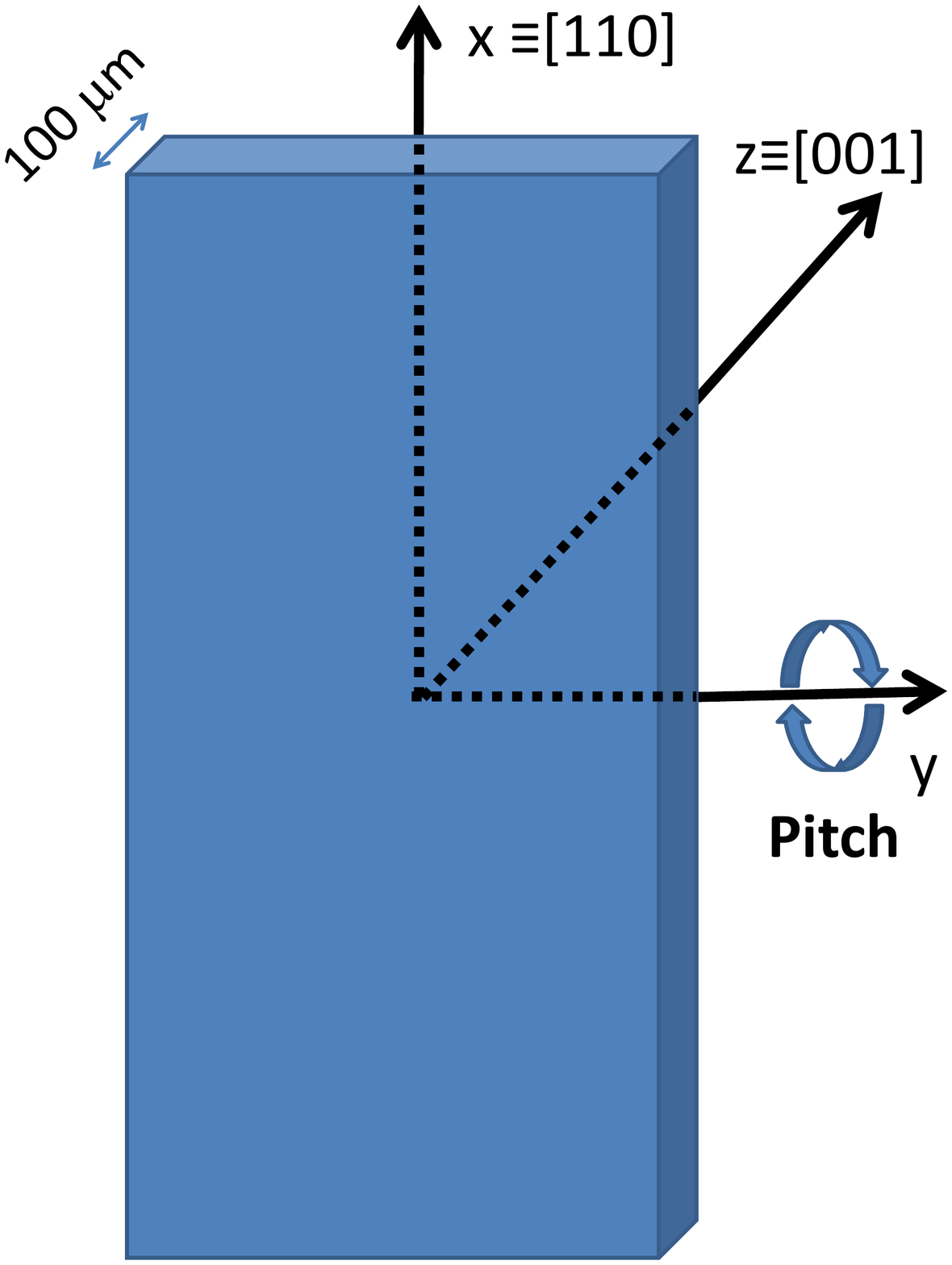}
\end{center}
\caption{Drawing of the orientation of the diamond crystal considered in the article.} \label{cryaxes}
\end{figure}

\begin{figure}
\begin{center}
\includegraphics[clip, width=0.5\textwidth]{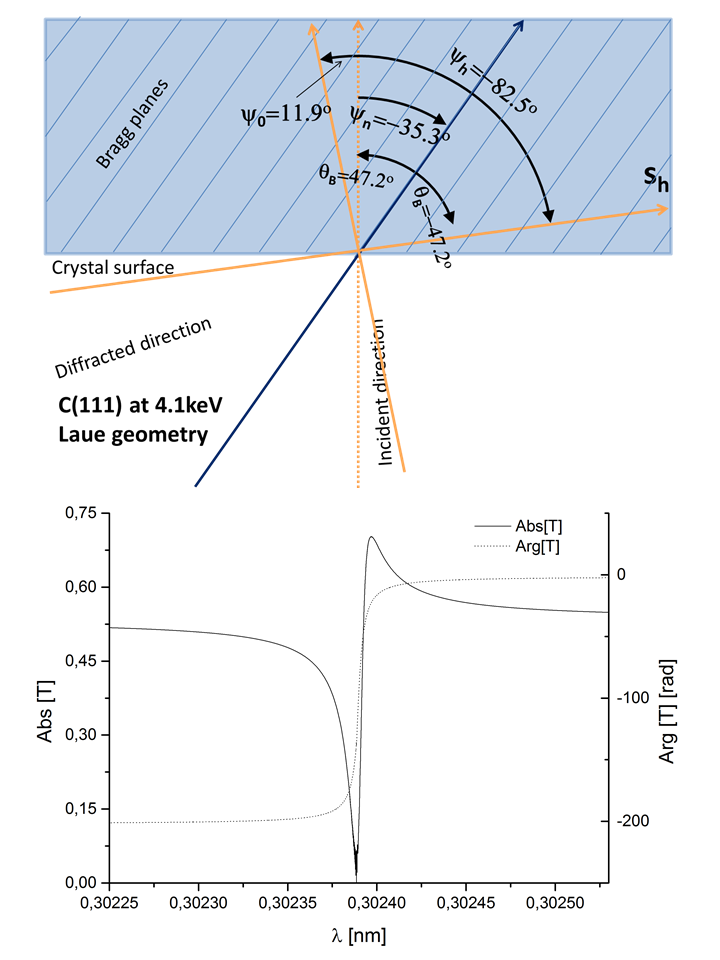}
\end{center}
\caption{Upper plot: scattering geometry (we are following the notation in \cite{AUTH}). Lower plot: modulus and phase of the transmittance for the C(111) asymmetric Laue reflection  from the diamond crystal in Fig. \ref{cryaxes} at 4.1 keV.} \label{C111}
\end{figure}
Our  design adopts the simplest self-seeding scheme, which uses the transmitted x-ray beam from a single crystal to seed the same electron bunch, Fig. \ref{cryg}. In the following we will consider a $100~\mu$m-thick diamond crystal. We  define a Cartesian reference system [$x, y, z$] linked with the crystal. The direction $z$ corresponds to the direction identified by the Miller indexes [0, 0, 1], while $x$ and $y$ are specified as in Fig. \ref{cryaxes}.  The crystal can rotate freely around the $y$ axis (pitch angle) as indicated in the figure. In this way we can exploit several symmetric and asymmetric reflections. By changing the pitch angle of the crystal in Fig. \ref{cryaxes} we are able, in fact, to cover the entire energy range between 3 keV and 13 keV \cite{SHVI}, \cite{ASYM}. In the low energy range between 3 keV and 5 keV we use a C(111) asymmetric reflection (in Bragg and Laue geometry, depending on the energy). For self-seeding implementation, we are interested in the forward diffracted beam. From this viewpoint, the crystal can be characterized as a filter with given complex transmissivity. In Fig. \ref{C111} we show scattering geometry, amplitude and phase of the transmittance for the C(111) asymmetric Bragg reflection at 4.1 keV.

Summing up, the overall self-seeding setup proposed here consists of three parts: a SASE undulator, a self-seeding single crystal monochromator and an output undulator, in which the monochromatic seed signal is amplified up to the 1.5 TW-level, Fig. \ref{layout}. Calculations show that, in order not to spoil the electron beam quality and to reach simultaneously signal dominance over shot noise, the number of cells in the first (SASE) undulator should be equal to 6. The output undulator consists of two sections. The first section is composed by an uniform undulator, the second section by a tapered undulator. The monochromatic seed signal is exponentially amplified passing through the first uniform part of the output undulator. This section is long enough, 7 cells, in order to reach saturation, which yields about 100 GW power. Finally, in the second part of the output undulator the monochromatic FEL output is enhanced up to 1.5 TW by taking advantage of the undulator magnetic field taper over the last 22 cells.

\section{\label{sec:SPBlayout} X-ray optics layout of the SPB beamline}

\begin{figure}
\begin{center}
\includegraphics[trim = 10 150 0 300, clip, width=1.0\textwidth]{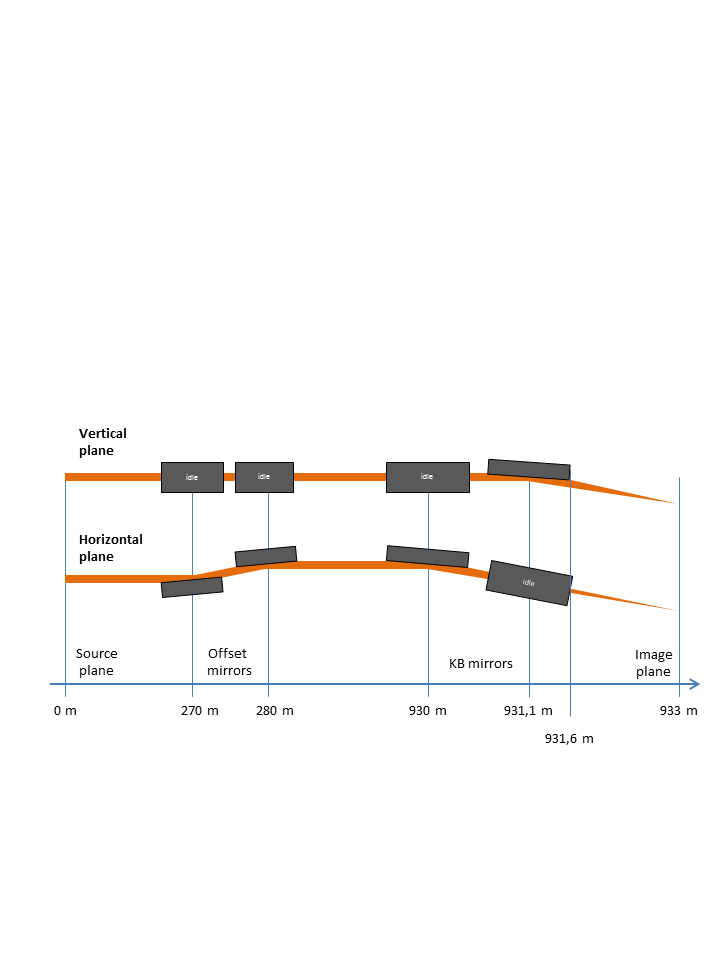}
\end{center}
\caption{Layout of the optical components for the SPB beamline.} \label{SPBsketch}
\end{figure}

\begin{figure}
\begin{center}
\includegraphics[trim = 0 0 0 0, clip, width=0.75\textwidth]{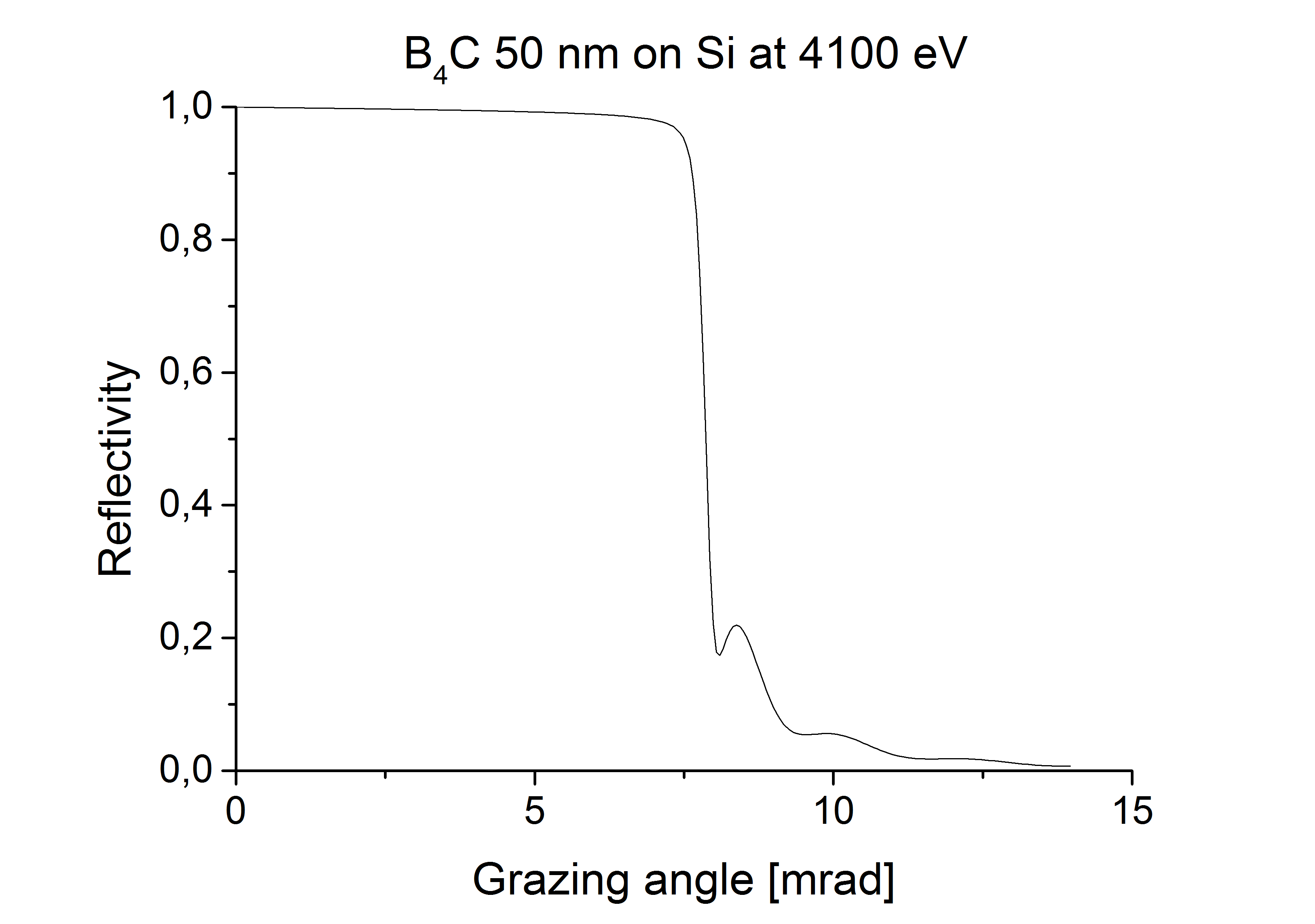}
\end{center}
\caption{reflectivity of a $50$ nm perfect $B_4 C$ substrate on $Si$ substrate at $4.1$ keV as a function of the grazing angle. Calculation are done via the web interface \cite{CXRO}.} \label{B4C}
\end{figure}
The SPB optical layout \cite{MANT} is sketched in Fig. \ref{SPBsketch}. The first upstream optical element, that is also the first optical element after the exit of the SASE1 undulator, is a Horizontal Offset Mirror (HOM) pair with a clear aperture of 800 mm \cite{XCDR}. This pair is used for avoiding high-energy Bremsstrahlung which propagates together with the FEL beam, and also remove spontaneous radiation outside the FEL bandwidth. The second HOM is adaptive; this allows, in principle, for both focusing and correcting wavefront aberrations in the horizontal direction. HOMs are to be coated with boron carbide ($B_4C$), as up to date it was demonstrated to be the best performing coating material during damage experiments. The HOM angles are adjustable, so that one can change the lateral aperture matching it to different photon energies; this implies different radiation beam size on the optics. For the maximal incident angle $\theta = 3.6$ mrad, one achieves an overall high-reflectivity close to $100 \%$ over the photon energy range between $3$ keV and $5$ keV. To be more specific, we show the reflectivity of a $50$ nm perfect $B_4 C$ substrate on $Si$ substrate at $4.1$ keV as a function of the grazing angle in Fig. \ref{B4C}, see \cite{CXRO}.

The source divergence is the most important parameter for the beam transport system, and is largest for the smallest photon energies and for the lowest electron charge\footnote{In these condition the gain length is at its shortest, so that the photon beam size is smallest and the divergence is largest.}. Therefore, fixing the photon energy in the baseline SASE mode of operation, a shorter photon pulse duration (corresponding to a lower charge) directly translates into a larger divergence of the radiation pulse. In particular, at a photon energy of $4$ keV, the highest FWHM divergence of $5~\mu$rad corresponds to the lowest charge for the baseline SASE mode of operation, that is $20$ pC. Ideally, any optics element should have a transverse aperture size of at least $4\sigma$ times the photon beam size, in order to minimize diffraction from the mirror aperture and consequently preserve the beam wavefront. If we accept the requirement of a transverse clear aperture of $4 \sigma$ times the beam size (or equivalently $1.7$ times the FWHM size), and we consider the HOM placed at about $300$ m from the source, then the desired lateral aperture is about $2.6$ mm for the $4$ keV case. If one considers $800$ mm clear aperture and $3.6$ mrad reflection angle, one obtains a lateral aperture of about $3$ mm, meaning that in our case of interest the HOMs are expected to preserve the wavefront.

\begin{figure}
\begin{center}
\includegraphics[trim = 10 300 0 250, clip, width=1.0\textwidth]{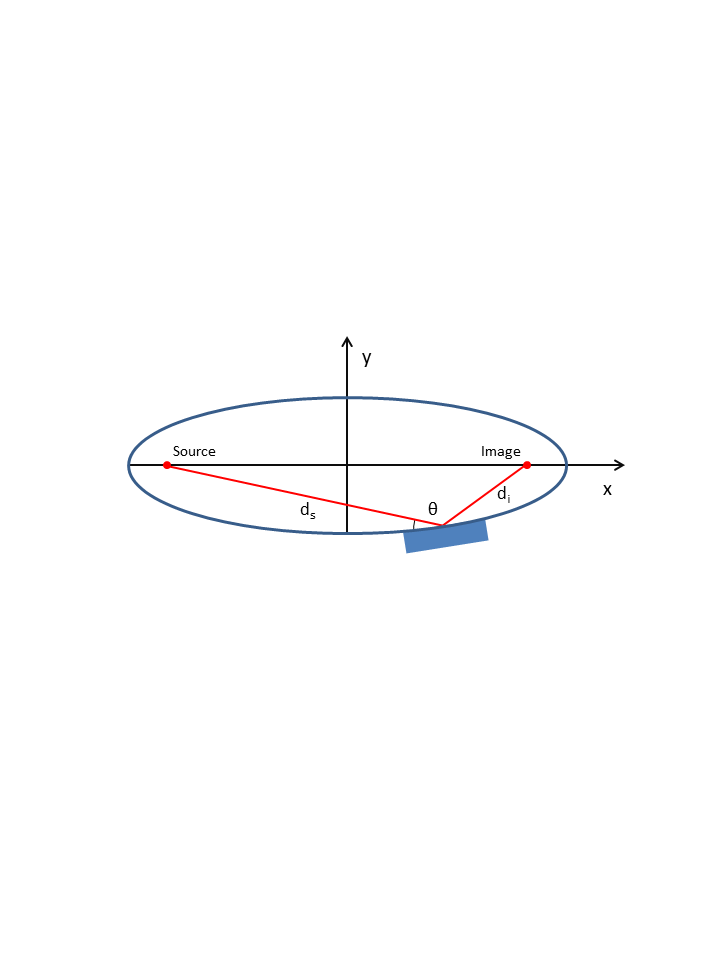}
\end{center}
\caption{Image formation by an elliptical mirror} \label{Ellip}
\end{figure}
Once the radiation pulse enters the experiment area, it is focused by a KB mirror system to about $100$ nm size\footnote{The SPB instrument design include two KB  systems focusing to $1 \mu$m and $100$ nm size respectively \cite{MANT}. In this work we only discuss about the 100 nm  KB system.}.  A baseline layout for KB system is shown in Fig. \ref{SPBsketch}. Two elliptical mirrors, Fig. \ref{Ellip}, with a $950$ mm clear aperture and a fixed incidence angle of $3.5$ mrad are assumed in the vertical and horizontal direction in order to achieve high efficiency at high photon energies\footnote{Due to the fact that the KB mirrors are elliptical mirrors, their angles must be fixed, and cannot be adjusted. This is in contrast with HOMs mirrors, that are plane. Their angle can therefore be adjustable.}.  The KB system reimages and demagnifies the source plane in the SASE1 undulator into the focal point in the SPB chamber. More in detail, in our case of interest the source plane of the SPB beamline is estimated to be located in the fifth SASE1 undulator module, starting counting from the last one. It should be remarked that this situation is different compared to the nominal mode of operation when source point is estimated to be located in the third SASE1 undulator module, starting counting from the last one \cite{MANT}. As the KB optics is elliptically polished, it is subject to defocusing and higher order aberrations if the source point is moved.  This effect was investigated in \cite{MANT}. The results show that a small tilt of the mirror system can compensate the 10 m difference in source position considered here.

The vertical KB mirror is closer to the sample to get a better demagnification. To be more specific, the ratio of the focal spot size in the vertical to that in the horizontal direction in the ideal case, i.e. without diffraction from the mirror aperture, is about $1.6$, and follows directly from the ratio between the distances from the KB mirrors to the image place, Fig.  \ref{SPBsketch}. Two different coatings, each with the same working angle of $3.5$ mrad but different working energy range have been selected \cite{MANT}. Boron-carbide will be used for the lower photon energy range between $3$ keV and $7$ keV, while for the higher energies between $7$ keV and $16$ keV the focusing system will rely on Ru coating.

Considering a $950$ mm clear aperture and a $3.5$ mrad  reflection angle, one straightforwardly obtains a lateral aperture of $3.3$ mm. However, the about $900$ m-long propagation distance from source to sample leads to a large lateral beam size at the focusing optics. In fact, for the photon energy  4 keV the desired lateral aperture for the ultra-short pulse case is about $8$ mm. As a result, due to the large divergence of a nominal X-ray pulse shorter then $10$ fs, one suffers major diffraction effects from the KB mirror aperture, leading to about a hundred-fold decrease in fluence at photon energies around 4 keV which, as was remarked before, are ideal for imaging of single biomolecules. Here we propose to overcome this obstacle exploiting a special mode of operation of the accelerator complex. In particular, we show how it is possible to obtain an X-ray source capable of producing X-ray pulses with small angular divergence of about $2 \mu$rad and, simultaneously, a few fs duration, by taking advantage of a minimal modification in the accelerator complex, amounting to the introduction of a slotted foil in the last electron bunch compressor.

\section{\label{sec:org} Organization of the experiment modeling program}

\begin{figure}
\begin{center}
\includegraphics[trim = 0 100 0 40, clip, width=1.0\textwidth]{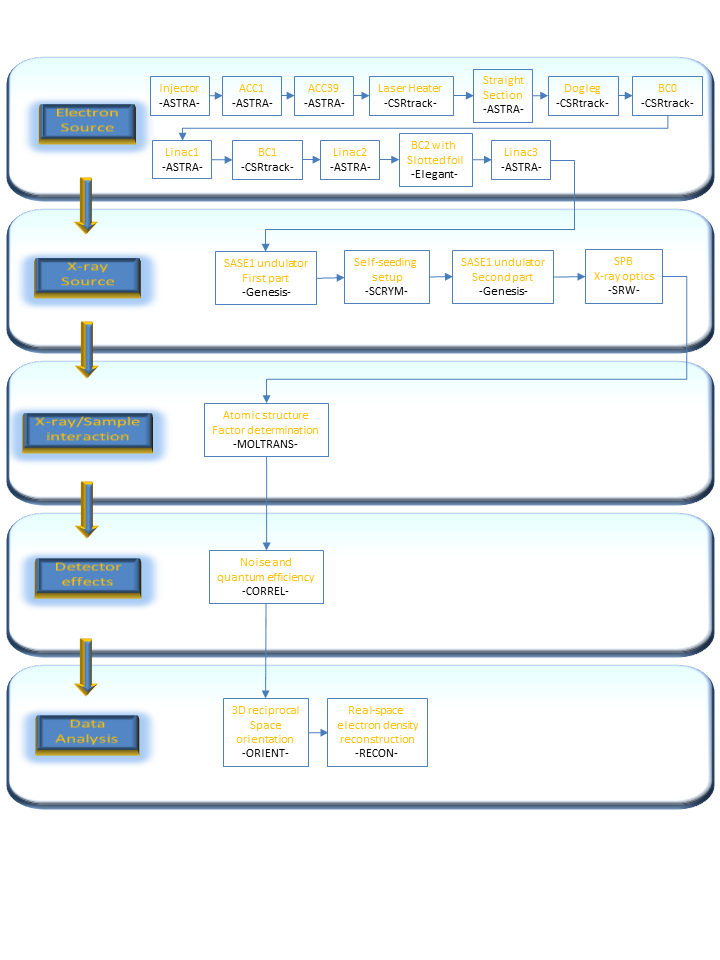}
\end{center}
\caption{Organization of start-to-end simulation program.} \label{org}
\end{figure}

As already discussed above, in this work we modeled a single biomolecule imaging experiment by performing start-to-end simulations, from the electron injector till the production of noisy diffraction data.  This helps understanding the feasibility of the overall proposed scheme. The experiment modeling program is organized using modular design, as shown in Fig. \ref{org}. As it can be seen from the figure, these modules focus on the following tasks:

\begin{itemize}
\item{Modeling the generation of the electron beam and its transport to the undulator entrance}

\item{Modeling the generation of the radiation and its transport to the interaction region}

\item{Modeling the interaction between the focused FEL beam and molecule}

\item{Modeling of the measurement of the diffracted radiation pattern by the detector system}

In the follow-up of the present article we will extend our analysis to the reconstruction of the electron density, which will need the final module:

\item{Interpretation the modeled data and reconstruction of the molecular electron density}
\end{itemize}
The upper block in Fig. \ref{org}, named `Electron source', shows the segmentation of the European XFEL accelerator complex that was implemented for modeling purposes, and describes the approach used to track an initial electron distribution through these segments. The set of codes ASTRA \cite{ASTR}, CSRtrack \cite{CSRT}, and ELEGANT \cite{ELEG} was used to perform beam dynamics simulation starting from the photocathode gun up to the undulator entrance. ASTRA is a macroparticle-based code that incorporates a space-charge algorithm. It was used for the injector and the acceleration regions. The simulations of the bunch compressor sections were performed with CSRtrack, a program that properly treats all bunch self-interactions via radiative effects. Our scheme relies on the introduction of a slotted foil in the BC2 section. The phase space distribution was tracked through that section with the code ELEGANT. Software have been written to properly pass the phase space distribution between the three different programs.

The 'X-ray source` simulation block delivers time-dependent simulations of the radiation field, to be used at the sample location, for the special mode of operation of the accelerator complex foreseen in our method.
Simulations up to the SASE1 undulator exit were performed with the help of the code GENESIS \cite{GENE}  and of the in-house code SCRYM \cite{SCRY} in the following way: first the 3D field distribution was calculated at the exit of the first part of the undulator using code GENESIS. Subsequently, the electromagnetic field file was dumped. Then we performed filtering through the monochromator using the code SCRYM. The electron beam microbunching is washed out by the presence of nonzero momentum compaction factor in the self-seeding chicane. Therefore, at the entrance of the second part of the undulator we used a beam file describing an electron beam distribution with no initial microbunching, but with the characteristics (mainly the energy spread) induced by the FEL amplification process in the first part of the SASE1 undulator. The amplification process in the second  undulator part starts from the seed field file and is simulated again with the help of the code GENESIS. Proper initial shot noise was included. Following the exit of SASE1, we performed the propagation of the output field up to the interaction region.  In particular, the X-ray optics and propagation code SRW \cite{CHUB} was utilized to simulate the FEL beam profiles after passing through the SPB beamline optics.

The `X-ray/sample interaction' step was exemplified by using RNA Pol II molecule. This molecule consists of about $31000$ atoms. We calculated the structure factor $F$ of the static molecule, without accounting for radiation damage within the structure. In Section \ref{sec:expr} we presented arguments for this assumption, based on the ultrashort pulse duration of $4$ fs \cite{CHAS}-\cite{SON1}. Such duration is sufficiently short to avoid damage problems due to atomic motion and direct damage changing the structure factor up to the fluence of $10^{23} \mathrm{ph}/\mathrm{cm}^2$ at the photon energy of $4$ keV. The atomic factors were calculated using MOLTRANS package \cite{MOLT}.

The `Detector effects' were simulated using a tool called CORREL \cite{CORR}. In particular we estimated the detector effects on the model data. Noise is only considered in terms of photon shot noise. No additional sources of noise such as detector noise or Compton background were considered. The quantum efficiency of detector was included.  Results of simulations presented in \cite{AGIP} allow an estimation of the total Equivalent Noise Charge (ENC) around $300$ electrons, and single photon energy deposition of about $1200$ electrons for the AGIPD detector at $4$ keV. In this case, detector noise is negligible compared with the photon shot noise.

In the follow-up of the present work, which will be published as a separate article, we will deal with the data processing, leading to the sample reconstruction. The final `Data analysis' step will be presented with the help of two in-house simulation tools, ORIENT \cite{ORIE} and RECON \cite{RECO}. They will be used in the final stage of our modeling program, in order to reconstruct images from the modeling data, and also to evaluate the fidelity of these reconstructions. The last step performed in this work was the production of $30000$  2D-diffraction  patterns of the molecule, with unknown orientations. Next, the relative orientations of the different images will be determined in order to assemble a 3D diffraction pattern in the reciprocal space. This task will be performed with the help of the code ORIENT. Finally, the 3D electron density of the molecule will be obtained by a phase retrieval method using the code RECON.

\section{\label{sec:SASE1} Radiation from the SASE1 undulator}

\begin{figure}
\includegraphics[trim = 0 310 0 0,clip, width=0.50\textwidth]{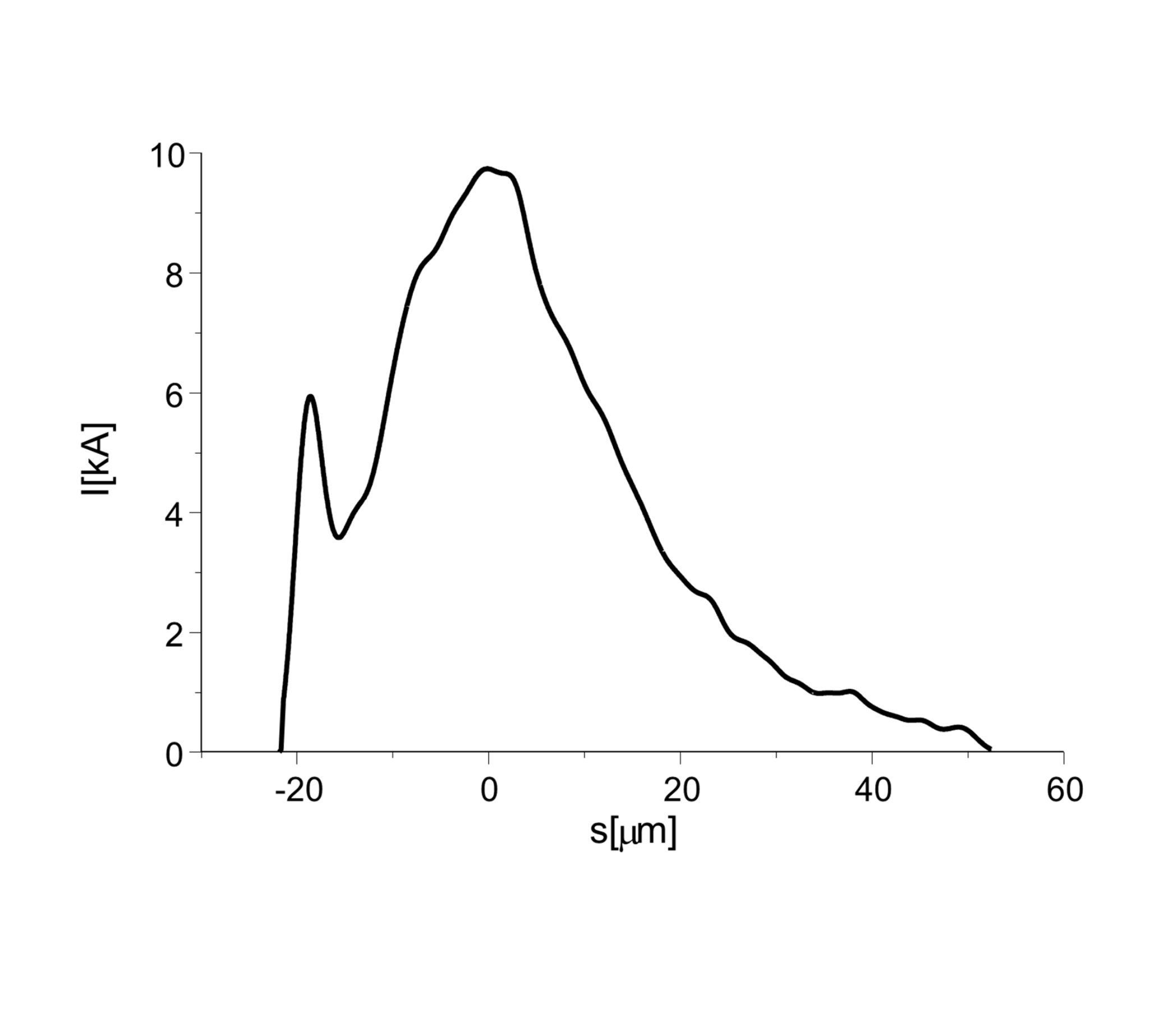}
\includegraphics[trim = 0 310 0 0,clip, width=0.50\textwidth]{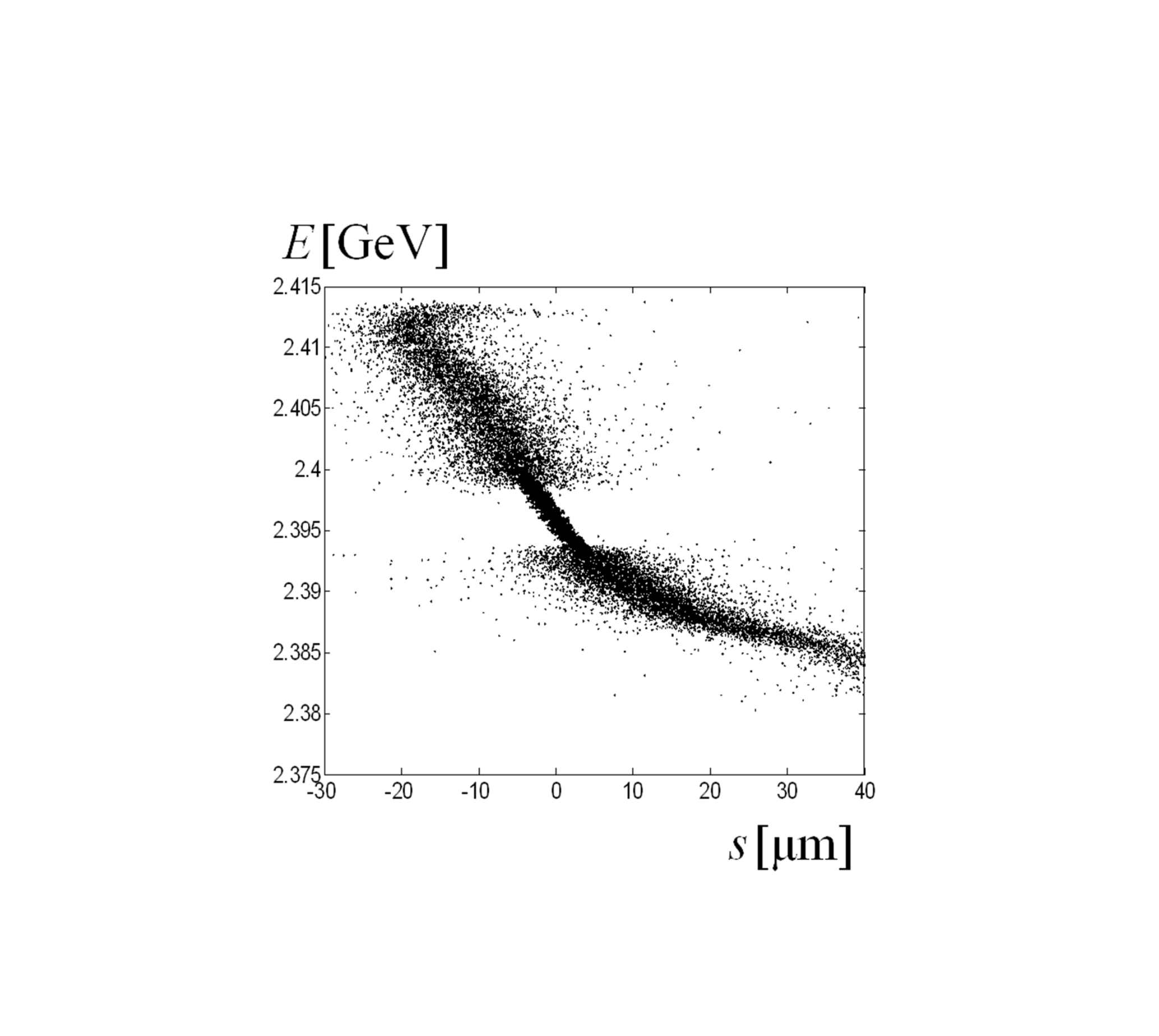}
\caption{Left plot: Current profile after BC2 without foil. Right
plot: Longitudinal phase space distribution of the particles after
BC2, with foil. The simulation includes multiple Coulomb scattering
in a $2\mu$m thin aluminum foil with a slot width of $0.7$ mm.}
\label{currsl}
\end{figure}

\begin{figure}
\includegraphics[trim = 0 310 0 0,clip,width=0.50\textwidth]{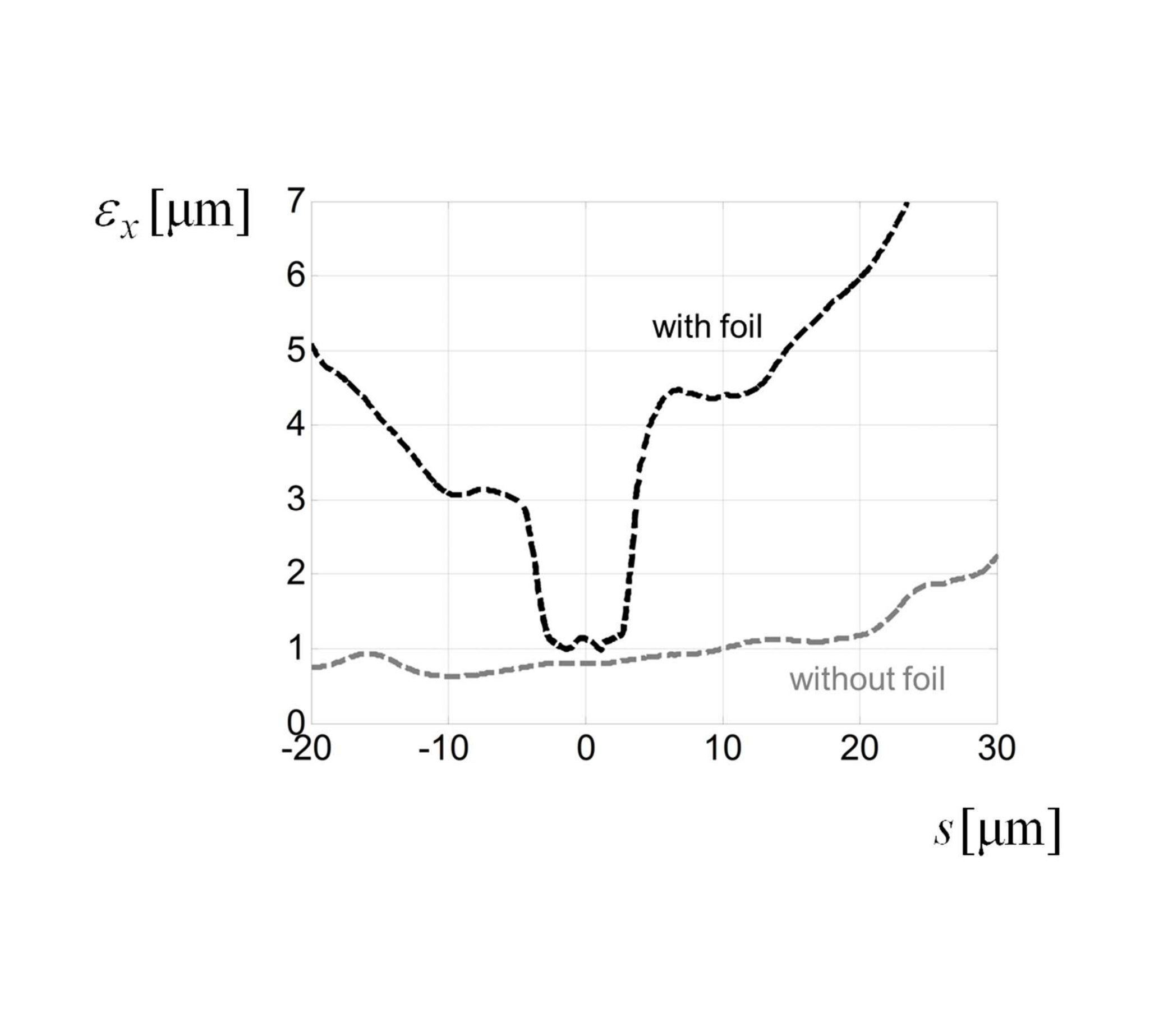}
\includegraphics[trim = 0 310 0 0,clip,width=0.50\textwidth]{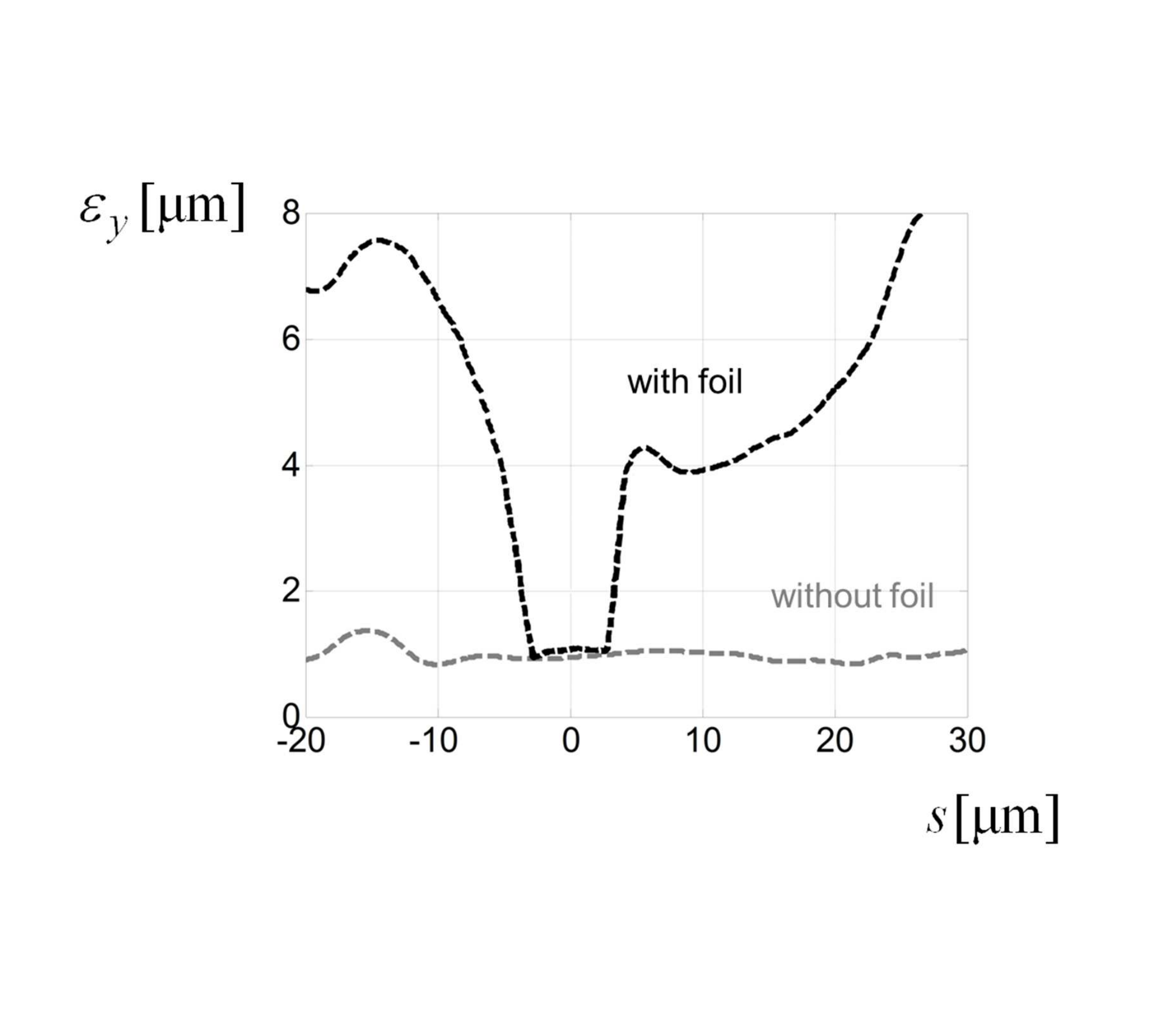}
\caption{Left plot: Vertical normalized emittance as a function of
the position inside the electron bunch after BC2. The grey dashed
curve is from particle tracking without foil. The black dashed curve
is from particle tracking with foil. Right plot: Horizontal
emittance as a function of the position inside the electron bunch
after BC2. The grey dashed curve is from particle tracking without
foil. The black dashed curve is from particle tracking with foil.
(In both plots we removed 6 $\%$ of strongly scattered particles
from the analysis.) } \label{emxy}
\end{figure}
\begin{table}
\caption{Parameters for the mode of operation at the European XFEL
used in this paper.}

\begin{small}\begin{tabular}{ l c c}
\hline & ~ Units &  ~ \\ \hline
Undulator period      & mm                  & 40     \\
Periods per cell      & -                   & 125   \\
Total number of cells & -                   & 35    \\
K parameter (rms)     & -                   & 3.217   \\
Intersection length   & m                   & 1.1   \\
Wavelength            & nm                  & 0.302 \\
Energy                & GeV                 & 14    \\
Charge                & nC                  & 1\\
\hline
\end{tabular}\end{small}
\label{tt1}
\end{table}

We use current profile, normalized emittance, energy spread profile, electron beam energy spread and wakefields from \cite{S2ER}. The electron beam charge is 1 nC, and the peak current is 10 kA, Fig. \ref{currsl} (left plot). Parameters for the mode of operation at the European XFEL used in this paper are summarized in Table \ref{tt1}.  Detailed computer simulations with $2\cdot 10^5$ macroparticles have been carried out to evaluate the erformance of the slotted spoiler using the tracking code ELEGANT \cite{ELEG}. They include multiple Coulomb scattering in a $2\mu$m thin aluminum foil. In order to demonstrate the duration tunability of the radiation pulse will present results for two case studies referring to pulse length of $12$ fs and $4$ fs.

We will begin studying the example for the longer pulse, that is the case for $12$ fs. The longitudinal distribution of the particles just after the BC2 chicane is shown in Fig \ref{currsl} (right plot). A slit full-width of $0.7$ mm selects a small fraction of electrons, about $20 \%$, and produces an unspoiled electron bunch slice after BC2, with a duration of about $18$ fs FWHM\footnote{As we will see, the FEL gain-narrowing allows an x-ray pulse duration of about $12$ fs}, Fig. \ref{emxy}.

\begin{figure}
\begin{center}
\includegraphics[width=0.50\textwidth]{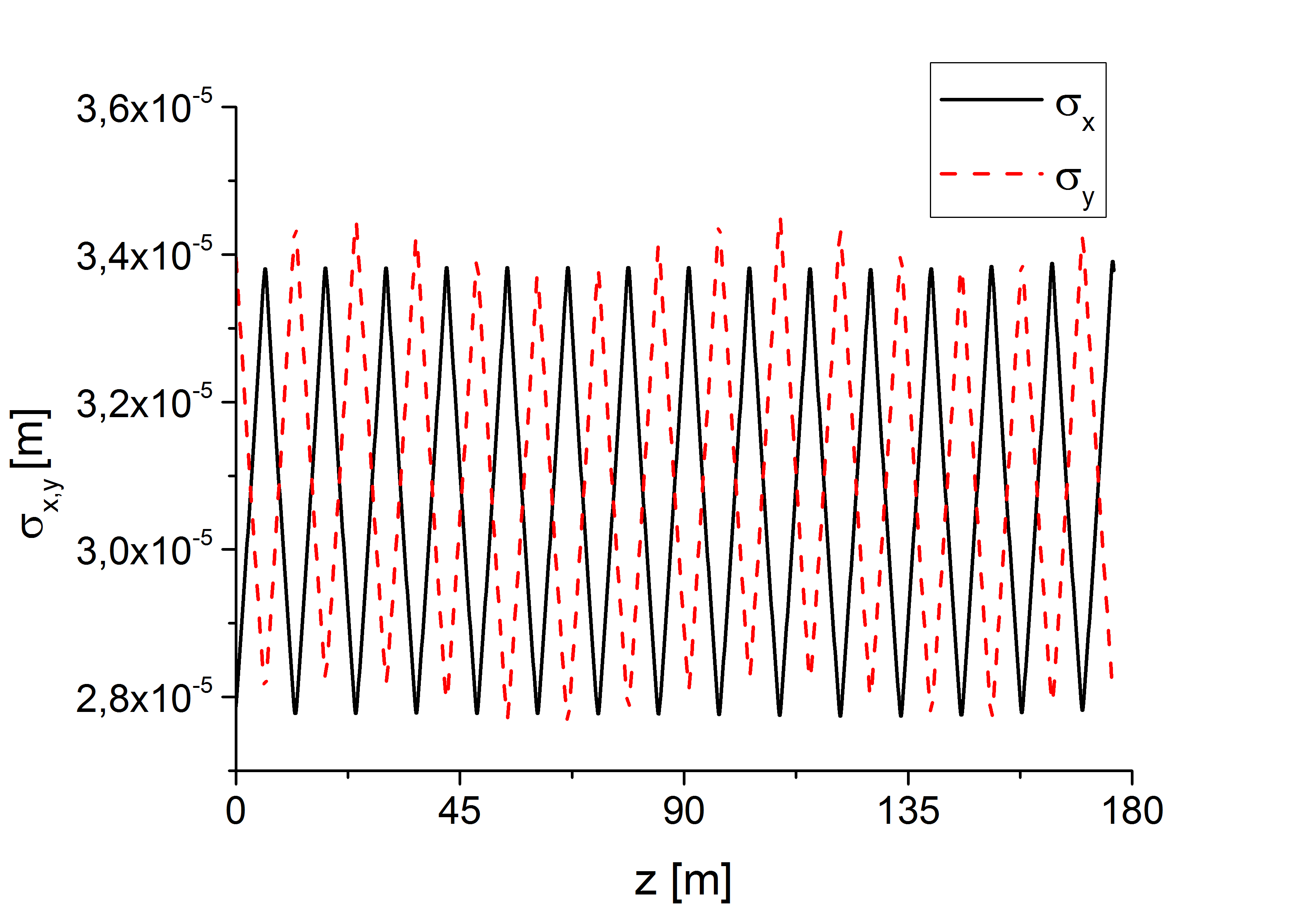}
\end{center}
\caption{Evolution of the horizontal  and vertical dimensions
of the electron bunch as a function of the distance inside the undulator. The plots
refer to the longitudinal position inside the bunch corresponding to the maximum
current value.} \label{focus_lo}
\end{figure}

\begin{figure}
\includegraphics[width=0.50\textwidth]{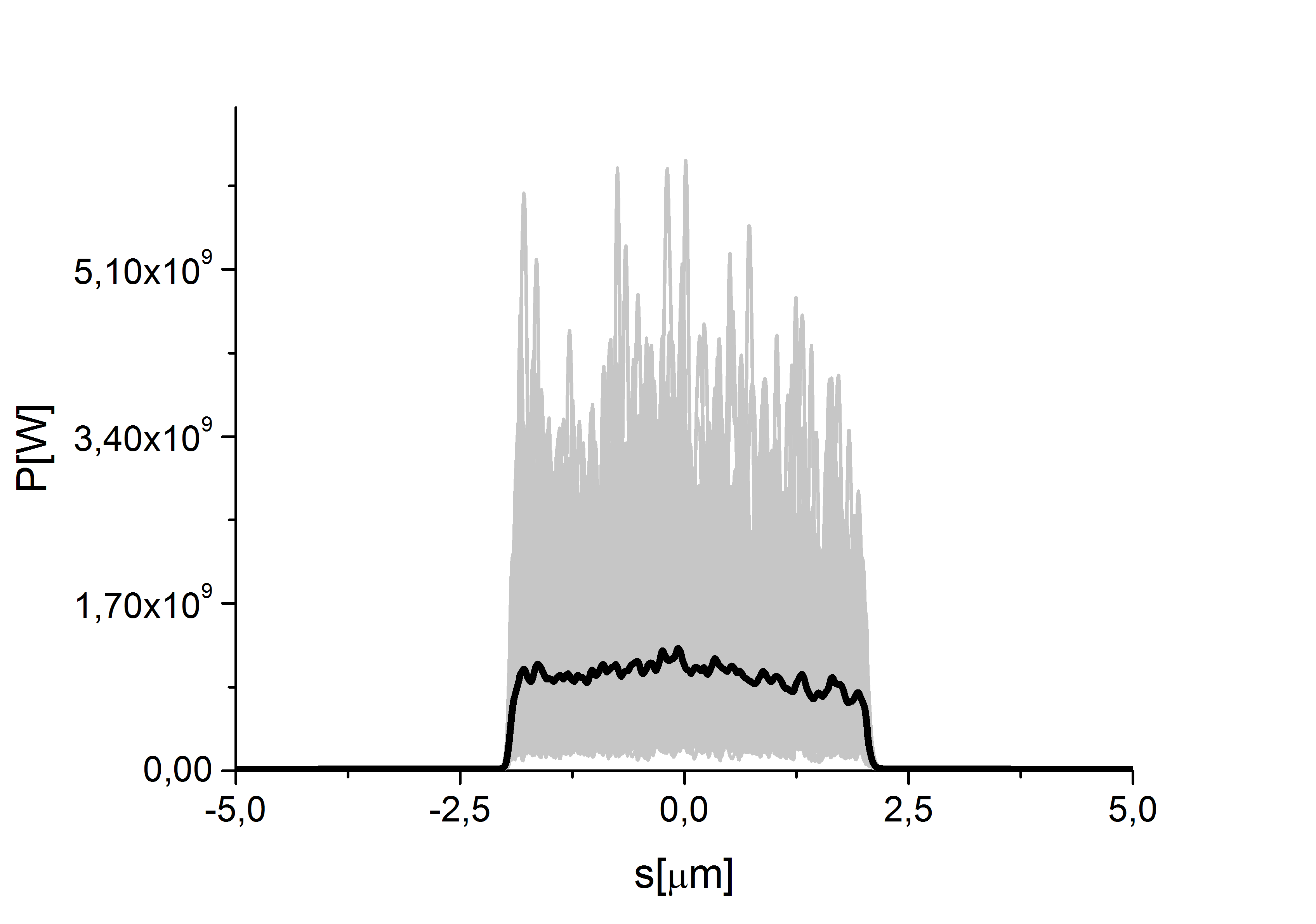}
\includegraphics[width=0.50\textwidth]{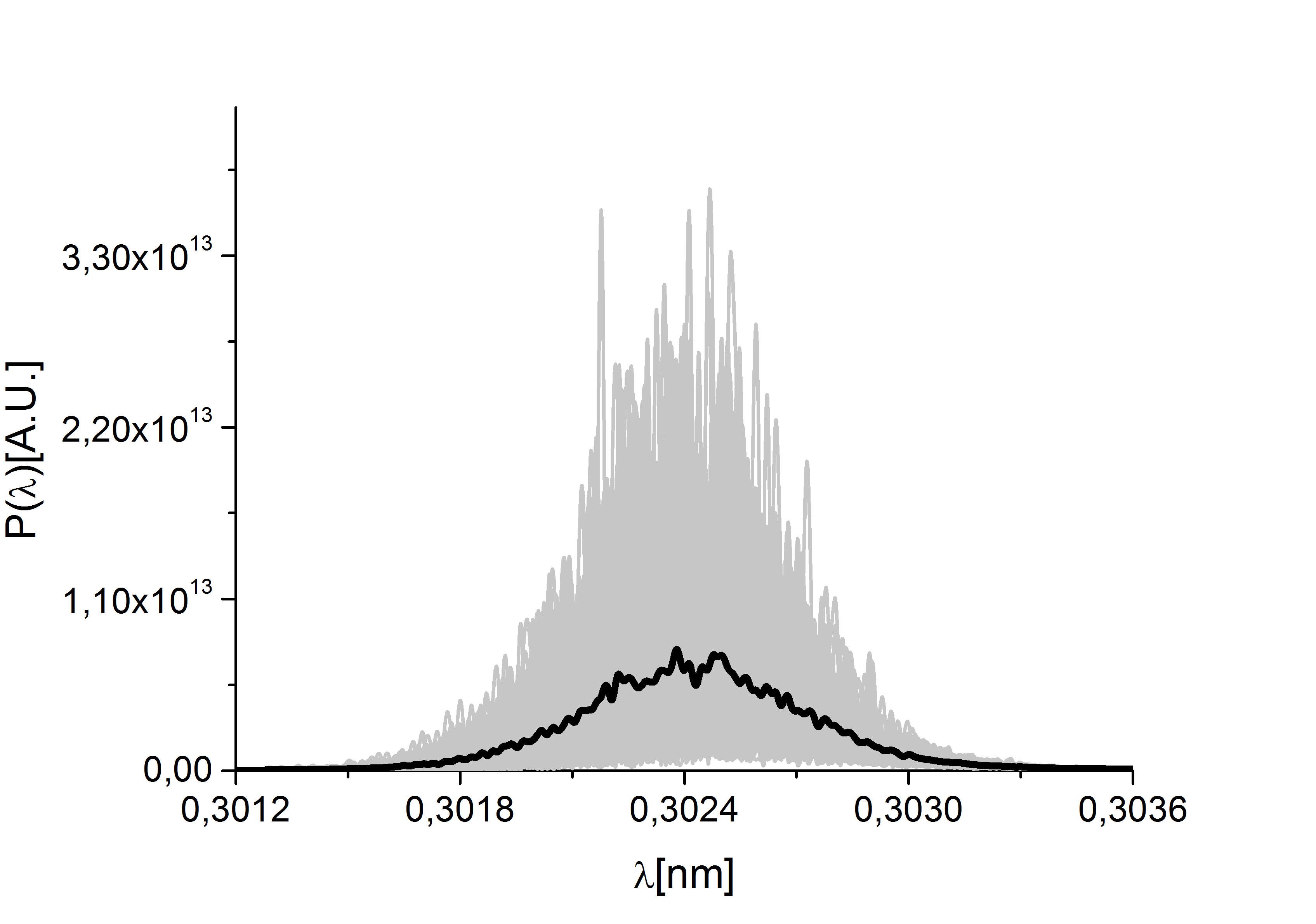}
\caption{Power distribution and spectrum of the SASE x-ray pulse at
the exit of the first undulator for the case of long ($12$ fs) pulse mode of operation.} \label{PSpin1}
\end{figure}

\begin{figure}
\begin{center}
\includegraphics[width=0.50\textwidth]{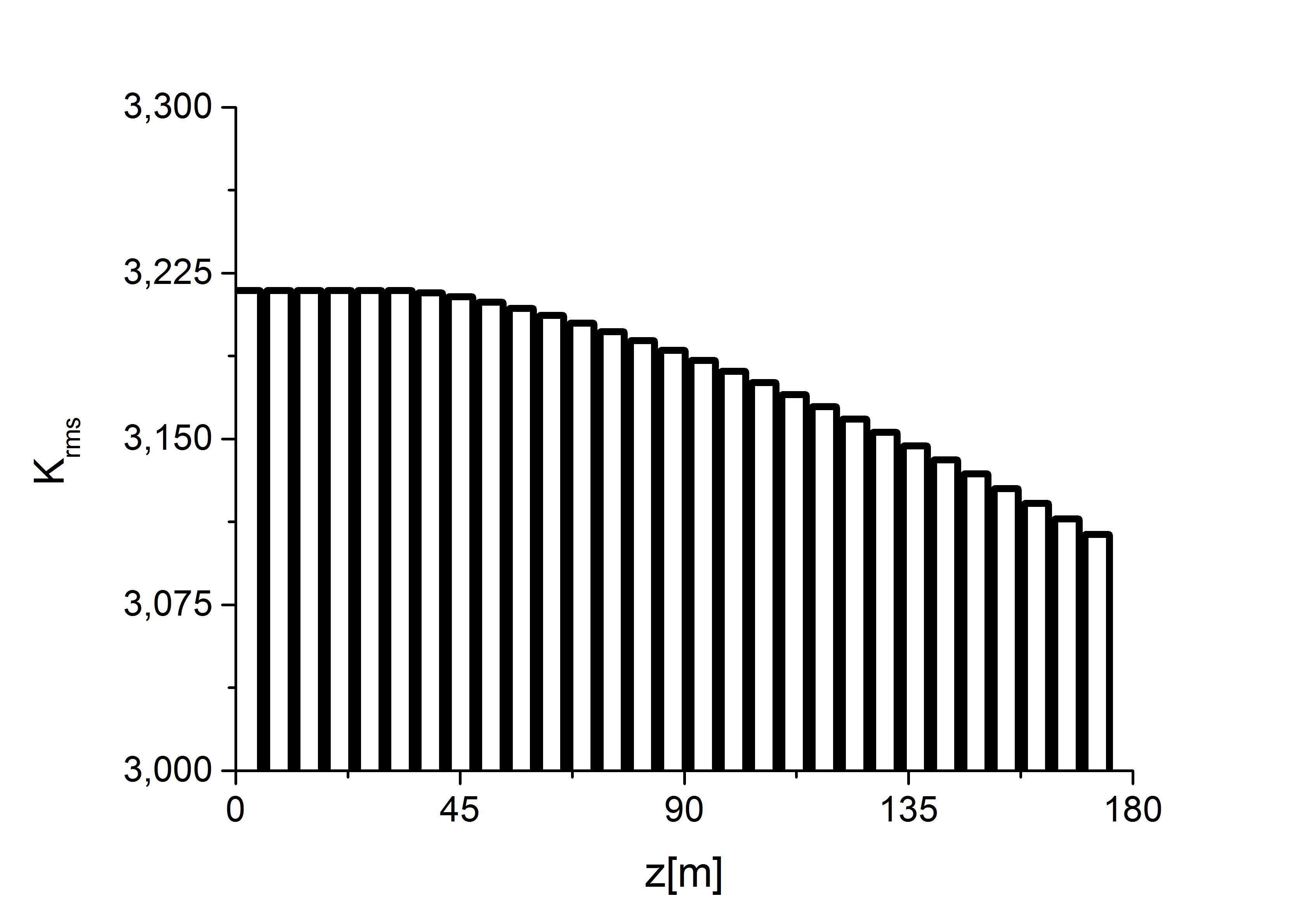}
\end{center}
\caption{Taper configuration for the case of long ($12$ fs) pulse mode of operation.} \label{Taplaw}
\end{figure}

\begin{figure}
\includegraphics[width=0.50\textwidth]{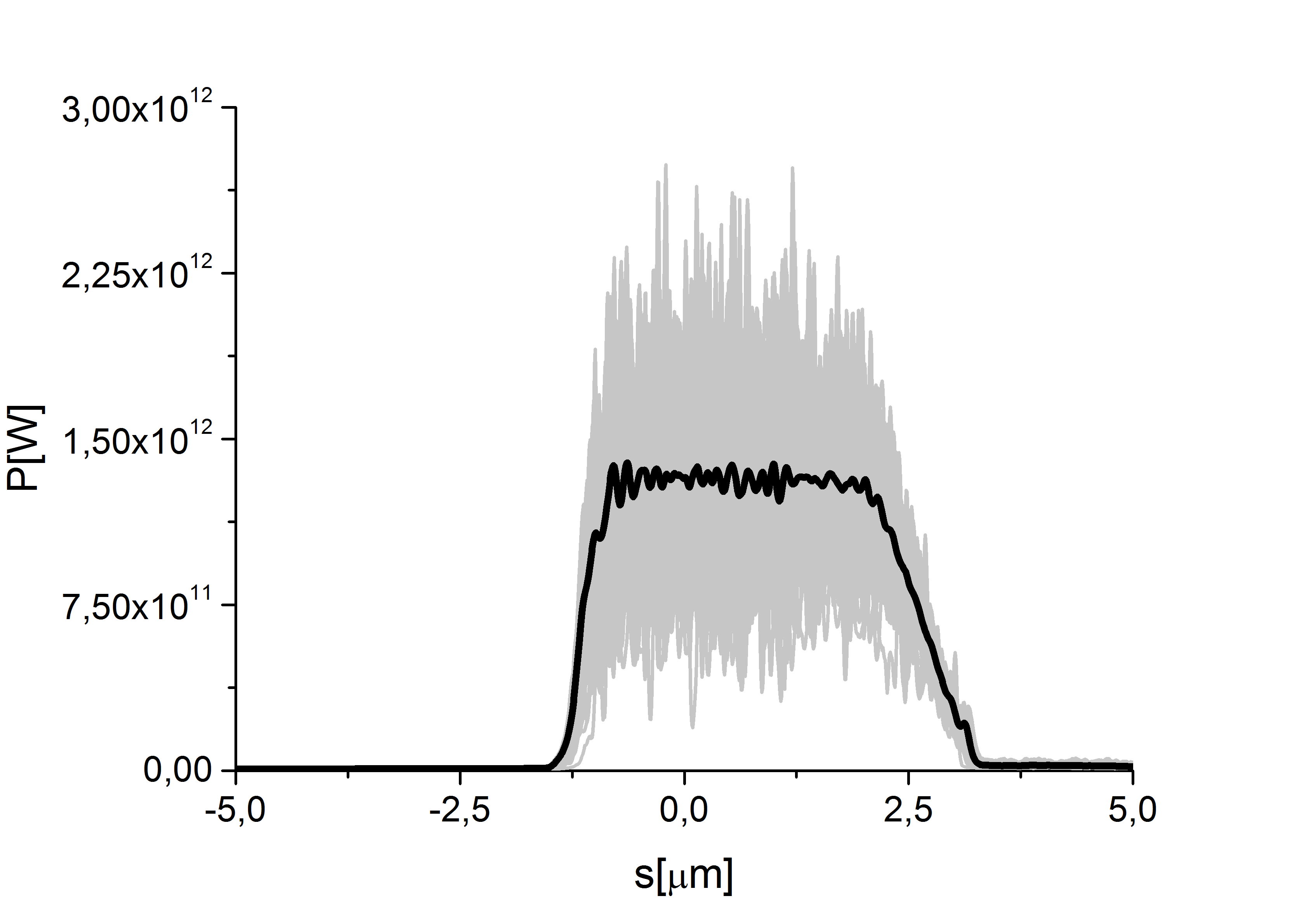}
\includegraphics[width=0.50\textwidth]{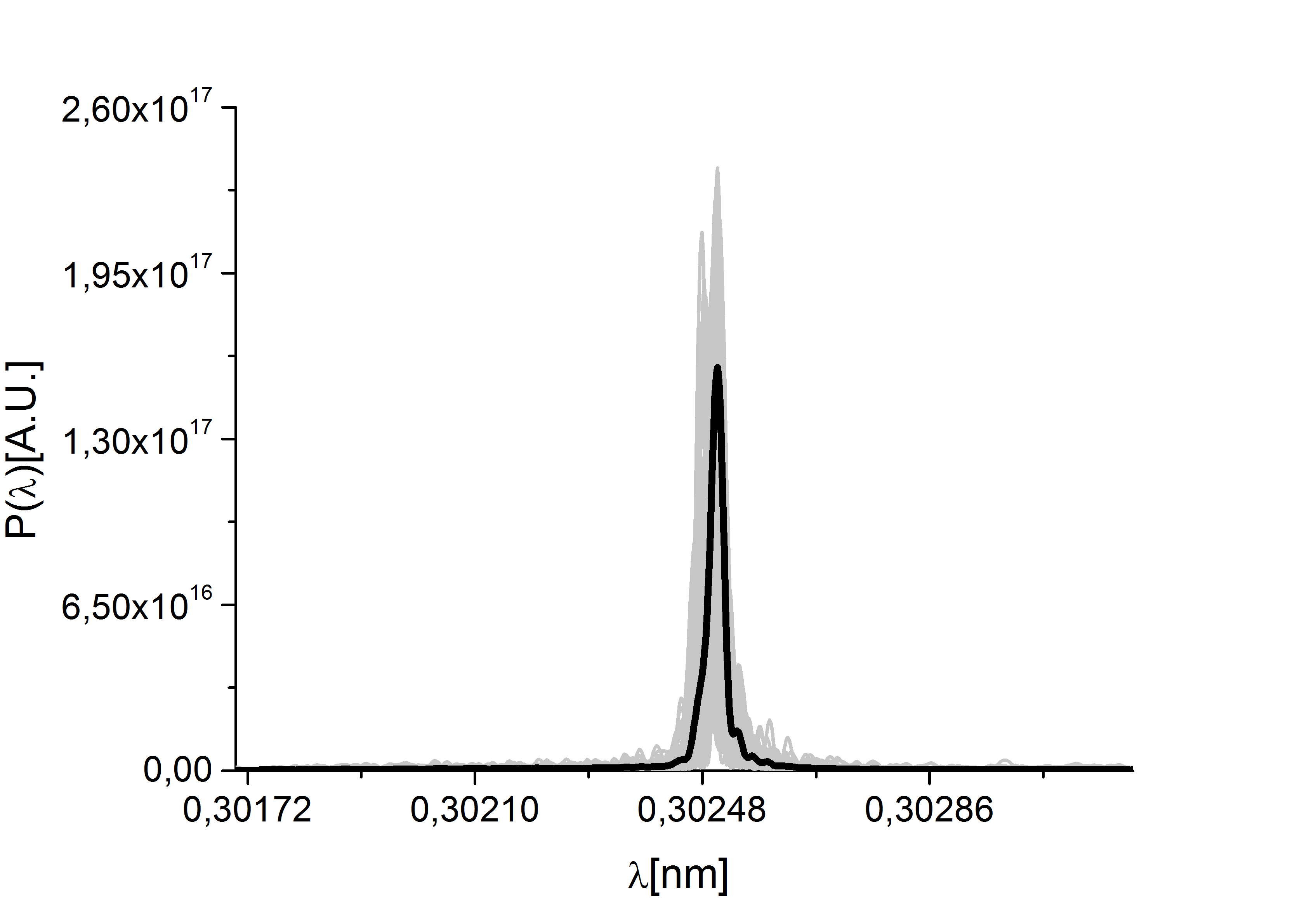}
\caption{Power distribution and spectrum of the output radiation
pulse for the case of long ($12$ fs) pulse mode of operation. } \label{PSpout1}
\end{figure}

\begin{figure}
\begin{center}
\includegraphics[width=0.50\textwidth]{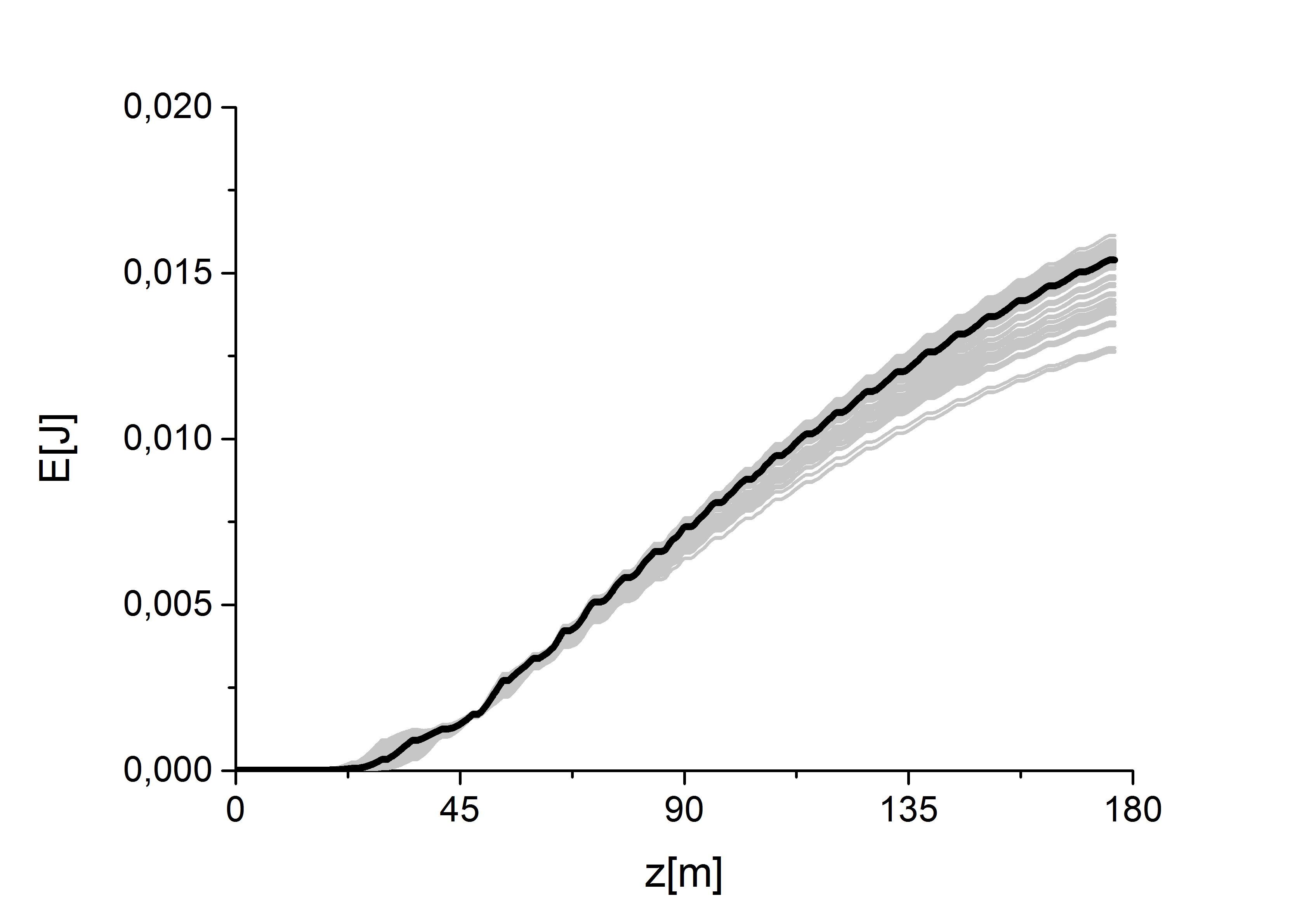}
\end{center}
\caption{Final output. Energy of the seeded FEL pulse as a function of the
distance inside the output undulator for the case of long ($12$ fs) pulse mode of operation.} \label{Eout1}
\end{figure}

Simulations were performed with the help of the Genesis code \cite{GENE} in the following way: first we calculated the 3D field distribution at the exit of the first undulator, and dumped the field file. Subsequently, we performed a temporal Fourier transform followed by filtering through the monochromator, by using the filter amplitude transfer function. The electron beam microbunching is washed out by the presence of a nonzero momentum compaction factor $R_{56}$ in the chicane. Therefore, for the second undulator we used a beam file with no initial microbunching, but with characteristics (mainly the energy spread) induced by the FEL amplification process in the first SASE undulator. The amplification process in the second undulator starts from the seed field file. Shot-noise initial condition were included. The evolution of the horizontal and vertical dimensions of the electron bunch as a function of the distance inside the undulator are presented in Fig. \ref{focus_lo}, which is valid for both the long ($12$ fs) and the short ($4$ fs) pulse case. The plots refer to the longitudinal position inside the bunch corresponding to the maximum current value. The output power and spectrum after the first SASE undulator tuned at 4.1 keV is shown in Fig. \ref{PSpin1}. The crystal acts as bandstop filter, and the signal in the time domain exhibits a long monochromatic tail, which is used for seeding. The electron bunch is slightly delayed by proper tuning of the self-seeding magnetic chicane, in order to superimpose the unspoiled part of the electron bunch with the seed signal. After saturation the undulator is tapered, i.e. the undulator $K$ parameter is changed section by section, following the configuration in Fig. \ref{Taplaw}.

The output power and spectrum of the entire setup, that is after the second part of the output undulator is shown in Fig \ref{PSpout1}.  The evolution of the output energy in the photon pulse is plotted in Fig. \ref{Eout1}.

\begin{figure}
\includegraphics[width=0.50\textwidth]{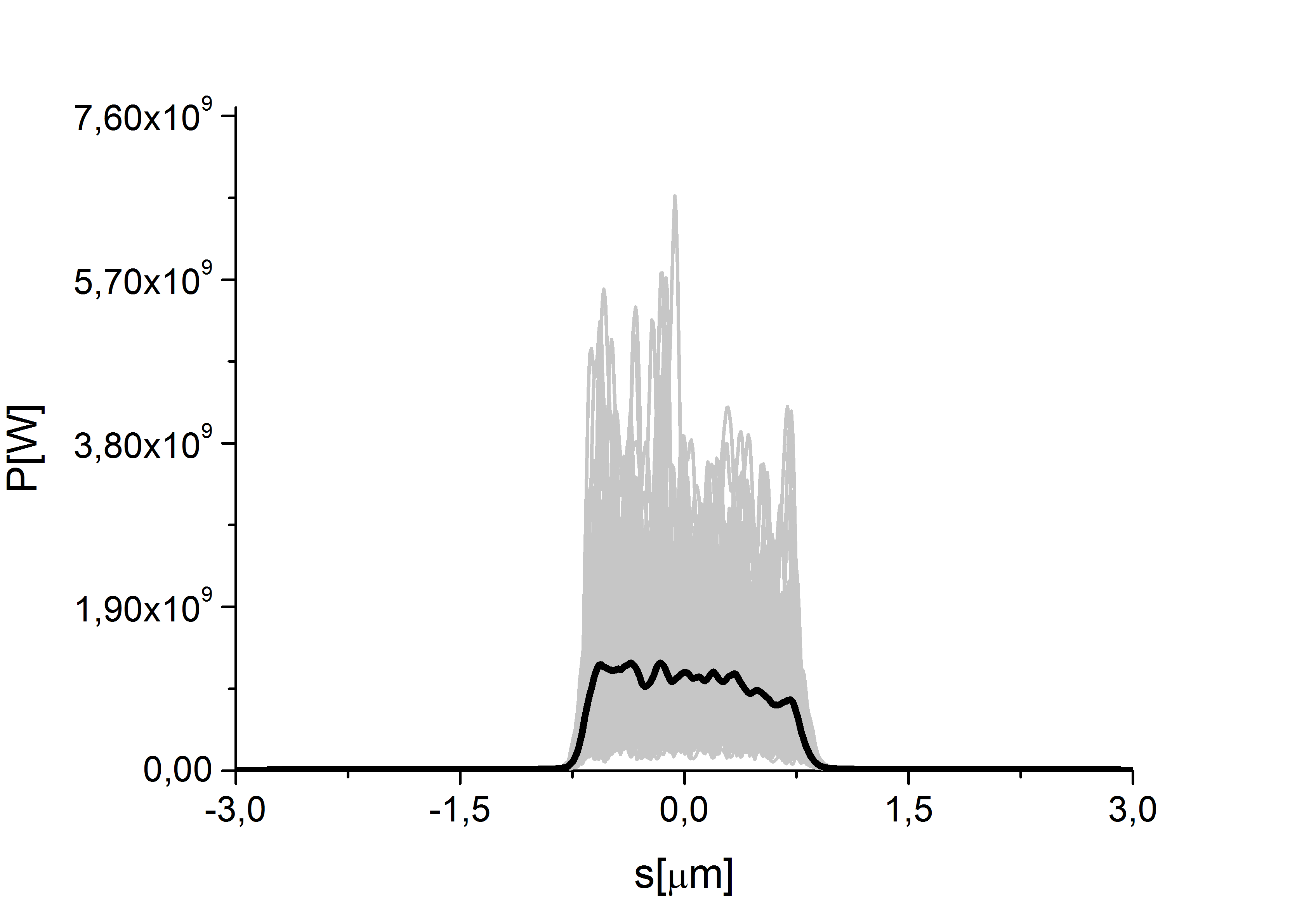}
\includegraphics[width=0.50\textwidth]{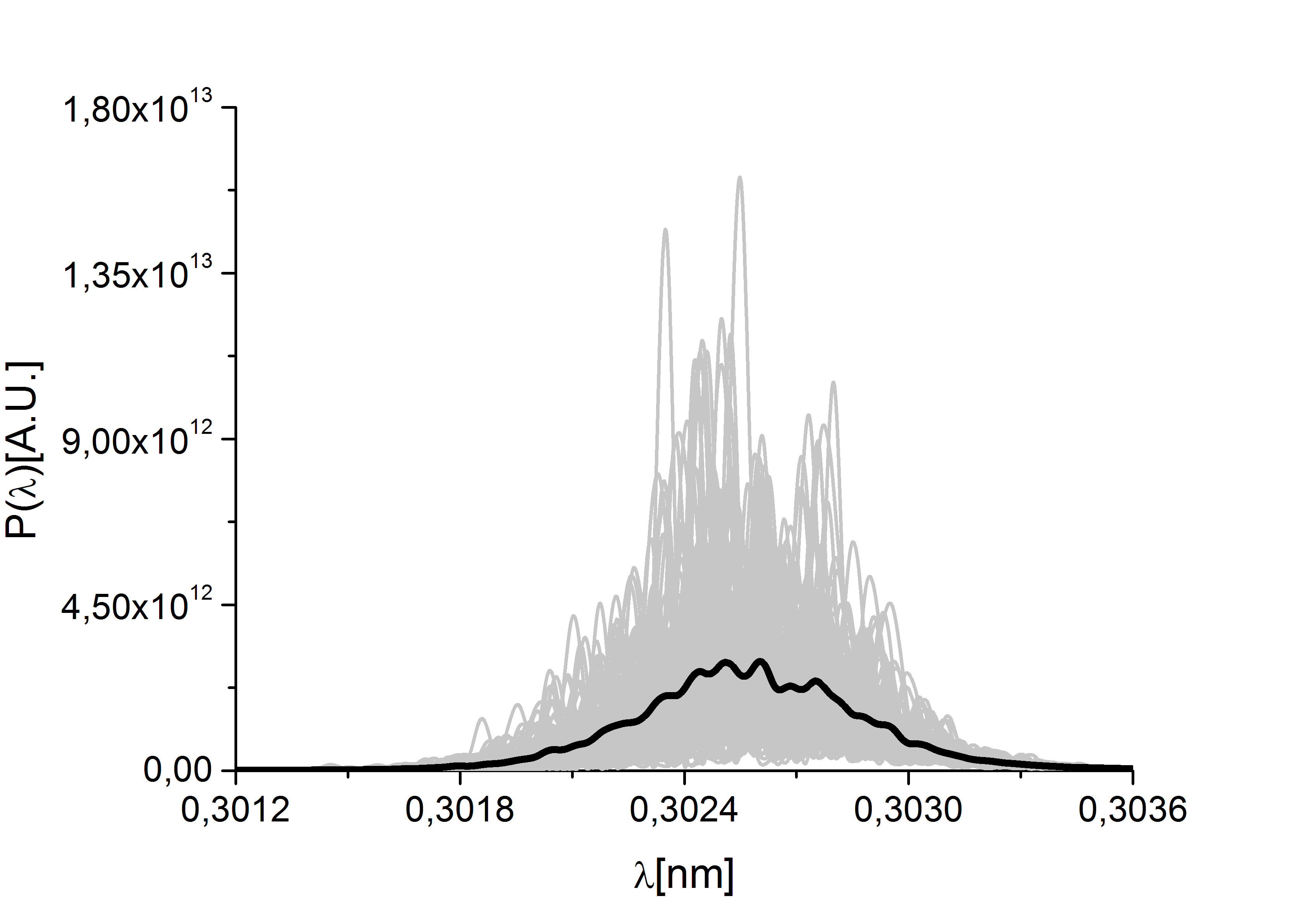}
\caption{Power distribution and spectrum of the SASE x-ray pulse at
the exit of the first undulator for the case of short ($4$ fs) pulse mode of operation.} \label{PSpin1_short}
\end{figure}

\begin{figure}
\begin{center}
\includegraphics[width=0.50\textwidth]{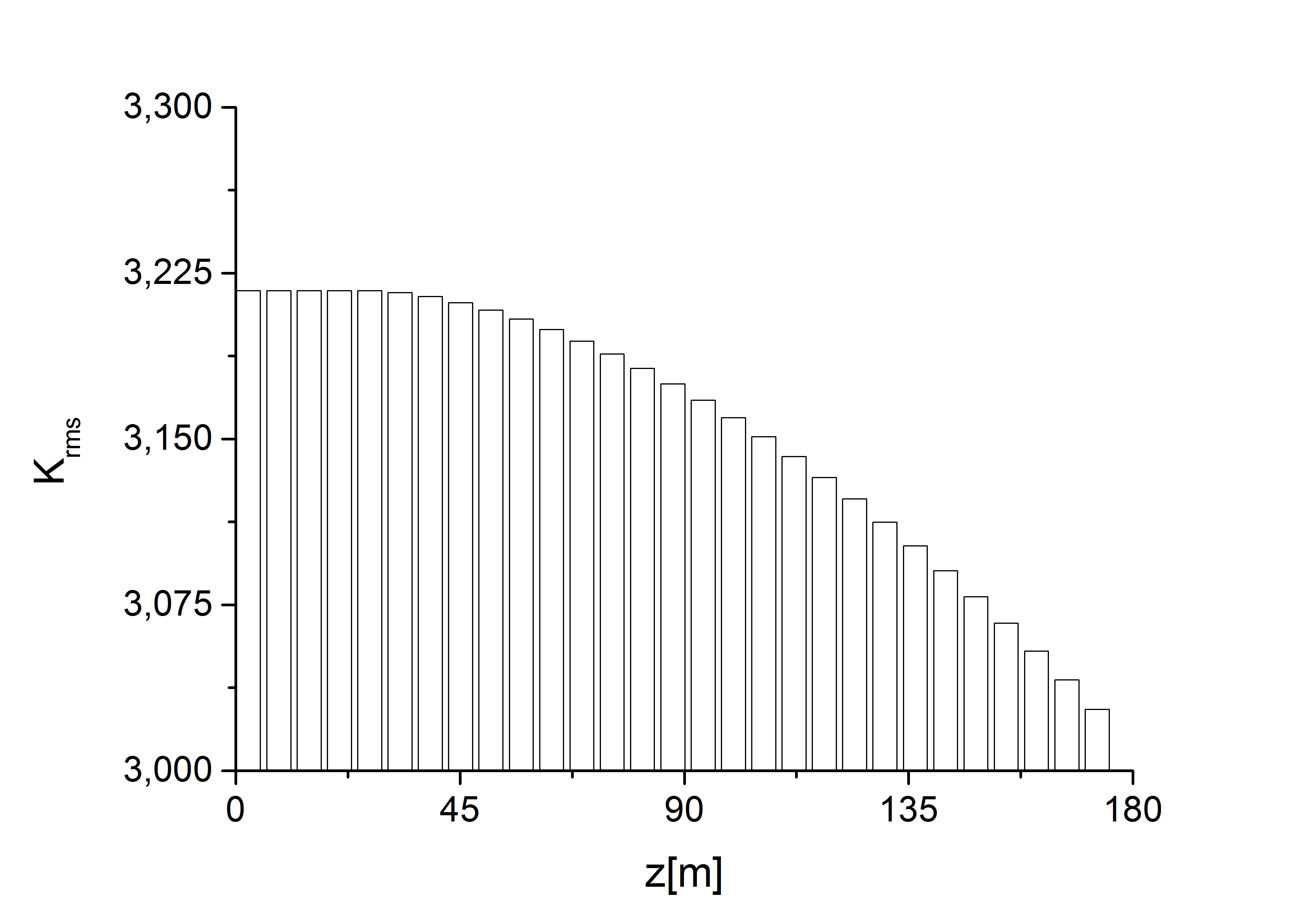}
\end{center}
\caption{Taper configuration for high-power mode of operation at
$0.302$ nm for the case of short ($4$ fs) pulse mode of operation.} \label{Taplaw_short}
\end{figure}

\begin{figure}
\includegraphics[width=0.50\textwidth]{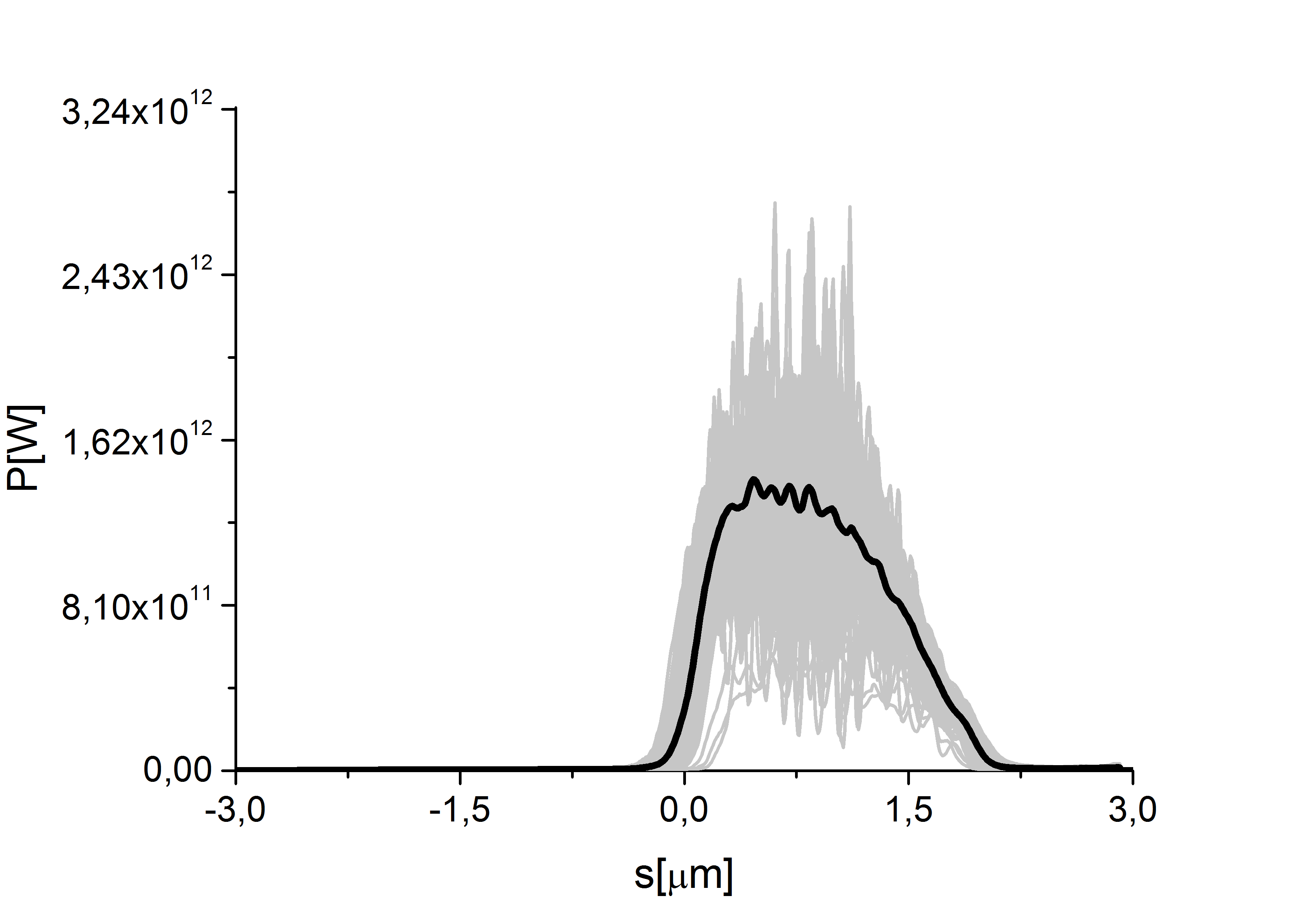}
\includegraphics[width=0.50\textwidth]{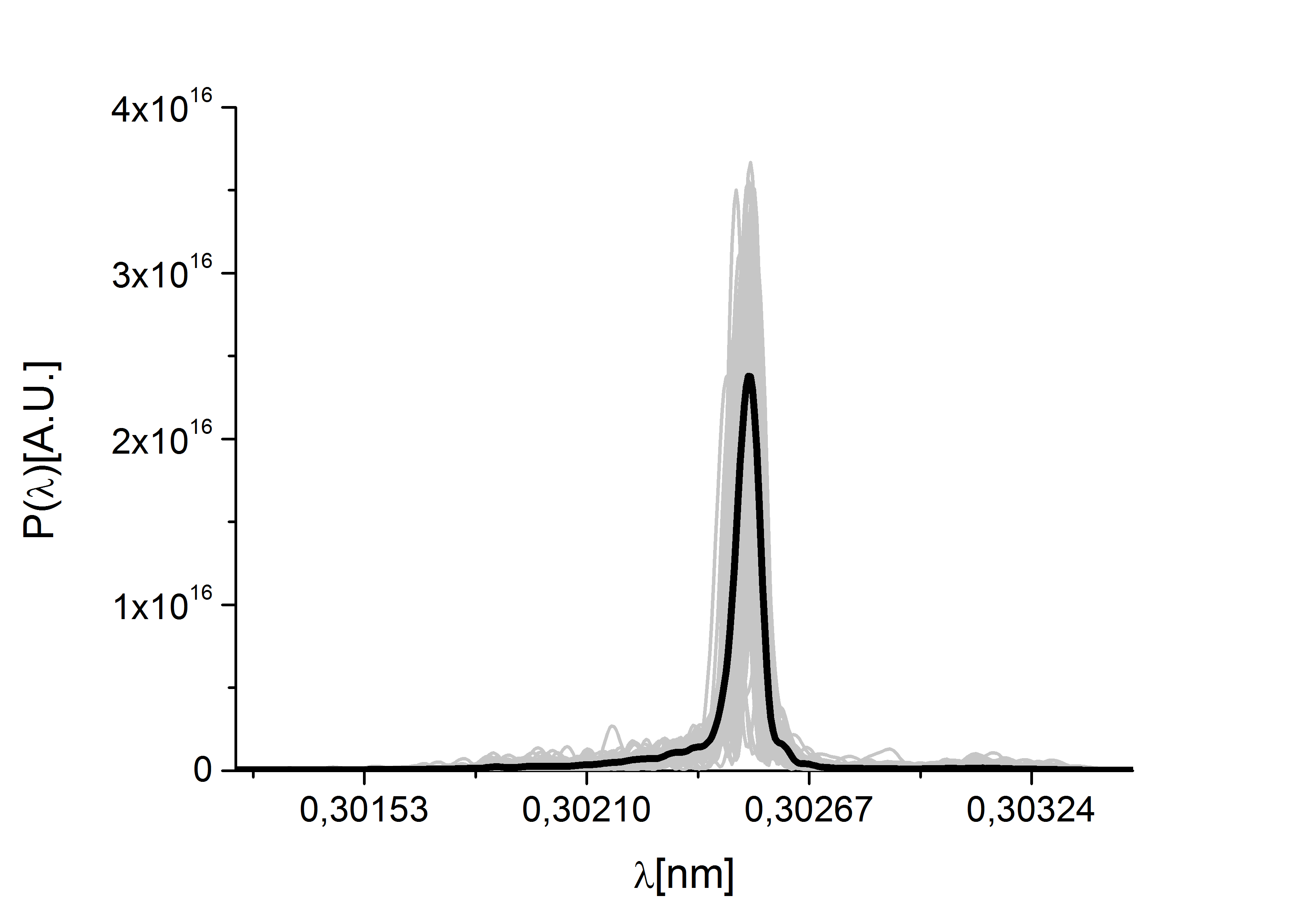}
\caption{Power distribution and spectrum of the output radiation
pulse for the case of short ($4$ fs) pulse mode of operation.} \label{PSpout1_short}
\end{figure}

\begin{figure}
\begin{center}
\includegraphics[width=0.50\textwidth]{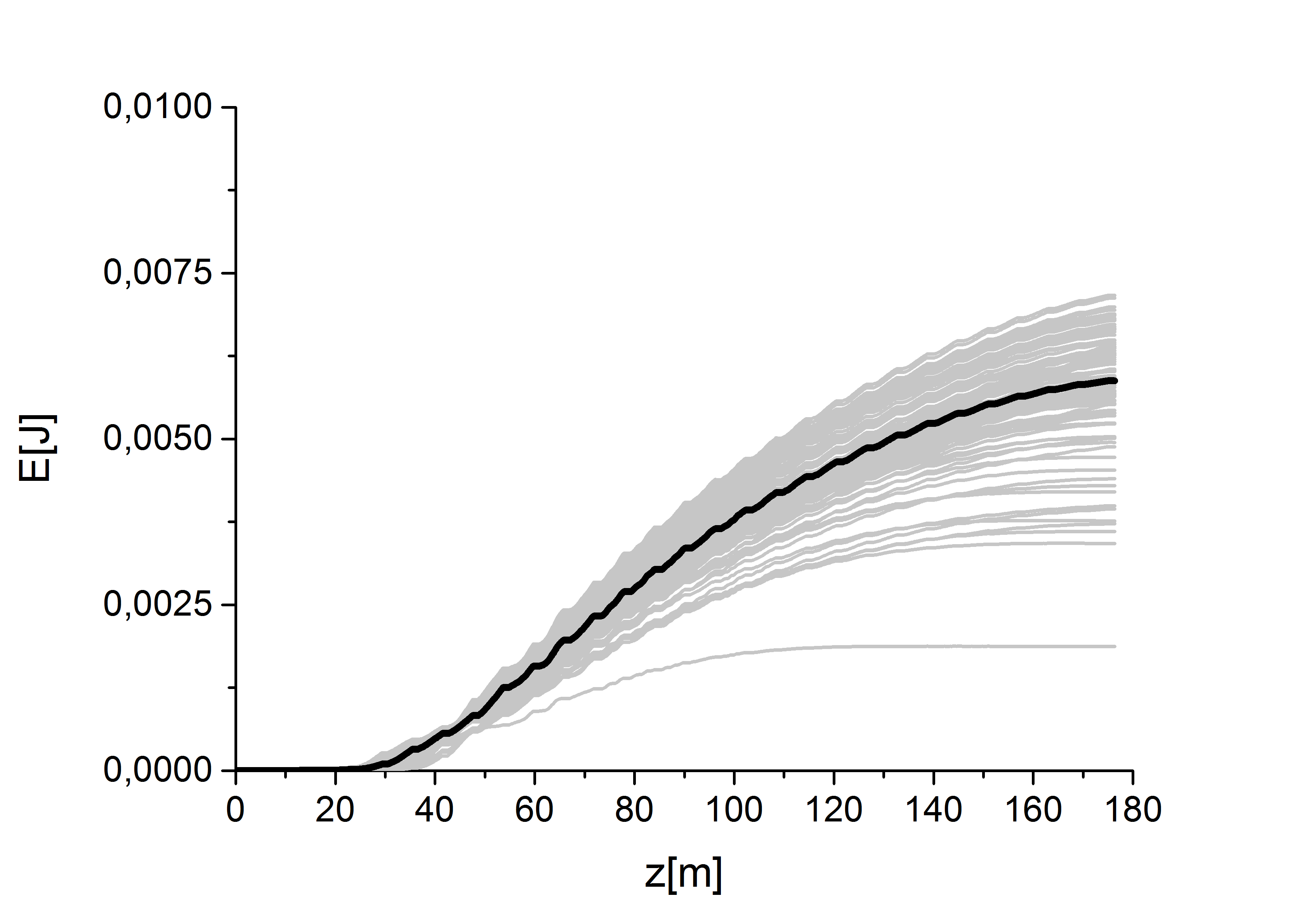}
\end{center}
\caption{Final output. Energy of the seeded FEL pulse as a function of the
distance inside the output undulator for the case of short ($4$ fs) pulse mode of operation..} \label{Eout1_short}
\end{figure}
The analogous results for the short pulse, that is for the $4$ fs case, are shown in Fig. \ref{PSpin1_short} - Fig. \ref{Eout1_short}.

\section{\label{sec:nano} Simulation of nano-scale focal spot}

The SPB optical layout used in our simulations was already presented in Fig. \ref{SPBsketch}. The KB mirror pair aims to produce a $100$ nm-scale spot from $B_4C$ coated mirrors with a length of $950$ mm. These mirrors can actually accept $4 \sigma$ of the X-ray beam from the SASE1 source at a photon energy around $4$ keV, under the assumption that the the accelerator complex and the SASE1 undulator operates in the special mode described above. In this section, results of wavefront propagation simulations are presented that have been carried out to investigate the evolution of the radiation beam profile through the SPB optics. In particular, we perform these simulations to investigate the KB system performance in terms of the beam profile at the sample position. In these calculations the source is described by the space-time radiation field distribution at the exit of the SASE1 undulator. We took advantage of the SRW code \cite{CHUB} in order to simulate the propagation of the radiation wave packet up to the sample position slice by slice. Our wave optics analysis takes into account all aberrations and errors from each optical element.

A very important issue that must be addressed is the preservation of the radiation wavefront from the source to the focal spot. The requirements on the X-ray mirrors are typically estimated based on ray-tracing codes for incoherent light sources. However, the XFEL beam will be almost diffraction-limited, and one needs to perform simulations of the effects of the mirror imperfections with the help of wavefront propagation codes. It is easy to demonstrate that an error $\delta h$ on the optical surface will perturb the wavefront of a phase $\phi$, according to

\begin{eqnarray}
\phi = \frac{4 \pi \delta h}{\lambda} \sin \theta_i~, \label{phidh}
\end{eqnarray}
where $\theta _i$ is the angle of incidence with respect to the surface.

In practice, $\phi$ represents the deformation of the wavefront in the propagation direction divided by the wavelength.

\begin{figure}
\begin{center}
\includegraphics[trim = 0 500 0 50, clip,width=0.75\textwidth]{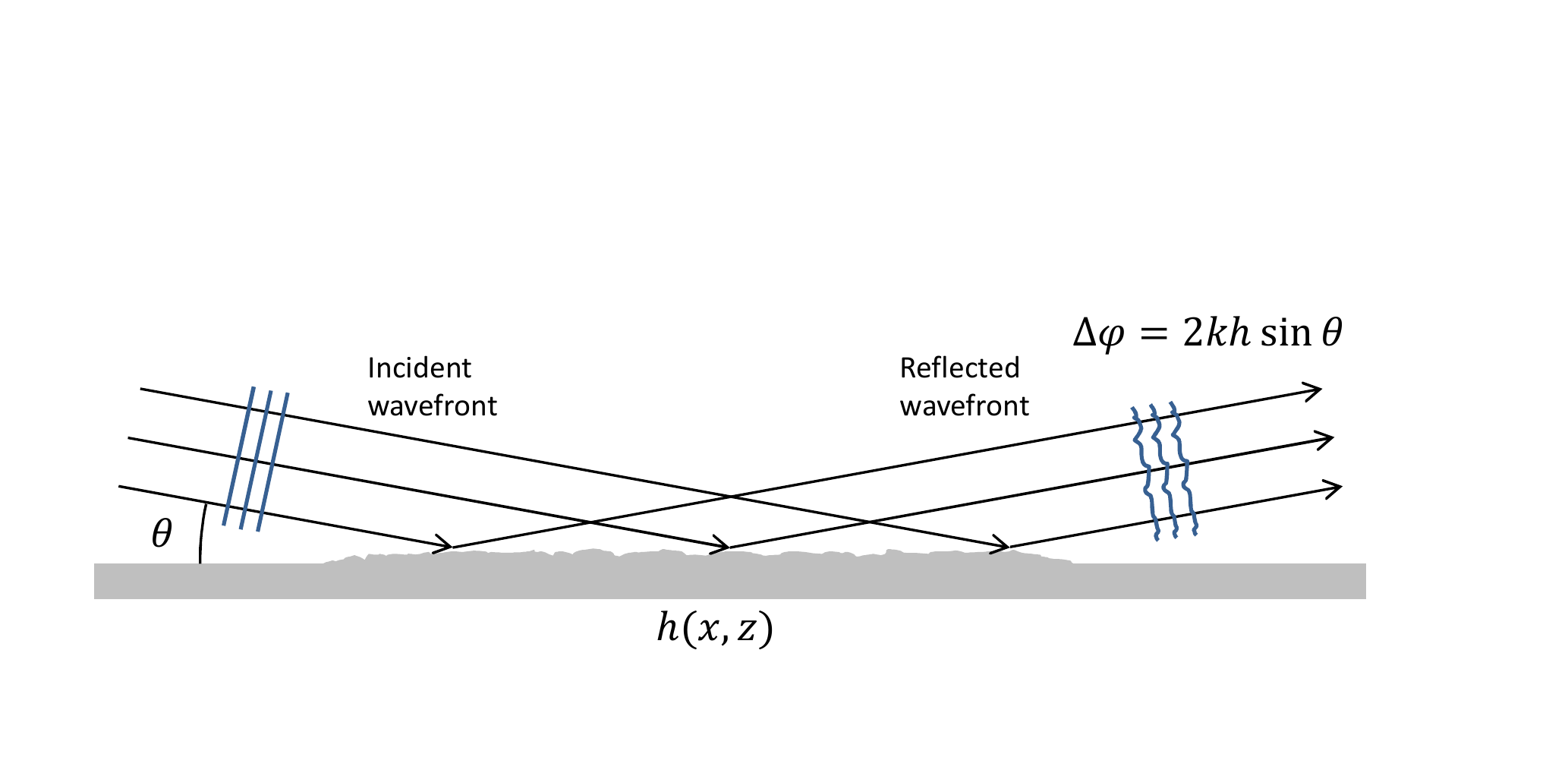} 
\end{center}
\caption{Thin-shifter-like behavior of surface roughness for small mean square of surface displacement, adapted from \cite{BART}.}
\label{roughness_scheme}
\end{figure}
A reflection from the mirror becomes similar to the propagation through a transparency at the mirror position, which just changes the phase of the reflected beam without changing its amplitude, \cite{BART} (see Fig. \ref{roughness_scheme}). For the shifter model to be applicable, the phase change must be small, i.e. $|\phi| << 1$.

\begin{figure}
\begin{center}
\includegraphics[trim = 0 0 0 0, clip,width=0.75\textwidth]{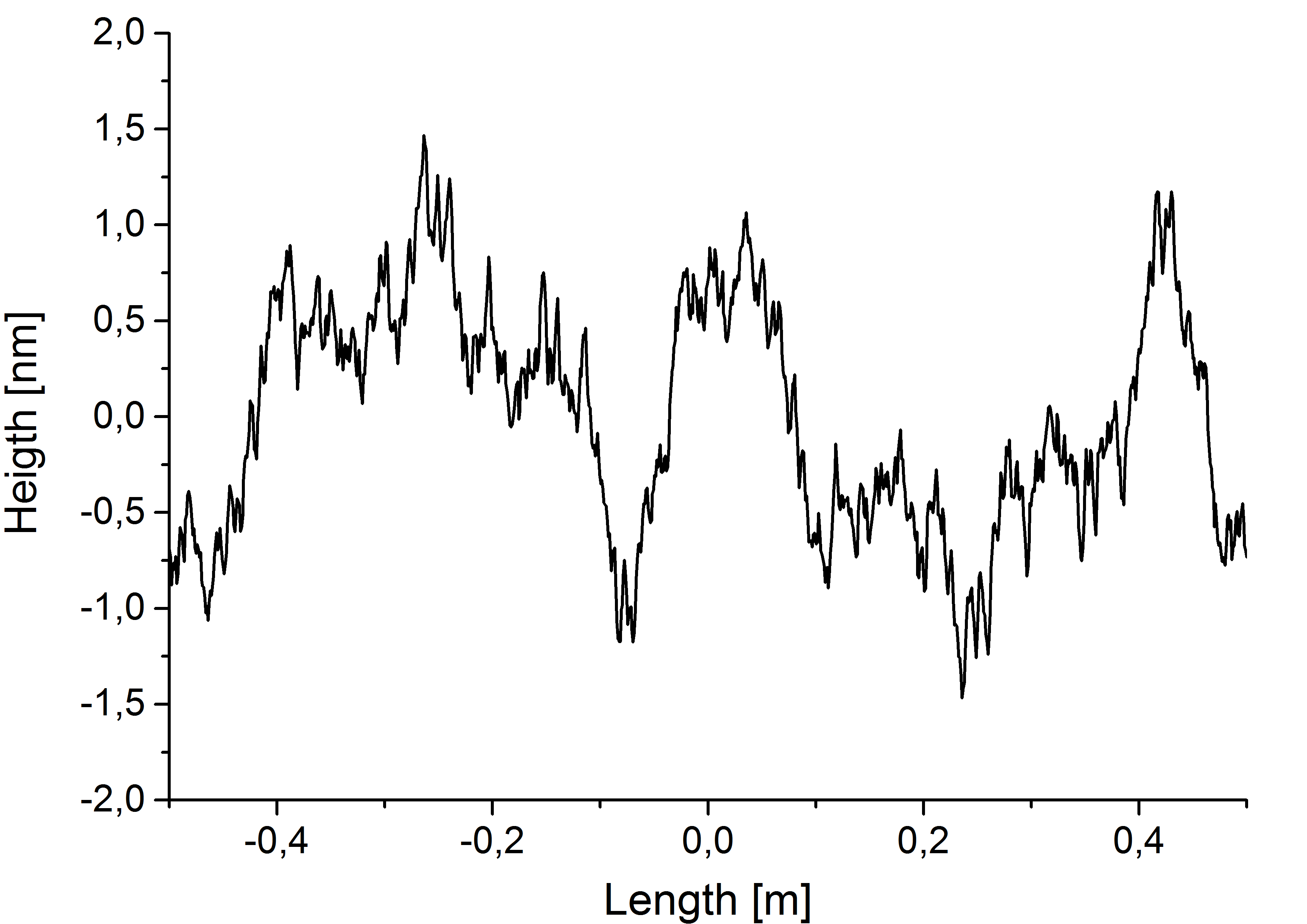} 
\end{center}
\caption{Simulation of residual height error of the mirror surface. The PSD data from \cite{XCDR} is scaled to produce a height error $\delta h_\mathrm{rms}  = 1$ nm.}
\label{roughness_example}
\end{figure}

The problem of simulating mirror surface errors it then reduced to the proper description of phase shifters.  Applying the Marechal criterion, i.e. requiring a Strehl ratio larger than $0.8$, and treating the errors from the different optics independently, we obtain the following condition for the rms height error $h_\mathrm{rms}$ \cite{HEIM}:

\begin{eqnarray}
2 h_\mathrm{rms} \theta_i \sqrt{N} < \lambda/14 ~, \label{Mare}
\end{eqnarray}
where N is number of mirrors. From Marechal criterion we conclude that an height error $h_\mathrm{rms} < 1.5$ nm should be sufficiently small for diffraction-limited propagation through the SBP beamline at a photon energy of $4$ keV. The most tight requirements correspond to the shortest wavelength, and the SPB beamline designers are planning to use mirrors with a height error $h_\mathrm{rms}  \sim 0.6$ nm in order to preserve the geometrical focus properties in the short wavelength range \cite{MANC,MANT}. Notwithstanding this fact, one should consider that  Eq. (\ref{Mare}) is an estimation only, and one needs to perform detailed simulations including surface error effects in order to understand the requirements on the roughness in our case of interest. The surface errors are usually generated from power spectral density (PSD) functions described  in the mirror specifications. The part of  the PSD that makes the most significant contribution to the overall height error, is the low spatial-frequency part. The PSD data from \cite{XCDR} has been used for these calculations. An example of height profile is shown in Fig. \ref{roughness_example}. Due to the small incident angle, the beam footprint is much larger in the tangential direction, compared to the sagittal direction.  It follows that the mirrors disturb the wavefront mainly in the tangential direction.

We preformed simulations using the SRW code \cite{CHUB}. The surface figure can be directly mapped onto the optical field coordinate system using the geometrical transformation described in Eq. (\ref{phidh}). Since the wavefront propagation results suggest that an rms height error of $0.6$ nm would increase the beam size at 4 keV by up to $5 \%$ only, we specify an rms height error of 1 nm for all mirrors.

The effects of the horizontal offset mirrors in the X-ray beam transport are modeled using the code SRW as a combination of two apertures with sizes determined by the mirror length ($800$ mm in our case) and two phase shifters describing the mirror surface errors\footnote{The second offset mirror is bendable. However, these simulations consider the case where the second mirror is bent in such a way to be plane.}.

The SRW code has further the capability of modeling the KB optics by elliptical mirrors  (with length $950$ mm in our case) and to account for all aberrations.  Geometrical ellipse parameters for each KB mirror are completely specified by the incident angle and two distances as illustrated in Fig. \ref{Ellip}. Our simulations consider only the baseline SPB optics distances presented in Fig. \ref{SPBsketch}. The KB mirror surface errors are simulated by two phase shifters, similar to the case of offset mirrors.

\begin{figure}
\begin{center}
\includegraphics[trim = 0 0 0 0, clip, width=0.9\textwidth]{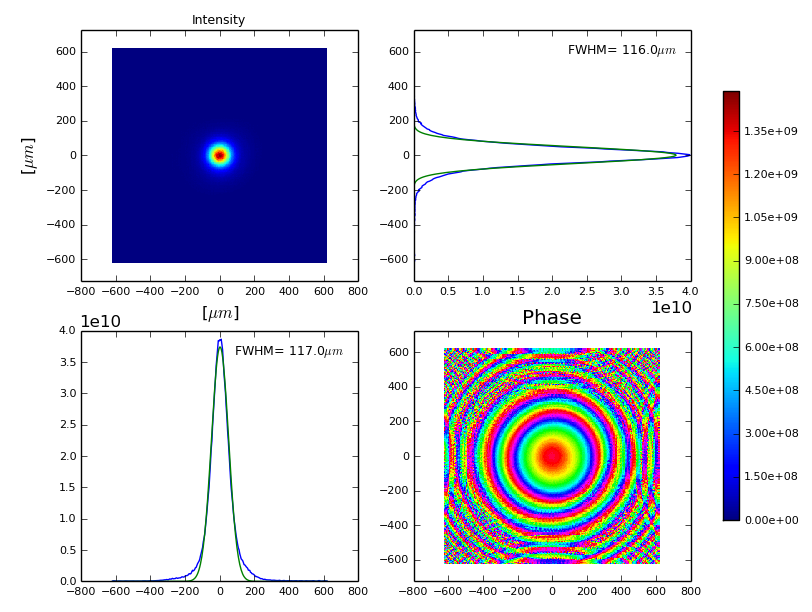}
\includegraphics[trim = 0 0 0 0, clip, width=0.9\textwidth]{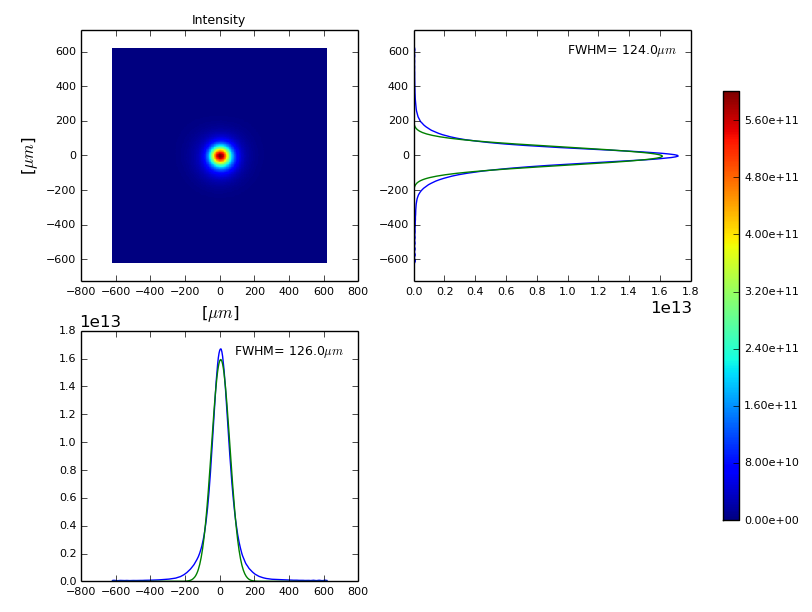}
\end{center}
\caption{Upper plot. Calculated intensity and phase distribution of the FEL beam at the exit of SASE 1 undulator. Results refer a longitudinal position inside the radiation pulse roughly corresponding to the maximum power value. Lower plot. Distribution of the radiation  pulse energy per unit surface at the exit of SASE1 undulator, integrated over the radiation pulse. The color scales are in arbitrary units.}
\label{slice_1}
\end{figure}

\begin{figure}
\begin{center}
\includegraphics[trim = 0 0 0 0, clip,width=0.9\textwidth]{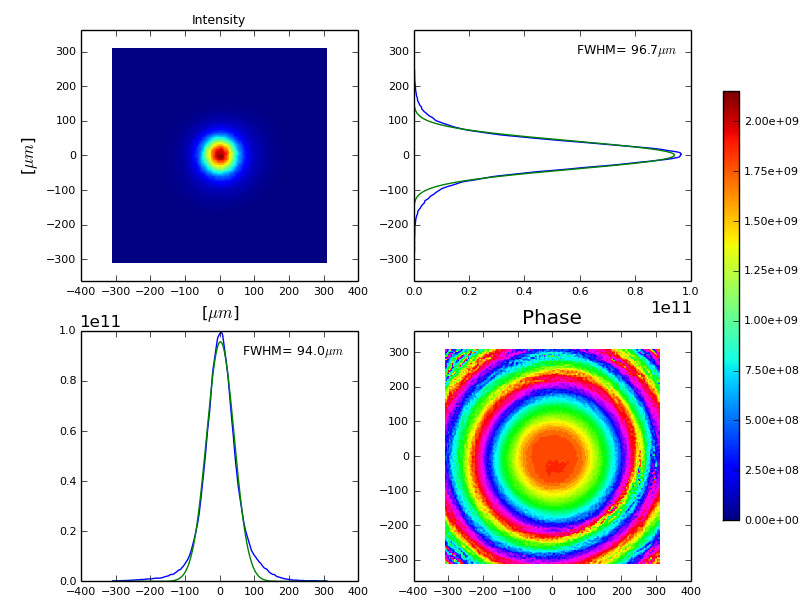}
\includegraphics[trim = 0 0 0 0, clip,width=0.9\textwidth]{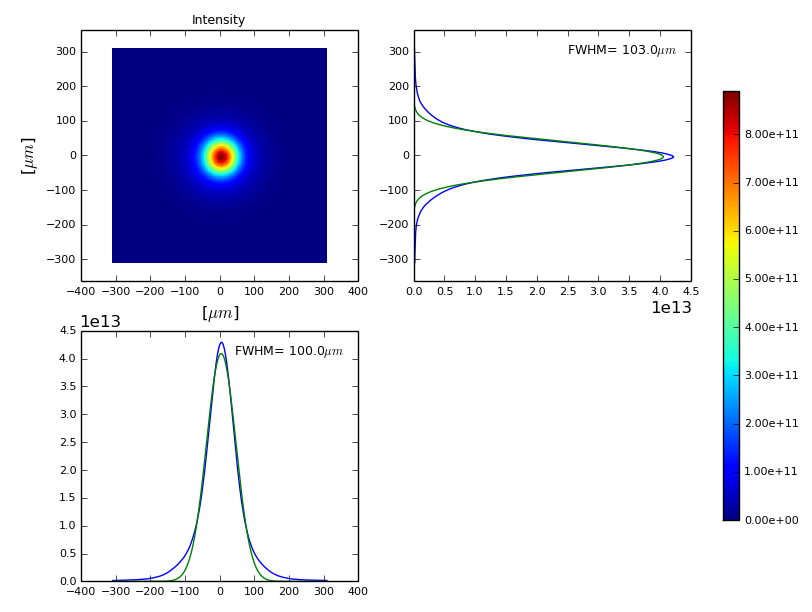}
\end{center}
\caption{Upper plot. Calculated intensity and phase distribution of the FEL source placed $26$ m upstream the undulator exit. Results refer a longitudinal position inside the radiation pulse roughly corresponding to the maximum power value. Lower plot. Distribution of the radiation  pulse energy per unit surface in the source plane, placed $26$ m upstream of the undulator exit, integrated over the radiation pulse. The color scales are in arbitrary units.}
\label{slice_2}
\end{figure}

\begin{figure}
\begin{center}
\includegraphics[trim = 0 0 0 0, clip,width=0.9\textwidth]{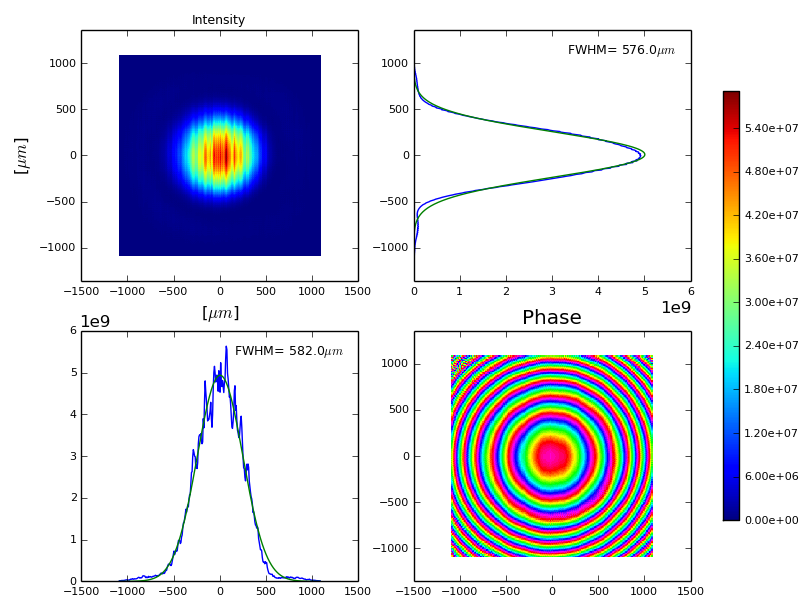}
\includegraphics[trim = 0 0 0 0, clip,width=0.9\textwidth]{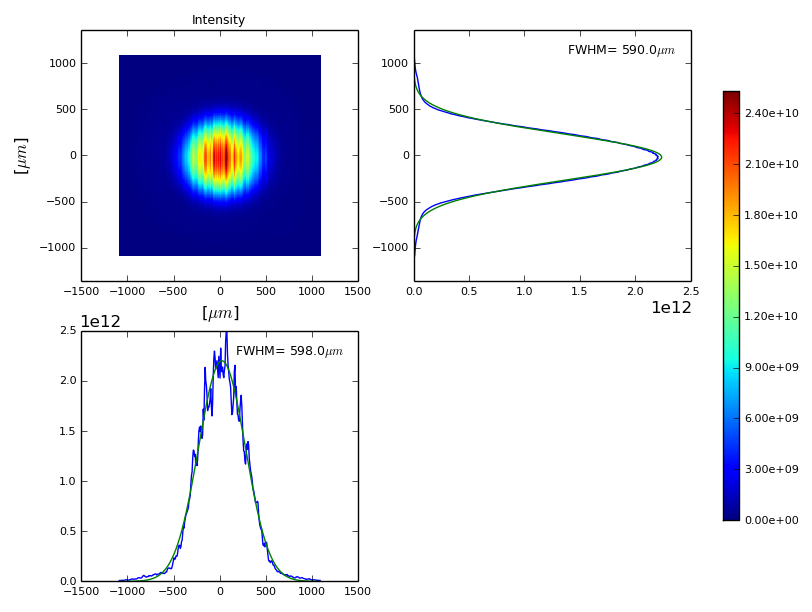}
\end{center}
\caption{Upper plot. Calculated intensity and phase distribution of the FEL beam at a distance of $5$ m downstream of the second offset mirror. Calculations take into account the roughness of both offset mirrors in the beam path. Results refer a longitudinal position inside the radiation pulse roughly corresponding to the maximum power value. Lower plot. Distribution of the radiation pulse energy per unit surface at a distance of $5$ m downstream of the second offset mirror, integrated over the radiation pulse. The color scales are in arbitrary units.}
\label{slice_3}
\end{figure}
\begin{figure}
\begin{center}
\includegraphics[trim = 0 0 0 0, clip,width=0.9\textwidth]{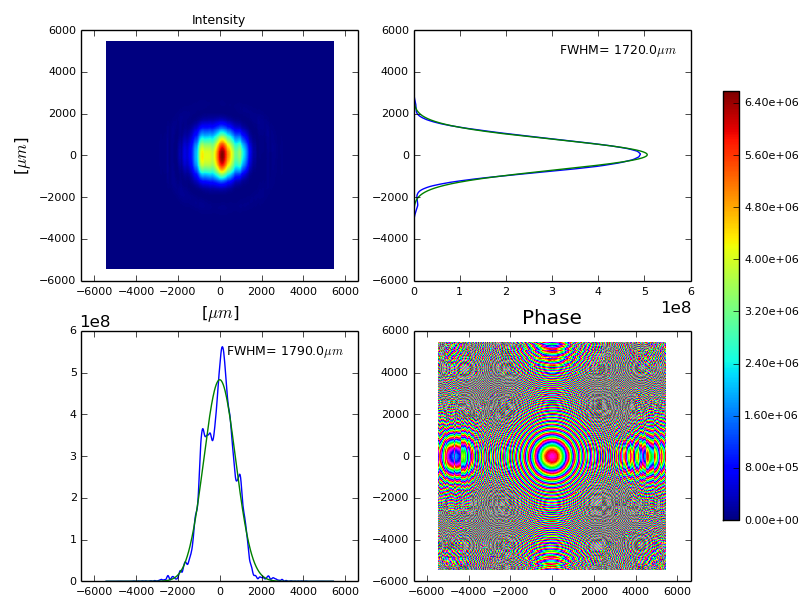}
\includegraphics[trim = 0 0 0 0, clip,width=0.9\textwidth]{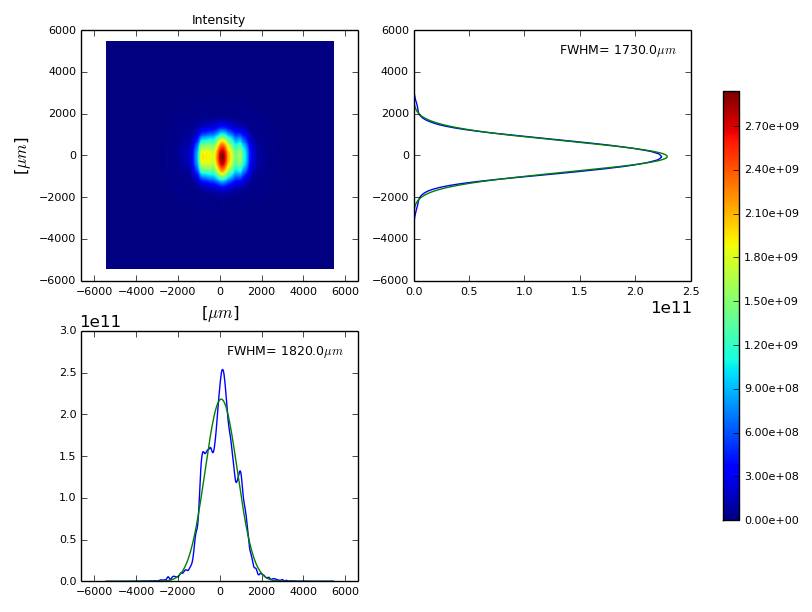}
\end{center}
\caption{Upper plot. Calculated intensity and phase distribution of the FEL beam in the plane immediately in front of first KB mirror. Results refer a longitudinal position inside the radiation pulse roughly corresponding to the maximum power value. Lower plot. Distribution of the radiation pulse energy per unit surface in the plane immediately in front of first KB mirror, integrated over the radiation pulse. The color scales are in arbitrary units.}
\label{slice_4}
\end{figure}
\begin{figure}
\begin{center}
\includegraphics[trim = 0 0 0 0, clip,width=0.9\textwidth]{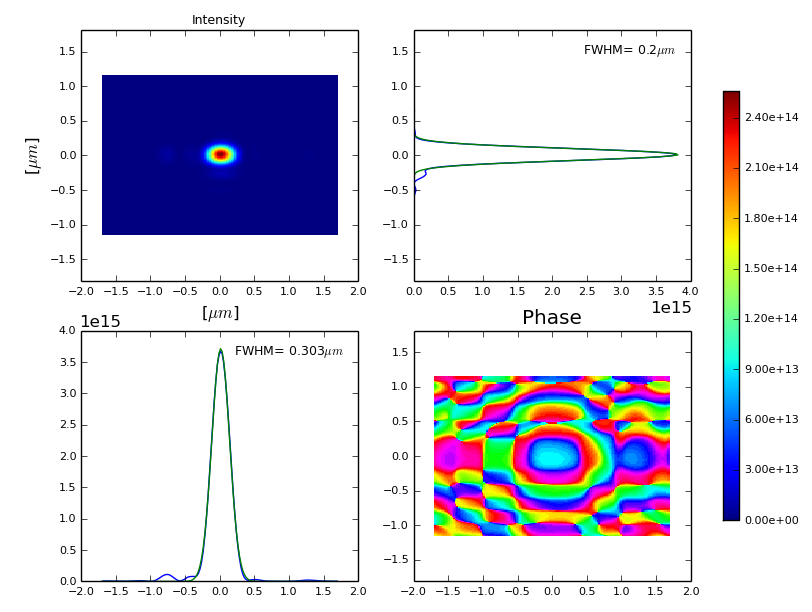}
\includegraphics[trim = 0 0 0 0, clip,width=0.9\textwidth]{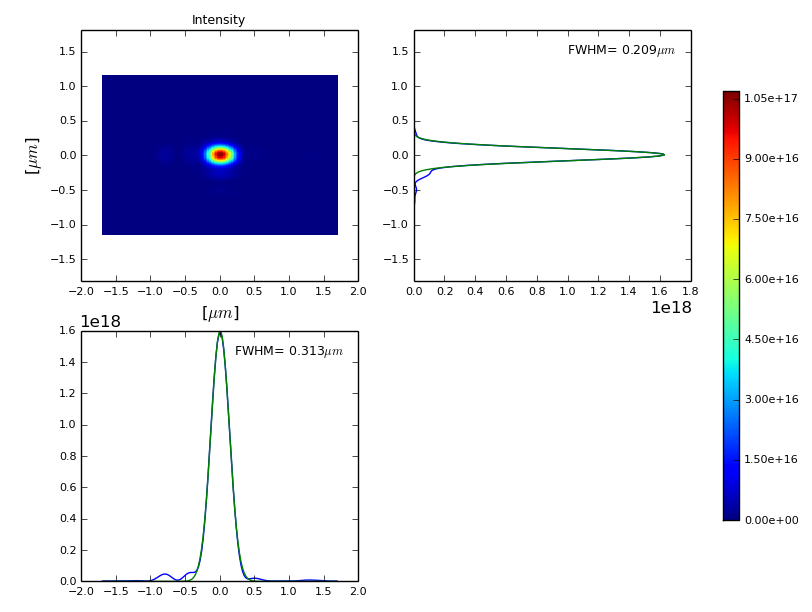}
\end{center}
\caption{Upper plot. Calculated intensity and phase distribution of the FEL beam in the focus. Results refer a longitudinal position inside the radiation pulse roughly corresponding to the maximum power value. Lower plot. Distribution of the radiation pulse energy per unit surface in the plane placed in the focus, integrated over the radiation pulse. The color scales are in arbitrary units.}
\label{slice_5}
\end{figure}
\begin{figure}
\begin{center}
\includegraphics[trim = 0 0 0 0, clip,width=0.9\textwidth]{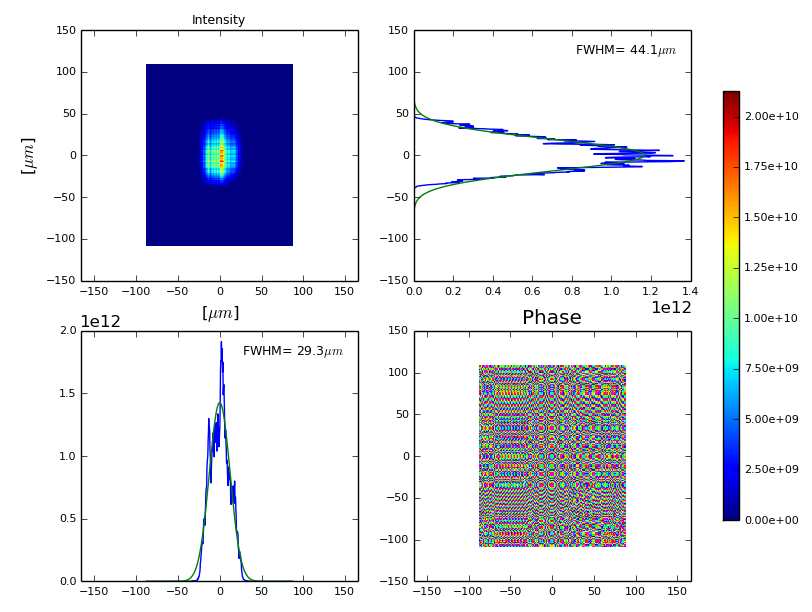}
\includegraphics[trim = 0 0 0 0, clip,width=0.9\textwidth]{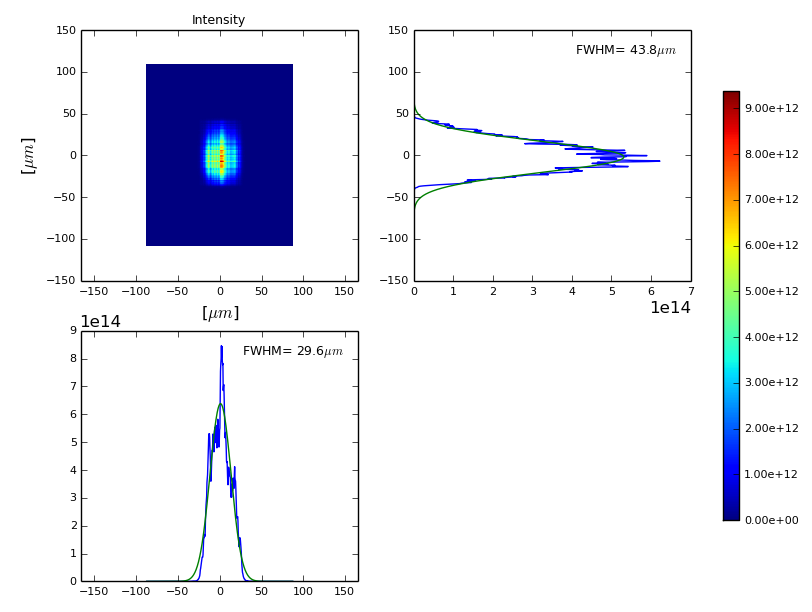}
\end{center}
\caption{Upper plot. Calculated intensity and phase of the FEL beam propagated $5$ cm out of the focus into the far zone. The stripped pattern originates from the roughness of all four mirrors in the beam path. Results refer a longitudinal position inside the radiation pulse roughly corresponding to the maximum power value. Lower plot. Distribution of the radiation pulse energy per unit surface in the plane placed $5$ cm downstream of the focus, integrated over the radiation pulse. The color scales are in arbitrary units.}
\label{slice_6}
\end{figure}
The initial field distribution consists of the SASE1 output radiation field distribution presented in Section \ref{sec:SASE1}. The upper plots in Fig. \ref{slice_1}-\ref{slice_6} show wavefront propagation results for a particular longitudinal position inside the radiation beam, which was chosen well within the 'flat-top' profile, so that it roughly corresponds to the maximum power values. As discussed above, we specified $1$ nm rms height errors for all mirrors at a photon energy of $4.1$ keV. The KB mirrors have incident angle is fixed to $3.5$ mrad. In the mode of operation proposed here, the source divergence is about $2 \mu$rad, and the KB mirrors accept $4 \sigma$ of the beam even at the design incident angle of $3.5$ mrad.  For our decreased source divergence, the issue of diffraction from mirror apertures does not exist at all.  The structures that can be seen in the transverse radiation profiles in the figures originate from the height errors of the mirrors only.

The lower plots in Fig.\ref{slice_1}-\ref{slice_6} show intensities at the same positions, but integrated over the radiation pulse. These plots are thus simulations of the energy profile per unit surface that can be measured by a detector that integrates over a single radiation pulse, placed in the plane of interest. Note that the vertical focus size is overall smaller than the horizontal one of about $1.6$ times. This is because the vertical KB mirror is closer to the sample to get a better demagnification. Simulations suggest, in agreement with the above estimations, that height errors of $1$ nm rms are tolerable for photon energies around 4 keV. Moreover, very high transmission close to $100 \%$ can be achieved with $B_4 C$ coating and, as already remarked, beam cutoff by mirror apertures is insignificant. This allows for a very high fluence. In fact, the maximal fluence in the focus may now be obtained from  $F = N_p/S$, where $N_p$ is the number of photons into the radiation pulse and $S$ is the effective focal spot squared, which is defined by

\begin{eqnarray}
\frac{1}{S} =  \frac{\mathrm{max}(W)}{\int W(x,y) dx dy}~  ,
\label{S}
\end{eqnarray}
where the integral is understood to span over the entire plane. Here  $W$ is the distribution of the pulse radiation energy per unit surface in the focal plane, Fig. \ref{slice_5}.

We found that  $1/S \sim 1.5\cdot10^9$ cm$~^{-2}$. For $10^{13}$ photons per pulse, which can be achieved as discussed in Section \ref{sec:SASE1}, this amounts to a fluence of about $1.5 \cdot 10^{22}$ photons/cm$~^2$. This result can be achieved without additional cost for the baseline optical layout of the SPB beamline, and with very moderate costs for the installation of the slotted foil setup into the beam formation system.

Since for our special mode of operation the source size is about $100 \mu$m, which is about two times larger than in the nominal mode of operation, we have about $200$ nm and $300$ nm vertical and horizontal spot sizes, respectively. These sizes are well within the geometrical and the diffraction limits. meaning that a significant improvement of the beam profile can not be achieved by only improving the optics quality, without changes in the optical layout. The X-ray optical layout of the SPB beamline provides an option of operation with an intermediate source point in the horizontal plane \cite{SPRI}. In this case, the image point of the second bendable offset mirror is the source point for the horizontal KB mirror, and a smaller horizontal focus at a sample, of about a factor 3, is in principle possible. However, the specifications for the bendable offset mirror and for the horizontal KB mirror should satisfy these refocusing conditions. We thus remark that with almost the same cost, but with significant mirror specification changes, a higher fluence of about $0.5 \cdot 10^{23}$ photons/cm$~^2$ is within reach.

Finally, we note that in order to achieve an additional tightening in the focusing of about a factor $2$ in the vertical plane, one can also additionally install two vertical  mirrors, one bendable and one plane, immediately after the offset mirrors, thus reaching an optical layout similar to that of the SASE3 beamline. These special refocusing mirrors can be used only in the photon energy range around 4 keV and should therefore be retractable. In this way, with a moderate additional cost, one may reach a fluence of about $1.0 \cdot 10^{23}$.

\section{\label{sub:noise} Simulated noisy coherent X-ray diffraction patterns}

\begin{figure}
\begin{center}
\includegraphics[trim = 0 0 0 0, clip,width=0.4\textwidth]{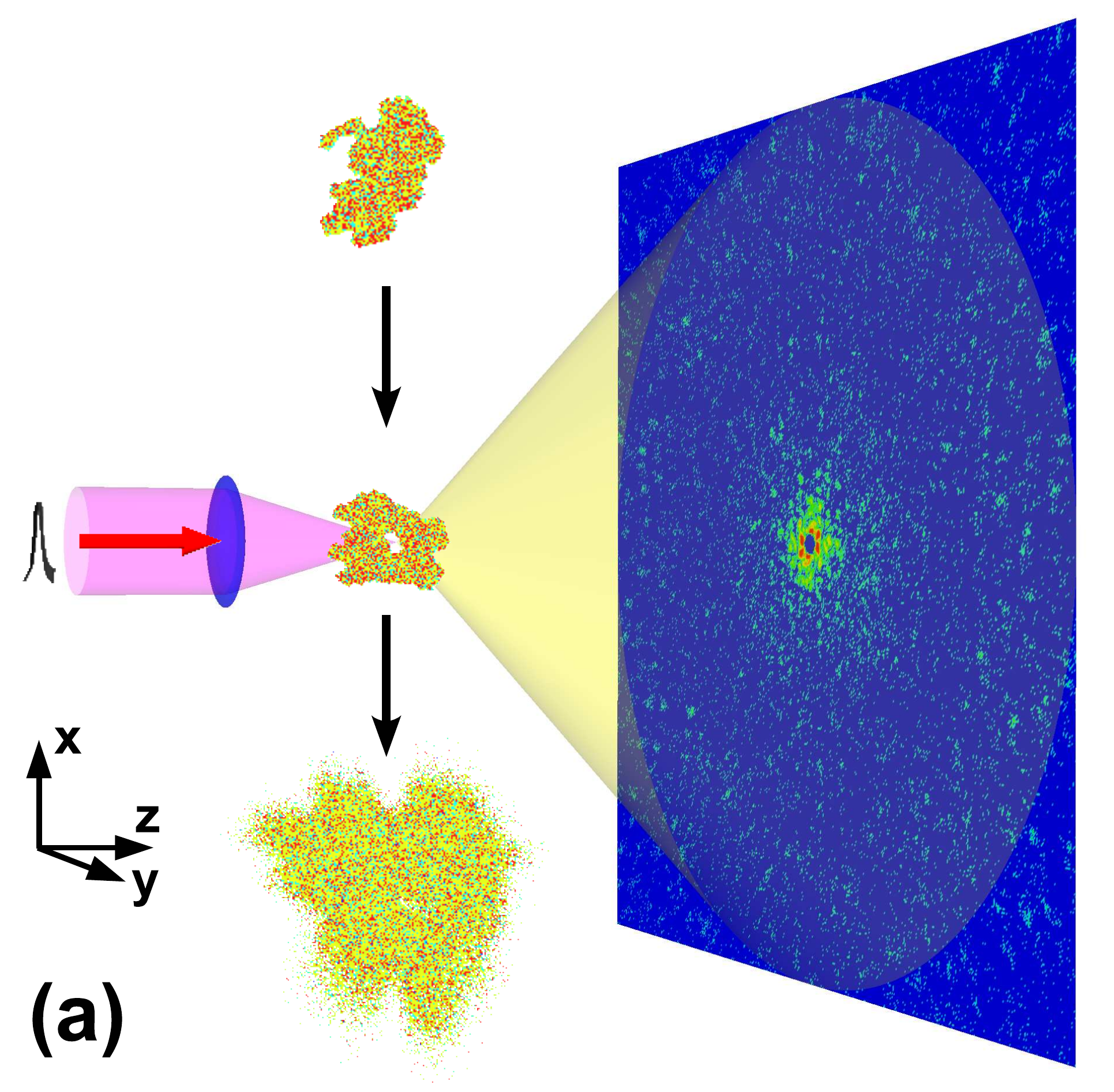}
\includegraphics[trim = 0 0 0 0, clip,width=0.4\textwidth]{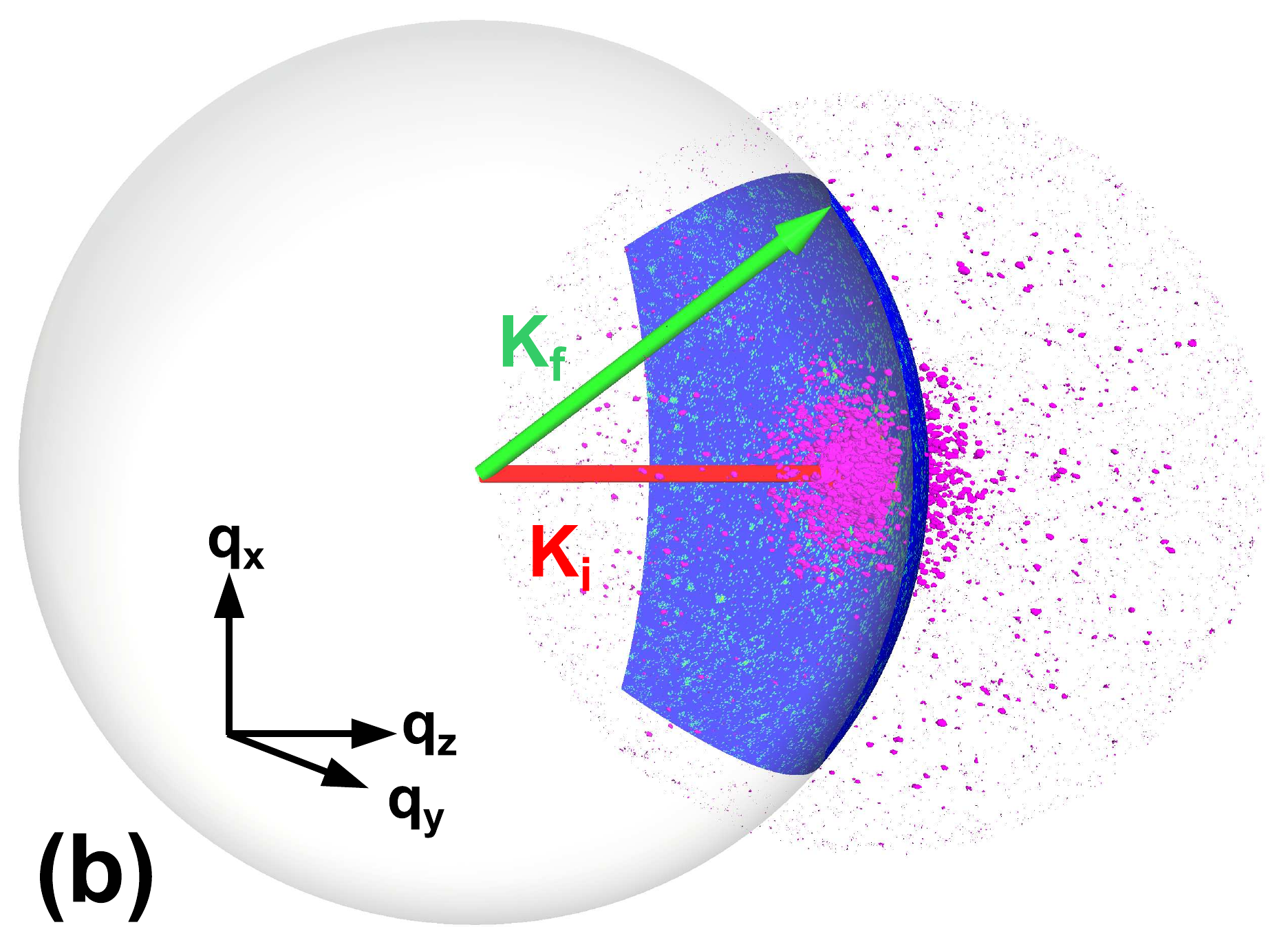}
\end{center}
\caption{(Color online) Schematic view of the experimental geometry. (a) In real space, diffraction
pattern from a sample in random orientation is measured by a single FEL pulse. (b) In reciprocal
space the measured diffraction pattern correspond to a cut of the 3D intensity distribution by an
Ewald sphere sector. Vectors $\vec{K}_i$ and $\vec{K}_f$ denote the incident and diffracted wavevectors. Adapted form \cite{YEFA}.}
\label{samsch}
\end{figure}
\begin{figure}
\begin{center}
\includegraphics[trim = 0 0 0 0, clip,width=0.75\textwidth]{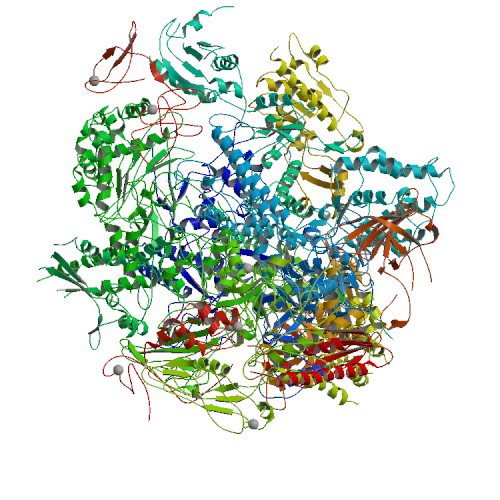}
\end{center}
\caption{Biological Assembly 1 of 1WCM - Complete 12-subunit RNA polymerase II at $0.38$ nm, from \cite{PDBR}.}
\label{bio}
\end{figure}
%

%

In the following we will discuss some details on 2D detector requirements in the context of our proposed single biomolecule imaging experiment at the SPB instrument. The results of some basic estimations lead to the specification of detector layout and geometry. We will conclude presenting the characteristics of simulated diffraction data, and showing that a sufficient number of average counts per pixel is expected to grant reconstruction.

Assuming with good approximation transverse coherence of the incident radiation, radiation diffraction from an object of maximum width $w$ produces fringes of finest period $2\pi/w$ in the reciprocal space.  The Fourier transform of the diffraction pattern, that is the autocorrelation function, has a maximum width $2 w$, that is twice the original object size. Then, from Shannon's sampling theorem follows that the intensity in reciprocal space is completely determined by sampling the diffraction pattern at least twice every $\pi/w$ period, in an equally-spaced way. It makes therefore sense to define the Shannon interval for an intensity pattern $\Delta q_s = \pi/w$. The choice of a certain a resolution $d$, requires furthermore that one  records the diffraction pattern up to $q_\mathrm{max} = 2 \pi/d$. It follows that one needs to measure a number of samples equal at least to $N_s = 2q_\mathrm{max}/\Delta q_s = 4w/d$, assuming that the diffraction pattern is measured from $-q_\mathrm{max}$ to $q_\mathrm{max}$. Fig. \ref{samsch} shows the experimental geometry in real and reciprocal space. If one calls with $2 \theta$ the scattering angle between the incident direction $\vec{K}_i$ and the diffraction direction $\vec{K}_f$, one obtains that the magnitude of the momentum transfer $\vec{q} = \vec{K}_f - \vec{K}_i$ is given simply by $q = 4\pi\sin(\theta)/\lambda$. Due to the conservation of the wavevector module $K = 2\pi/\lambda$, which is enforced due to the fact that we are considering elastic scattering, the momentum transfer vector $\vec{q}$ connects two points of the surface of a sphere, universally known as the Ewald sphere, Fig. \ref{samsch}(b). The maximum achievable resolution at a certain wavelength $\lambda$ and at a distance $z$ between sample and detector is limited by the geometry and the size of the detector. Assuming a plane geometry and a total detector size $D$, one easily obtains that the  geometrically-limited resolution  $2\pi/d$ satisfies the equation

\begin{eqnarray}
\frac{2\pi}{d} = q_\mathrm{max} = \frac{4\pi}{\lambda} \sin(\theta\mathrm{max}) = \frac{4\pi}{\lambda} \sin\left[\arctan\left(\frac{D}{2z}\right)\right]~,
\label{reso}
\end{eqnarray}
The calculations in this article were carried out for a photon energy of $4.1$ keV, corresponding to a wavelength $\lambda = 0.3 \mathrm{nm}$. For reference, note that in order to reach the resolution $2\pi/(0.4 ~\mathrm{nm})$ with a $200$ mm by $200$ mm detector, the sample to detector distance needs to be as short as $10$ cm. For this resolution, and for a molecule size of $w= 10 ~\mathrm{nm}$, we estimate the requirement for the number of Shannon pixels as $N_s = 4w/d = 100$. In this case, the average size of a Shannon pixel can be estimated as $D/N_s = 2$ mm.

The Adaptive Integrating Detector (AGIPD) \cite{AGIP}, which will be installed at the SPB instrument, will serve as main 2D detector from the beginning of the operation phase \cite{MANC,MANT}. The AGIPD detector features a pixel size of 0.2 mm and frame of 1 megapixel, amounting to a 200 mm by 200 mm size.  Summarizing, the geometrical  parameters of the baseline detector for the SPB station are well within the requirements for single biomolecule imaging experiments with resolution $2\pi/(0.4 ~\mathrm{nm})$. However, our calculations assume that the detector can be placed at the necessary propagation distance ($10$ cm) to realize the desirable resolution.

The numerical simulations carried out here are based on some simplifying assumptions. In particular, noise is only considered in terms of photon noise, i.e. we assume Poisson shot noise. No additional sources of noise such as detector readout noise are considered. A quantum efficiency of about $85\%$ is assumed for the AGIPD detector, for standard window and at the photon energy of $4$ keV.

\begin{figure}
\begin{center}
\includegraphics[trim = 0 0 0 0, clip,width=0.30\textwidth]{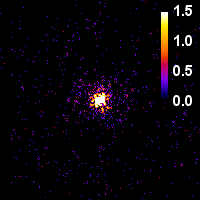}
\includegraphics[trim = 0 0 0 0, clip,width=0.30\textwidth]{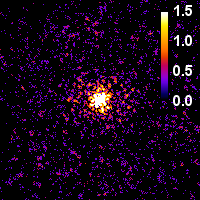}
\includegraphics[trim = 0 0 0 0, clip,width=0.30\textwidth]{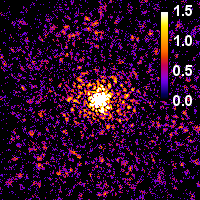}
\end{center}
\caption{Simulated diffraction data from the RNA pol II test object. The simulations were performed at a photon energy of $4.1$ keV. The detector considered here is $200$ mm by $200$ mm in size, and is located at a distance of $100$ mm from the sample. The detector rim corresponds to $0.39$ nm resolution. The Shannon pixel size is $2$ mm. The pixel array has dimensions 200 by 200, corresponding to sampling $s = 2$ and binning $b = 5$. The intensities are displayed on a logarithmic scale. Left plot: Diffraction pattern of the test object at a fluence of $0.15 \cdot 10^{23} \mathrm{photons}/\mathrm{cm}^2$, which can be achieved without changes to the SPB beamline design. Center plot: Diffraction pattern of the test object at a fluence of $0.5 \cdot 10^{23} \mathrm{photons}/\mathrm{cm}^2$, which can be achieved by changing the mirror specification in the SPB beamline. Right plot: Diffraction pattern of the test object at a fluence of $1.0 \cdot 10^{23} \mathrm{photons}/\mathrm{cm}^2$, which can be achieved with the installation of additional refocusing mirrors in the SPB beamline.}
\label{diffp}
\end{figure}

\begin{figure}
\begin{center}
\includegraphics[trim = 0 0 0 0, clip,width=1.0\textwidth]{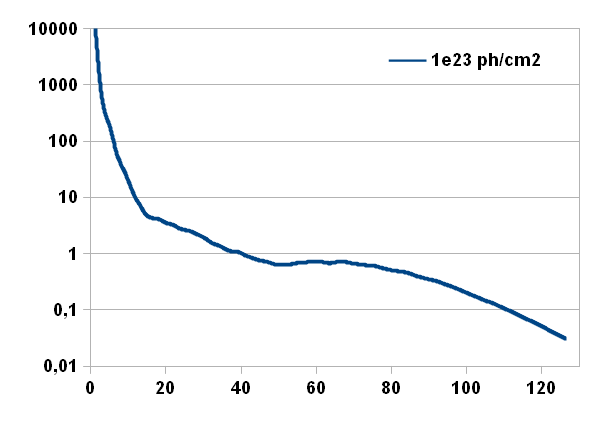}
\end{center}
\caption{Simulated diffraction data from RNA Pol II test object. The plot show the radial average
of the photon counts per pixel (vertical axis) plotted against the position along the detector in $mm$, starting from the center of the detector. The simulations were performed at a photon energy of $4.1$ keV. The detector considered here is $200$ mm by $200$ mm in size, and is located at a distance of $100$ mm from the sample. The detector rim corresponds to $0.39$ nm resolution. The pixel size is $1$ mm. The pixel array has dimensions 200 by 200, corresponding to sampling $s = 2$ and binning $b = 5$. The data correspond to $10^{23} \mathrm{photons}/\mathrm{cm}^2$, a fluence which can be achieved with moderate modifications of the SBP X-ray optics.}
\label{phcount1}
\end{figure}

\begin{figure}
\begin{center}
\includegraphics[trim = 0 0 0 0, clip,width=1.0\textwidth]{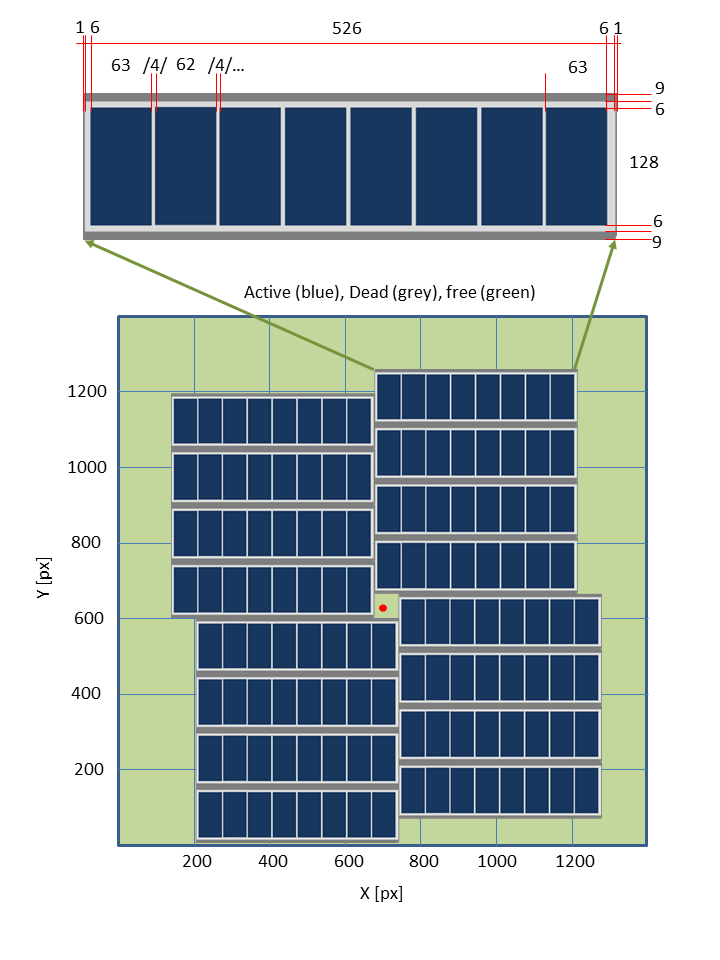}
\end{center}
\caption{Schematic of the layout of the AGIPD detector. Here we consider a basic pile design with four independently movable detector quadrants, each adjustable in the vertical and horizontal directions. Each quadrant consists of four modules aligned with their long side in the horizontal direction. The central hole is assumed to remains square and can be opened from 1 mm up to 1 cm. All units are in detector pixels (adapted from \cite{MANT}).}
\label{mask}
\end{figure}

\begin{figure}
\begin{center}
\includegraphics[trim = 0 0 0 0, clip,width=0.75\textwidth]{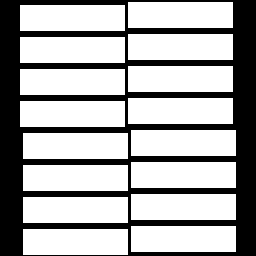}
\end{center}
\caption{A numerical mask was introduced in all simulated diffraction patterns in order to account for the modular layout of the AGIPD detector.}
\label{mask2}
\end{figure}
\begin{figure}
\begin{center}
\includegraphics[trim = 0 0 0 0, clip,width=0.75\textwidth]{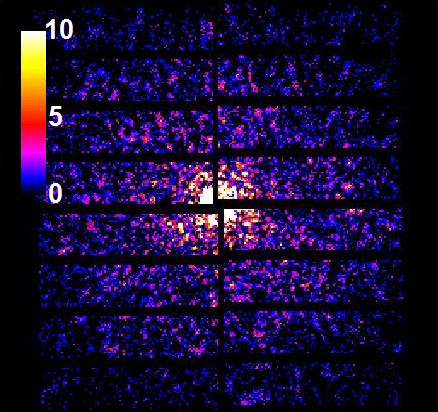}
\end{center}
\caption{Simulated diffraction pattern from the RNA pol II test object
as seen by AGIPD detector at a fluence of $10^{23} \mathrm{photons}/\mathrm{cm}^2$.}
\label{diffdet}
\end{figure}
We simulated $30000$ randomly oriented diffraction patterns for the RNA Pol II structure. The original structure is shown in Fig. \ref{bio}. Typical diffraction patterns from a single FEL pulse, simulated for the experimental conditions described above, are shown in Fig. \ref{diffp}. The left, center and right plots respectively refer to fluences of $0.15 \cdot 10^{23} \mathrm{photons}/\mathrm{cm}^2$, $0.5 \cdot 10^{23} \mathrm{photons}/\mathrm{cm}^2$, and  $1.0 \cdot 10^{23} \mathrm{photons}/\mathrm{cm}^2$. The first fluence can be achieved without changes to the SPB X-ray optics. The second fluence can be achieved by changing the mirror specification in the SPB beamline. The third fluence can only be achieved with additional installation of special refocusing mirrors in the SPB beamline.

The plot in Fig. \ref{phcount1} shows the radial average of the photon count. The simulated array size was $200$ by $200$ pixels, with sampling ratio per dimension of $s = 2$. Note that larger molecules do not necessary give larger signals; there are a fixed number of photons per pulse and larger molecules require a proportionally larger focal spot size, hence giving a lower fluence. The plot in Fig. \ref{phcount1} demonstrates that a signal of the order of 0.1 photons per pixel, corresponding to $0.4$ photons per Shannon angle can be achieved even without changes to the SPB beamline design.

The basic layout of the AGIPD detector is illustrated in Fig. \ref{mask}. In practice, some "dead" area is present in  any detector.  As a result, it is important to answer two main questions concerning the retrieval process with the AGIPD detector, namely, how well does our 3D assembly and retrieval algorithm work  in the presence of the detector mask in Fig. \ref{mask2}, and what error can be expected in  the retrieved electron density.  For the present detector layout, almost $30 \%$ of the detector area is in the "dead" area. A central hole with a size of about $3$ mm, covering $1.5$ speckles, was introduced in all simulated diffraction patterns, and this region was excluded from the calculation, Fig. \ref{mask2}. A typical diffraction pattern from a single FEL pulse as seen by AGIPD detector is shown in Fig. \ref{diffdet}. Unfortunately, the $6$ mm gap width between two detector modules  corresponds, in our case of single biomolecule imaging, to the size of three Shannon pixels. In the follow-up of this work we will include the effect of the mask in the 3D assembly and retrieval. In this way we will be able to study the algorithm performance when the mask is present, and to determine the price to be paid  in accuracy and statistics.

The main challenge related with detector electronics is the ability of distinguishing small signals from noise. Charge amplification is always accompanied by noise, and the main source of irreducible noise is thermal noise. Noise is usually expressed in terms of equivalent noise charge (ENC), which is defined as the hypothetical charge that, produced in the detector, would give a peak output response equal to the rms value of the actual noise. Results of simulations for the AGIPD detector presented in \cite{AGIP} allow an estimation of the total Equivalent Noise Charge (ENC) around 300 electrons, and single photon energy deposition $N_e$ of about $1200$ electrons at $4$ keV photon energy. Paraphrasing \cite{AGIP}, for a Gaussian-distributed random noise in a pixel, a reasonable approximation for the expected number of false positives per pixel, is given by

\begin{eqnarray}
N_\mathrm{false} = \frac{1}{2} \mathrm{erfc}\left(\frac{\mathrm{Th} \cdot N_e}{\sqrt{2} \mathrm{ENC}}\right) ~.
\label{Nfalse}
\end{eqnarray}
Here $\mathrm{Th}$ is the threshold of the discriminator, relative to the single photon electron deposition $N_e$. In other words if, for example, $\mathrm{Th}$ is set to unity, only an electron deposition of $N_e$ electrons or higher is considered a photon hit by the detector electronics. Qualitatively, if the threshold is set too low, many false events due to noise are recorded. If it is set too high, many actual events are not recognized due to both photon shot noise and detector amplifier noise. Therefore, one should set $\mathrm{Th}$ as low as possible, but still keeping $N_\mathrm{false}$ well below the average photon count per detector pixel, of size $0.2$ mm by $0.2$ mm.

\begin{figure}
\begin{center}
\includegraphics[trim = 10 330 0 0, clip,width=0.75\textwidth]{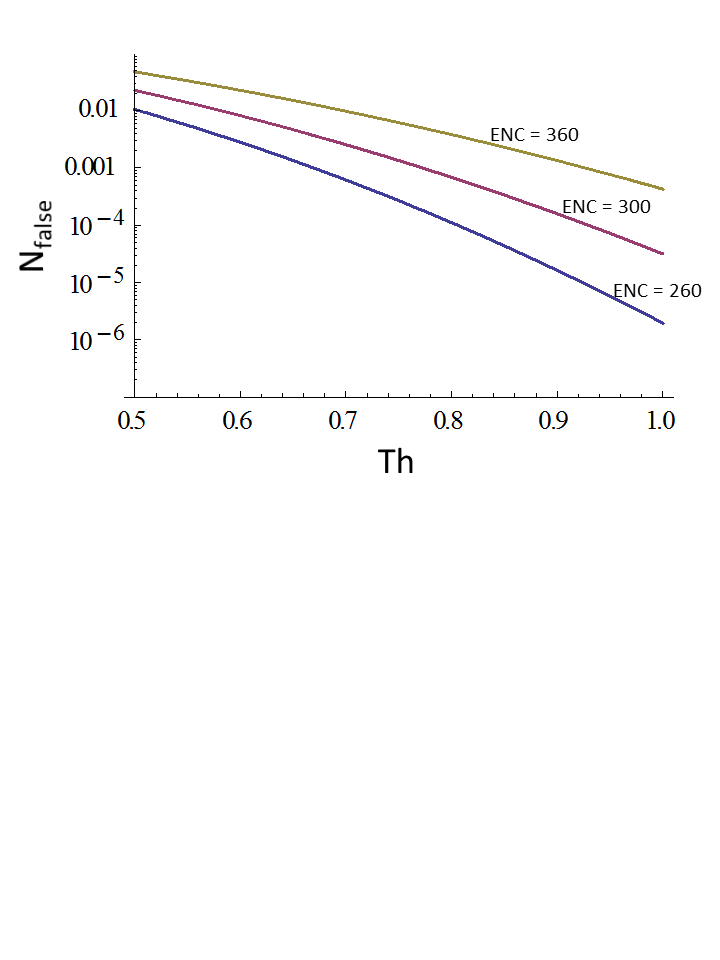}
\end{center}
\caption{Number of false events as a function of the relative threshold $\mathrm{Th}$, estimated according to Eq. (\ref{Nfalse}). Different curves refer to different values of ENC. Here we assumed $N_e = 1200$.}
\label{Falsehits}
\end{figure}
We plot number of false events as a function of the relative threshold $\mathrm{Th}$, estimated according to Eq. (\ref{Nfalse}). Different curves refer to different values of ENC. Here we assumed $N_e = 1200$.
For example, setting $\mathrm{Th} = 0.9$, one sees that the expected number of false hits per detector pixel ($0.2$ mm by $0.2$ mm) can be estimated as $N_\mathrm{false} \sim 10^{-5}$, and can be neglected.

\section{\label{sec:test} Outlook for data processing. Reconstruction algorithm for single-molecule diffraction imaging experiments}

In this work we considered the start-to-end simulations for a single-molecule imaging experiment up to the production of diffraction data. In a natural follow up of this work, which will be published separately, we will deal with the data processing, leading to the sample reconstruction. In this Section we give an Outlook of the final data analysis step.

In a typical single-molecule imaging experiment, a sample with unknown orientation is injected into the focused, coherent X-ray beam of an XFEL. The scattered radiation pattern is then measured in the far field by a 2D detector.  The diffraction pattern can be considered as a perspective projection of an Ewald sphere sector onto the 2D detector plane as viewed from the  sample position. Each measured diffraction pattern corresponds to the unknown molecule orientation. Image reconstruction requires (1) image orientation and assembly of the three-dimensional data set, (2) signal averaging to reduce the effects of noise, and (3) phase retrieval. Several approaches have been proposed so far to find an unknown molecule orientation in such experiments.

One is based on common arc algorithm \cite{HULD}, originally developed for electron microscopy \cite{FRANK,HEEL}. The main problem of this method is its demand for a high signal-to-noise ratio, which is difficult to satisfy even exploiting the scheme for an high-power XFEL source proposed in this paper. It was suggested to overcome this limitation by an additional classification step \cite{HULD,BORT}, in which diffraction patterns arising from similar molecule orientations are averaged prior to orientation determination. This step improves the statistics of each averaged diffraction pattern, but at the same time reduces its contrast and, as a result, decreases the achievable resolution. Other methods are based on generative topographic mapping \cite{FUNG,SCHW} and expectation maximization technique \cite{LOHN}. These approaches work well for a low signal-to-noise ratio, but are expensive as concerns requirements of computational time and memory.

In order to proceed with the electron density reconstruction, we can use a recently proposed method of orientation determination, based on an improved common arc algorithm \cite{YEFA}. Instead of performing the additional classification step described above, we make a simultaneous analysis of common arcs between many diffraction patterns. In order to improve the quality of the orientation determination, a 3D angular refinement procedure is applied at the final step. This method is inspired by a technique employed in single-particle electron cryo-microscopy \cite{VAIN,HEL2} and works well even with a low photon signal, down to $0.5$ photons per Shannon pixel. When the relative angular orientation of all diffraction patterns is determined, the full 3D intensity distribution in reciprocal space can  be obtained. The structural information, or electron density of the sample can be finally determined by phase retrieval \cite{FINE,MARC}.

\subsection{\label{sub:3D} 3D assembly of diffraction data  }

When two independent measurements of identical molecules with different orientations are considered, the orientation of the first molecule can be fixed as known. The orientation of the second molecule can be uniquely described relative to the first one. Alternatively, two measured diffraction patterns could be considered to originate from the same molecule in different experimental geometries.

\begin{figure}
\centerline{\includegraphics[width=0.75\textwidth]{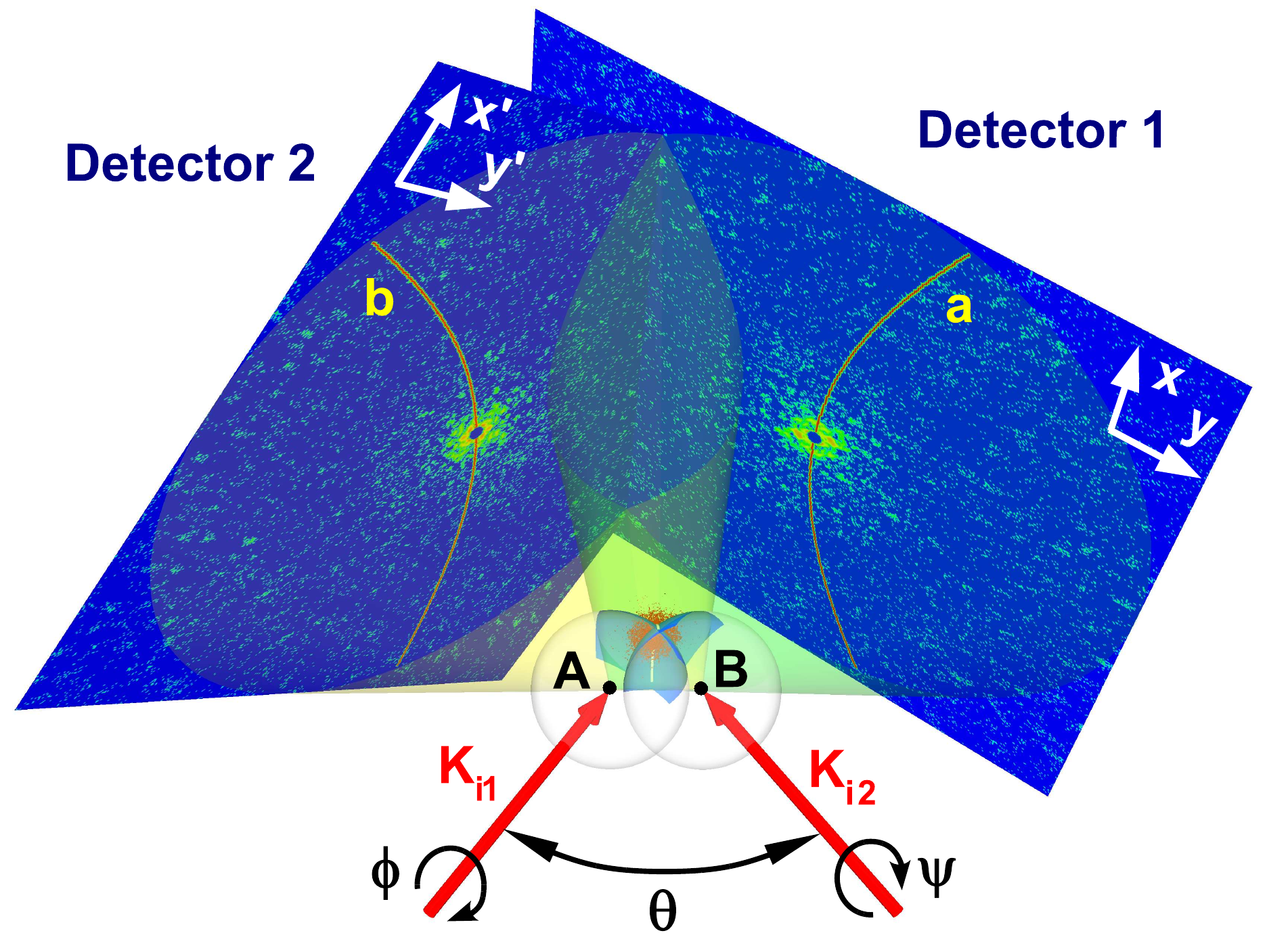}}
\caption{(Color online) Measurements of two reproducible samples at random orientation can be considered as two measurements of the same sample with two different incident beam directions indicated by vectors $\vec{K}_{i1}$ and $\vec{K}_{i2}$. Angles $\phi, \theta, \psi$ are Euler's rotation angles. Points A and B are the centers of the corresponding Ewald's spheres. Coordinates on the first and second detector are indicated as $x ,y$ and $x', y'$, respectively. Adapted form \cite{YEFA}.}
\label{2Ewalds}
\end{figure}
In this case, the molecule orientation is fixed, but the direction of the incident beam and the detector orientation are different for each measurement as shown in Fig. \ref{2Ewalds}. For the first measurement, the incident beam direction, given by its wave vector $\vec{K}_{i1}$, can be taken along the $q_z$ axis in the reciprocal space coordinate system shown in Fig. \ref{samsch}(b).  The direction of the incident
beam for the second measurement is given by its wave vector $\vec{K}_{i2}$, Fig. \ref{2Ewalds}. The
relative orientation of the second geometry with respect to the first one can be described by the three Euler angles $\phi, \theta, \psi$. With reference to Fig. \ref{samsch}(a), and Fig. \ref{2Ewalds}, in real space $\phi$, $\theta$ and $\psi$ represent three consecutive rotations around the $z$ axis, the line of nodes (not shown in figure) and the new $z$-axis (defined after the second rotation) respectively. Note that in reciprocal space, rotations of angles $\phi$ and $\psi$ are equivalent to rotations by the same angles of the detectors in real space, Fig.  \ref{2Ewalds}.  The Ewald sphere has radius $K = 2\pi/\lambda$, where $\lambda$ is the wavelength of the incident radiation. Both  spheres corresponding to the two measurements pass through the origin of the reciprocal space coordinate system, Fig.  \ref{2Ewalds}. The center  of the first sphere is at point A, and has coordinates $(0,0,-K)$. This means that the center of the first detector also corresponds to the origin $(0,0,0)$ of the reciprocal space. The center of the second sphere is at point $B$, and has coordinates $(q_{x0}, q_{y0}, q_{z0})$, which are determined by a rotation of the point $(0, 0, -K)$ around the origin of the reciprocal space $(0, 0, 0)$ by the angles $\phi$ and $\theta$. The intersection of the two spheres is a common arc that also passes through the origin of the reciprocal space. This common arc can be projected on the two detectors (curves a and b in Fig. \ref{2Ewalds}). It is clear that the intensity along these arcs must be the same at both detectors. By analyzing the intensity correlations along all possible common arcs, the unique relative orientation of the two measurements can be determined.

\begin{figure}
\centerline{\includegraphics[width=0.75\textwidth]{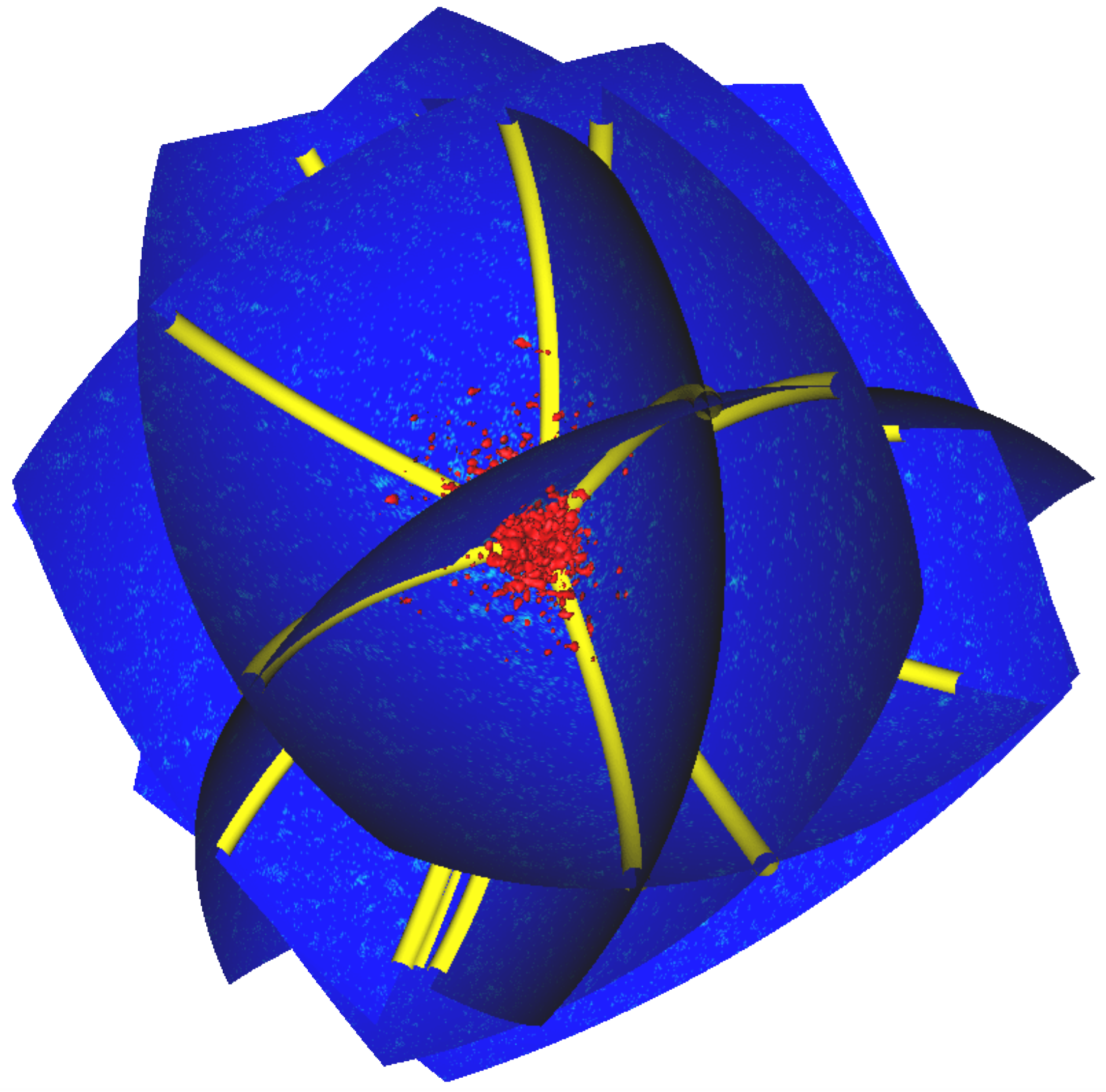}}
\caption{(Color online) Few Ewald sphere sectors intersecting the 3D intensity distribution of the sample in reciprocal space. Yellow lines indicate common arcs between different patterns. Adapted form \cite{YEFA}.}
\label{ManyEwalds}
\end{figure}

In our case we deal with $N_d \simeq 3\cdot 10^4$ diffraction patterns. One can proceed similarly as described above. A first pattern, which we define as base pattern, is fixed. Then, performing the steps described above $N_d-1 \simeq N_d \simeq 3\cdot 10^4$ times, one can find the orientation of all other patterns with respect to the base. This is the essence of the common arc algorithm. This algorithm, however, performs well only for diffraction data sets with high signal-to-noise ratio, where the analysis of the intensity correlation gives clear results. One way to overcome this limitation is to exploit an advanced algorithm for orientation determination in the presence of noise \cite{YEFA}. First we should note that each diffraction pattern has a common arc with any other diffraction pattern (see Fig. \ref{ManyEwalds}). Therefore, common arcs between all patterns could, in principle, be analyzed simultaneously.  Such analysis can significantly improve the fidelity of the orientation; however, in practice it requires an increase in computation resources. In fact, one now needs $N_d \cdot (N_d -1) \simeq N_d^2 \simeq 9\cdot 10^{8}$ common-arc calculations. A compromise can be found by implementing the following strategy. As a first step, elements belonging to a certain subset of $N_b$ base diffraction patterns are analyzed with respect to each other using the common arc algorithm, to determine the correct orientations of these chosen patterns. We consider, for example, the case $N_b = 64$. This step takes $N_b \cdot (N_b - 1) \simeq 4\cdot 10^3$ common-arc calculations. In the next step all other patterns are oriented with respect to each of the patterns in this subset. The second step thus takes $(N_d - N_b) \cdot N_b \simeq 1.9 \cdot 10^6 $ common-arc calculations. The total number of common-arc calculation amounts, therefore, to $N_b \cdot (N_d -1) \simeq N_b \cdot N_d \simeq 1.9 \cdot 10^6 $. Summing up, this implementation requires $N_b$ times more calculations compared to a single base pattern. In the final step all intensities are mapped to a 3D array of voxels in reciprocal space by 3D gridding and averaging procedure. The benefit of this approach is the possibility of solving the orientation problem for noisy data.

\begin{figure}
\centerline{\includegraphics[width=0.75\textwidth]{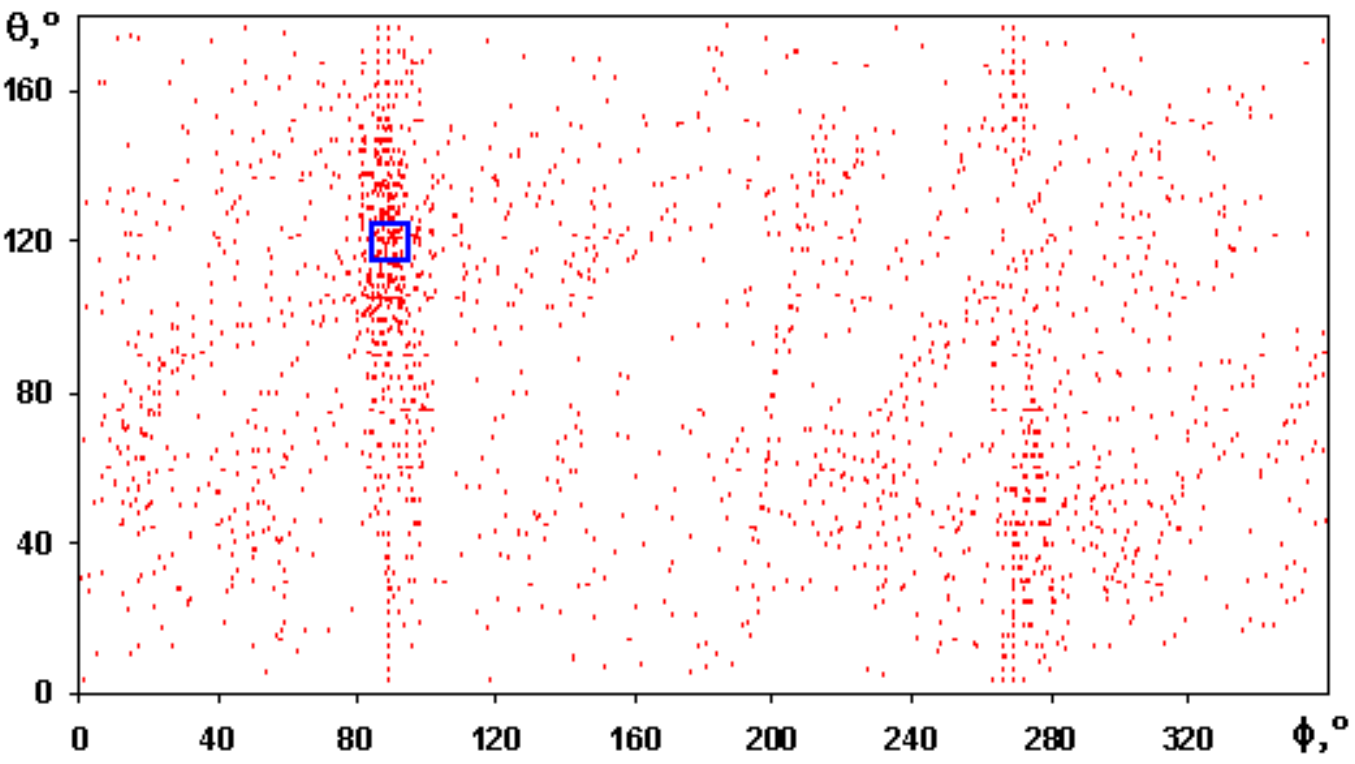}}
\caption{Two-dimensional $(\phi, \theta)$ distribution of angles corresponding to good correlation between a pattern and all base patterns. A blue box correspond to the region with highest density of good orientations. The number of base patterns here is $N_b = 64$, while $N_a = 30$. Adapted form \cite{YEFA}.}
\label{Plot2Dangles}
\end{figure}

It should be noted that the best correlation coefficient between two noisy patterns can correspond to a completely wrong orientation. Therefore, in practice, several orientations $N_a$ corresponding to the best set of correlations must be
stored. Each pattern has therefore $N_b \cdot N_a$ "best" angles with respect to all other base patterns. Then, all these angles are recalculated with respect to one selected pattern (with  all angles set equal to zero). Finally, all these angles determined for each pattern can be plotted in a 3D space of rotation angles $(\phi, \theta, \psi)$, and the angular region with the maximum density of points can be selected as the best angle. In Fig. \ref{Plot2Dangles}, an example of such operation in the case for varying angles $\theta$ and $\phi$ and with fixed $\psi$ is shown. The best angle corresponds to the middle of the bue box in Fig. \ref{Plot2Dangles}.

In real experimental situations, all diffraction patterns have different intensities due to the fact that each injected molecule is hit by a different part of a focused beam. As a consequence, all measured diffraction patterns have to be rescaled. This is implemented in the algorithm discussed in \cite{YEFA} by exploiting once more the fact that each pair of diffraction patterns has a common arc, and that the intensities along this arc must be equal. The scaling factor for the intensities between two patterns can thus be determined by taking the ratio of the intensities along such common arc. Having all information about the experimental geometry, orientation and scaling factor for each pattern, the 3D intensity distribution in reciprocal space can be constructed.

Independently of the method used to reconstruct the 3D intensity distribution in the reciprocal space, the orientation determination can be significantly improved by an additional refinement that is based on the correlations between an individual pattern and the whole 3D intensity distribution. It can be implemented in the following way. First, the 3D intensity distribution is obtained from all diffraction patterns except one. Then, the orientation of the selected diffraction pattern is varied in a small angular range and the correlation between this 2D pattern and the whole 3D intensity distribution is analyzed. The orientation corresponding to the highest correlation value is considered to be the correct one. Finally, the rescaled intensity of the selected pattern with the refined orientation is included, and another diffraction pattern is excluded, thus obtaining a new 3D intensity distribution. The refining procedure is then repeated. Proceeding iteratively and refining the orientation of all diffraction patterns, the final 3D intensity distribution is finally obtained.

\subsection{\label{sub:rec} Reconstruction of the electron density  }

The 3D electron density is related to the measured intensities in reciprocal space. However, while the modulus of the Fourier amplitudes of the electron density can be directly obtained from measured intensities, the phases are missing. This lack of information is a well-known issue in diffraction imaging, and goes under the name of phase retrieval problem. To solve it, some additional information is required.

The phase retrieval problem in single molecule imaging can be described as a search for an object that fulfills two constraints. The first constraint is given by the experimental data. The amplitudes of the Fourier transform of the object should match the square root of the measured intensities. This constraint is referred to as the "Fourier-space constraint". The second constraint is given by the oversampling in Fourier space. In real space, this constraint is simply a limit to the size of the molecule, that is its support. The real space can therefore be split in two regions: one region where the object is allowed to have non-zero density, and one region that must be empty. Because of this fact, the oversampling limit is called "the real-space constraint".

Moreover, if the imaging experiment occurs at photon energies far from atomic resonances, Friedel's law applies, and the object can be assumed to be purely real with very good approximation. This extra constraint can aid phase-retrieval algorithms. In fact, in this case the retrieved object in real space corresponds to the electron density function. This means that, when Friedel's law applies, one can automatically impose positivity of the retrieved object in real space. A common way to estimate the shape of the support of an unknown object is through calculations of the autocorrelation of the object, which is the object cross-correlation  with itself. It can be calculated from the diffracted intensities alone. In fact, using the the convolution theorem one can show that the object autocorrelation is just the inverse Fourier transform of the diffraction intensities, and gets an envelope that is twice as large as the object.

Reference \cite{MAIA} provides a comprehensive review of the methods of solution of the phase retrieval problem in single particle diffraction analysis, which we summarize down here.

The phase retrieval is typically based on a kind of Iterative Transform Algorithms (ITA). ITA were originally proposed by Gerchberg and Saxton \cite{GERC} as a way to retrieve the phase from two intensity measurements (in real and reciprocal space). The work by Gerchberg and Saxton was used and modified by Fineup \cite{FIN1}, who introduced the so-called error reduction algorithm  to retrieve real space data from the knowledge of the intensity in reciprocal space, and of a support function in real space. This kind of algorithms cycles between real and reciprocal space, enforcing known constraints. Once a support is constructed, for instance but not necessarily from the object autocorrelation, and an initial random phase set is added to the measured intensity in reciprocal space, one can construct a first guess of the object by Fourier transforming the data from reciprocal space to real space, subsequently setting pixels outside of the support to zero, and possibly using positivity. Then, transformation to reciprocal space follows, and one loops further until the object is recovered.  Since only the modulus of the Fourier transform of the object is measured, the convergence of the algorithm to a unique solution, that is the global minimum of a properly defined error metric, is a question of fundamental importance. When the object is complex, the reconstruction is not trivial. In fact, in that case, the algorithm usually gives good results, but the uniqueness of the solution is not granted. However, when an object is endowed with the property of nonnegativity, the application of iterative methods is drastically simplified. Electron density is obvious a a real positive function. In this case negative electron density, and/or its imaginary part is set to zero inside the support, iteration by iteration. In general, the main limit of the error reduction algorithm, is that it sometimes stagnates near local minima of the error metric, failing to find the global minimum.

This limitation is avoided by the so-called Hybrid Input-Output (HIO) algorithm \cite{FINE}, which is probably the most popular algorithm for phase retrieval. Following Levi and Stark \cite{LEVI}, one can interpret images in real space as vectors in a high-dimensional Hilbert space, with one dimension per pixel. Then, iterations in real and reciprocal space of any phase retrieval algorithm can be interpreted as particular projections onto hypersurfaces, in the Hilbert space, defined by the constraints in the real and in the reciprocal space. Adopting this point of view, the HIO algorithm differs from the error reduction algorithm for a different real-space projection.

As stated above, the HIO algorithm is quite good at getting out of local minima of the error metric, and returns very satisfactory reconstructions when a tight support is known. However, for our application one can only provide, a priori, a rough guess of the support. The most promising approach to overcome this obstacle is the so-called Shrinkwrap algorithm,  introduced by Marchesini in \cite{MAR2}. In this case the initial support function, derived from the autocorrelation, is refined during the reconstruction procedure every few iterations. In other words, the Shrinkwrap algorithm exploits the object features recovered during these few iteration in order to improve the guess on the support.

There is undoubtedly a very large number of methods of image reconstruction from diffraction patterns. However, as remarked in \cite{MAIA}, there seems to be, to date, only one open-source software for single-molecule diffraction analysis, the package Hawk \cite{HAWK}. Yet, in our case, such package is not suitable. In fact, the final step of the advanced common arc algorithm, all intensities are mapped to a 3D array of voxels in reciprocal space by 3D gridding and averaging procedure. Therefore we will recover the phases  exploiting the Shrinkwrap algorithm combined with the HIO algorithm using the in-house code RECON \cite{RECO}. The reconstructed object will be constrained to be real and positive. The volume constraint will be based on an estimate from the object autocorrelation.

\section{\label{sec:conc} Discussions and Conclusions}

The imaging method "diffraction before destruction" promises to be a revolutionary technique for protein determination, capable of resolving the structure of molecules that cannot crystallize.  Nevertheless, being a novel concept, it should be first demonstrated experimentally, before it can be actually relied upon in the upgrades of the European XFEL and other XFEL facilities. Here we propose a cost-effective proof-of-principle experiment, aiming to demonstrate the actual feasibility of a single molecule diffraction experiment using the baseline European XFEL accelerator complex and the SPB beamline hardware, with minimal modifications only. More specifically, we want to determine the structure of a relatively small (about $30000$ non-hydrogen atoms), well-known protein molecule and compare it with results in the protein data bank. As a first step towards the realization of an actual experiment, we developed a complete package of computational tools for start-to-end simulations predicting the performance of this experiment. The problem can be naturally separated in two parts. Part I, described in this work, deals with the simulation of the experiment from the photocathode of the European XFEL injector up to the collection of noisy 2D diffraction data set.  Part II, which will be published in a separate work, will deal with the data processing, leading to the sample reconstruction.

The main complication related to the proof-of-principle experiment that we are proposing is already known from bio-imaging experiments at the LCLS:  even without accounting for non-ideal effects like background and intrinsic detector noise,  the diffraction signal for a typical baseline mode of operation of an XFEL  is very small. In other words, experiments at the LCLS confirmed that the number of photons per Shannon pixel per single shot is too small to perform 3D assembly of diffraction data of a single biomolecule imaging.  There are no big differences
between the baseline output characteristics of the LCLS and of the European XFEL except the time diagram. However, the higher repetition rate of the European XFEL hardly helps in the assembly process of 3D diffraction data. A variety of methods has been proposed to overcome this difficulty. One of these consists in the development of a software capable of working even
with a very low photon signal, down to $0.05$ photon per Shannon pixel \cite{LOH1}. At this moment, such software is not publicly available, and in this article we refer to an in-house software based on a variant of the common arc algorithm, which works well down to 0.5 photon per Shannon pixel only.

The largest diffraction signals are achieved at the longest wavelength that supports a certain required resolution, which should be better than 0.3 nm for imaging of single molecules. This suggests that the ideal photon energy range for a proof of principle experiment spans between 3 keV and 5 keV. In this article we show how small modifications of the European XFEL baseline mode of operation allow to reach, for our proof-of-principle experiment, a diffraction signal up to 0.5 photons per Shannon pixel at the SPB beamline assuming realistic estimates for beamline and focusing efficiency. In order to overcome difficulties with the low diffraction signal, we propose here a straightforward method  to increase the XFEL peak output power, based on  self-seeding and undulator tapering, which greatly improves the FEL efficiency: parameters of the accelerator complex and availability of long  baseline (SASE1) undulator at the SPB beamline offers the opportunity to build 1.5 TW power source around $4$ keV. This method has become the foundation for the proposed proof-of-principle experiment. Further small changes to the SPB line are expected to lead to an increase in the diffraction signal up to 3-4 photons per Shannon pixel.

Start-to-end simulations allowed us to identify in advance potential bottlenecks for the proposed proof of principle experiment. The first issue is related with the SPB focusing efficiency, and is intrinsic to the design of the SPB beamline. The SPB instrument is designed to work in the photon energy range between 3 keV and 16 keV, but the 100 nm X-ray focusing system is optimized for operation in the hard photon energy range, between 8 keV and 16 keV.  At photon energies around 4 keV, one suffers diffraction effects from mirror apertures due to the large divergence of nominal X-ray beams, leading to a 100-fold decrease in fluence. We propose to overcome this issue  based on a special mode of operation of the accelerator complex that relies on the insertion of a slotted foil into the last bunch compression chicane. This will allow us to avoid the substantial decrease in focusing efficiency around 4 keV without modification of the SPB beamline optics. Nevertheless, our special mode of operation yields respectively about 200 nm and 300 nm vertical and horizontal spot sizes. In other words, compared with the ideal case of a 100 nm focus, we still have a 6 times smaller fluence in the spot area on the sample. As a result, without additional design changes only a maximum detector count of 0.5 photons per Shannon pixel is within reach.  In order to achieve an additional tightening of the focus of about factor 2-3, one needs a slight re-design of the SPB X-ray optics. In this way, with moderate additional costs, one may reach an upper limit for the fluence of about $10^{23} \mathrm{photons/cm}^2$, corresponding to a diffraction signal of 3-4 photons per Shannon pixel. This is the best optimization that can be made on the setup in order to maximize the diffraction signal. Further increase in the fluence will rapidly increase the damage-induced background contribution to the diffraction pattern.

A second issue is related with the minimum achievable distance between sample and detector. On the one hand, the achievable resolution for our proof-of-principle experiment is geometrically limited by the numerical aperture covered by the detector at the certain distance to the sample. For the AGIPD detector, the distance between the outermost pixels, that is the detector size, is of order of 25 cm. On the other hand, a resolution better than 0.3 nm should be provided. This leads to a requirement on the minimum sample-detector distance, which should be shorter than  7 cm at  photon the energy of 4 keV. However, with the current design of the sample chamber, the minimal sample-detector distance achievable is about 13 cm.  As a result, the geometrically-limited resolution of our proof-of-principle experiment with the baseline design of the SPB instrument cannot be better than 0.4 nm.  In contrast, the bio-imaging instrument at the LCLS  operates at significantly shorter sample-detector distance, down to 5 cm and, as a result, provides better resolution even with a smaller detector size of $10$ cm. Our conclusion is that, in order to achieve a resolution better than 0.4 nm in single biomolecule imaging at the SPB beamline, one forcefully needs to upgrade the sample chamber.

A third bottleneck is related with the detector mask. There is a 6 mm gap between two detector modules corresponding to the size of 3 Shannon pixels in our case of a small (10 nm-scale) molecule. To put this issue into context, one shoul remember that all protein molecules with undetermined structure have a size larger than 100 nm. Clearly, for such molecular size, the gap width will correspond to 30 or more Shannon pixels, and retrieval algorithms will hardly convergence in this situation. In other words, while in earlier sections we showed that the AGIPD detector is suitable for the proposed proof-of-principle experiment, we also maintain that its applicability for obtaining new results in the field of single biomolecular imaging remains problematic \footnote{The mask of the AGIPD detector is not an issue, instead, for nanocrystallography applications.}.

Finally, our studies have shown that the diffraction signal contributes only for about 0.005 to 0.05 counts per detector pixel\footnote{Our simulations correspond to a sampling $s =  2$ and a binning $b = 5$.}. This requires that external and intrinsic detector noise must be less than this level. We demonstrated that the intrinsic detector noise is not issue, and that the expected false positives per detector pixel is below $10^{-5}$. In contrast, external background is an issue. From LCLS experiments one can estimate that the external background contributes  for about 1 count per detector pixel. Such background is not a problem for nanocrystallography, but it is unacceptable for single biomolecule imaging. The background at the LCLS mostly originates from photons scattered in the diamond window used for separating the the sample chamber from the KB mirror chamber, and from the gas jet used for sample delivery\footnote{There are two options for sample delivery discussed at the LCLS: gas or liquid jet. The background level discussed here was observed in experiments with gas jets. We thank Anton Barty (CFEL) for discussing these details with us.}. In the SPB instrument  a differential pump will be used in order to separate the sample chamber from the 100 nm KB chamber. The absence of a window can eliminate much of the photon background. However, we may find it necessary  to avoid background from the gas jet even with the use of differential pumping. In this case, it may be feasible to use laser beams for sample injection.

We confirmed by simulations that using the SPB baseline design one can achieve diffraction signal of about $0.5$ photons per Shannon pixel for a relatively small (15 nm size) protein molecule. At this signal level one needs only  a few ten thousands diffraction patterns with different molecule orientations to get a geometrical resolution and achieve full 3D information (assuming negligible detector background). The high repetition rate at European XFEL gives the possibility to accumulate these data during about an hour, even assuming a relatively low hit probability\footnote{The hit probability is the probability for a randomly injected single molecules to be hit by X-rays.} of about a percent.

We emphasize the need for a better theoretical and experimental understanding of background conditions at the SPB  instrument. The results of our proof-of-principle experiment aim to support a future, moderate upgrade of the SPB beamline enabling to increase the diffraction signal of a further factor $6$, in order to obtain novel results in the field of single biomolecule imaging.

The image reconstruction up to the determination of the electron density distribution, to be described in the follow-up of this article, requires image orientation and assembly of the 3D diffraction data set. Several approaches have been proposed so far to find the orientation of an unknown molecule. All these approaches assume that injected molecules are identical, that is the sample is reproducible.  We do rely on the assumption that our choice of a small protein molecule  satisfies well this software-reconstruction constraint.

\section{Acknowledgements}

We are grateful to Massimo Altarelli, Reinhard Brinkmann, Henry Chapman, Janos Hajdu, Viktor Lamzin, Serguei Molodtsov and Edgar Weckert for their support and their interest during the compilation of this work. We wish to thank Andrew Aquila, Anton Barty, Julian Becker, Winfried Decking, Heinz Graafsma, Adrian Mancuso, Filipe Maia, David Pennicard, Liubov Samoylova, Robin Santra, Harald Sinn and Ulrich Trunk for useful discussions. Finally, we are grateful to Edgar Weckert for providing the codes MOLTRANS and CORREL to one of us (O.Y.).

\end{document}